\documentclass[a4paper,11pt]{article}
\pdfoutput=1 

\usepackage{jinstpub} 

\newcommand{\percms}{$\text{cm}^{-2} \text{s}^{-1}$}
\newcommand {\roots}{\mbox{$\sqrt{s}$}}

\title{\boldmath The Potential of a TeV-Scale Muon-Ion Collider}

\author[a,1]{D. Acosta,\note{Corresponding author.}}
\author[b]{E. Barberis,}
\author[b]{N. Hurley,}
\author[a]{W. Li,}
\author[a]{O. Miguel Colin,}
\author[a,c]{Y.  Wang,}
\author[b]{D. Wood}
\author[a,2]{and X. Zuo\note{Now at Karlsruhe Institute of Technology, Germany}}

\affiliation[a]{Rice University,\\ Physics \& Astronomy Department, Rice University, Houston, Texas 77251, USA}
\affiliation[b]{Northeastern University, \\Physics Department, Boston, MA 02115, USA}
\affiliation[c]{University of Science and Technology of China,\\  Department of Modern Physics, University of Science and Technology of China, Hefei, 230052, China}

\emailAdd{dea6@rice.edu}

\abstract{
We propose the development of a novel muon-proton and muon-nucleus collider facility at the TeV scale that is capable of performing precision deep inelastic scattering measurements in new regimes 
and providing a rich program in nuclear and particle physics. Such a facility could seed, or leverage, the development of a muon-antimuon  collider and make use of the existing hadron accelerator infrastructure when sited at a facility such as Brookhaven National Laboratory, Fermilab, or CERN. We discuss the possible energy and luminosity design parameters for several collider configurations, and illustrate the science potential with several studies on deep inelastic scattering kinematics, Higgs and vector boson production, top quark production, and beyond Standard Model leptoquark production. Detector design considerations and a possible road map toward development are also given.
}


\arxivnumber{2203.06258} 

\proceeding{2021 US Community Study on
the Future of Particle Physics\\
  July 17-26, 2022\\
  University of Washington}

\begin{document}
\maketitle
\flushbottom

\section*{Executive Summary}

We propose the development of a muon-proton and muon-nucleus 
collider, referred to as a ``muon-ion collider'' (MuIC),
that utilizes the existing hadron accelerator facilities at Brookhaven National Laboratory (BNL), Fermilab, or CERN while seeding, or leveraging, the development of a high-energy muon storage ring at the same site. A center-of-mass energy at the TeV scale is achieved when a TeV muon beam is brought into collision with a hadron beam of hundreds of GeV to several TeV.
Muon collider technology has been considered as an avenue toward 
reaching the next high energy frontier of particle physics with a relatively compact machine footprint. 
Despite its advantages, it remains challenging to achieve a
high energy, high-luminosity muon
collider because of the short muon lifetime. 
The efficient muon cooling in six dimensions 
(spatial and momentum) and the fast ramping of the muon energy
are among the most critical elements for its realization, necessitating an R\&D program in the U.S and at CERN through the Muon Accelerator Program (MAP) and the international muon collider collaboration (IMCC), respectively.

This proposal enables deep inelastic scattering  measurements in new regimes at low parton momentum fraction $x$ and high squared four-momentum transfer $Q^2$, which will further elucidate the structure of the proton and nuclei as well as provide precision QCD and electroweak measurements. We note that these measurements also lay the groundwork for precision measurements made at future high energy hadron colliders, such as the FCC-hh, much as HERA data improved LHC calculations. 
The TeV energy scale at the MuIC also allows for the direct production of Standard Model vector gauge bosons, Higgs bosons, and top quarks in a way complementary to hadron colliders, which provides further sensitivity to the electroweak sector. 
While similar in scientific potential to the proposed Large Hadron electron Collider (LHeC), the MuIC offers complementary scattering kinematics and complementary sensitivity  with a muon beam to beyond Standard Model processes. The MuIC also could provide polarization of both beams (when utilizing the BNL facility) and provide lepton-proton and antilepton-proton collisions with similar luminosity. 

The possible configurations and design parameters of muon-ion colliders are explored here along with some representative physics process studies. For muon-proton collisions, a  center-of-mass energy of up to 1~TeV at BNL and 6.5~TeV at CERN can be achieved, with a estimate on the achievable luminosity ranging from 10$^{33}$ to 10$^{34}$~\percms.  This should yield enough integrated luminosity to explore QCD phenomena at extreme parton densities, the electroweak and QCD couplings up to the highest $Q^2$ reach in deep inelastic scattering, and to measure the Higgs boson cross sections in its largest branching fraction decay modes through the vector boson fusion processes. Sensitivity to physics beyond the Standard Model is also feasible, particularly for lepton-flavor violating processes, of which we study  leptoquark interactions as a case study. 

Experimentally, the final state products other than the scattered lepton tend to be centrally distributed in the experiment at MuIC and LHmuC for processes at the electroweak scale and above, more so than for the same processes at the LHeC. The scattered muon, however, peaks in the muon beam direction for low-$Q^2$ DIS and for vector boson fusion processes, necessitating the need for a muon spectrometer along the beamline. But such a spectrometer design also may prove useful for experiments at a $\mu^+\mu^-$ collider.

In summary, the goals and merits of our proposal are as follow:
\begin{itemize}
    \item Open a unique new frontier in particle and nuclear physics, ranging from the partonic structure of matter, precision QCD and electroweak interactions, Higgs bosons, and searches for physics beyond the Standard Model.
    \item Serve as a scientific target for a muon collider demonstrator to establish sustained muon collider R\&D, and serve as a stepping stone toward the ultimate ${\cal O}(10+)$ TeV $\mu^+\mu^-$ collider.
    \item Provide an affordable option by re-using established infrastructure and leveraging funding resources from both the particle and nuclear physics communities to realize a (first?) muon-based collider in the U.S. and/or CERN in the next 20--25 years. 
\end{itemize}

\section{Introduction}

Lepton-hadron (nucleus) deep inelastic scattering (DIS) 
has been a powerful tool to understand the fundamental 
structure of nucleons and nuclei. Decades of DIS
experiments have revealed the point-like substructure of
quarks and gluons inside the nucleon, and how they share
the longitudinal momentum of a fast-moving nucleon.
To develop a deeper understanding of the quark-gluon
structure and dynamics (especially in three dimensions) 
of matter, governed by
quantum chromodynamics (QCD), a high-energy and 
high-luminosity polarized electron-ion collider (EIC) has recently
been endorsed to be built at Brookhaven National Laboratory 
(BNL) by the late 2020s~\cite{NAP25171} as a high priority
on the agenda of the US nuclear physics community. The EIC 
is capable of carrying out deep inelastic electron-proton 
and electron-nucleus collisions with polarized beams 
at a center-of-mass energy ($\roots$) up to
140~GeV~\cite{Accardi:2012qut,NAP25171}. It will 
establish a new QCD frontier to address key open questions
such as the origin of nucleon spin, mass, and the emergence of QCD
many-body phenomena at extreme parton densities. At CERN, 
the Large Hadron-electron Collider
(LHeC)~\cite{Agostini:2020fmq} at CERN 
has been proposed as a possible 
extension to the Large Hadron Collider (LHC)
to explore the TeV energy regime of DIS with high luminosities. As a potential long-term step beyond the LHC at CERN, 
the Future Circular Collider (FCC) proposed to be built in a new 100~km tunnel 
also includes a mode of electron-hadron collisions 
(FCC-he) at $\roots$ = 3.5~TeV~\cite{Abada:2019lih}
by utilizing the LHeC's electron beam.

We propose an alternative approach to achieve the next-generation
lepton-hadron (ion) collider at TeV scales using high-energy
muon beams based on existing hadron collider facilities:
(1) a Muon-Ion Collider (MuIC) at BNL to succeed the EIC after its mission is completed by 2040s; and/or (2) a Large Hadron-Muon Collider (LHmuC) at CERN that can operate concurrently with
hadron collisions at the LHC.
Possibilities of muon-hadron colliders and their scientific 
potential have been discussed previously, 
for example in Refs.~\cite{Shiltsev:1997pv,Ginzburg:1998yw,doi:10.1063/1.56424,Sultansoy:1999na, Cheung:1999wy, Acar:2016rde,Canbay:2017rbg,Acar:2017eli, Caliskan:2018vep,Ketenoglu:2018fai, Kaya:2019ecf, Ozansoy:2019rmu, Aydin:2021iky, Cheung:2021iev}. 
A MuIC at BNL is first proposed in Ref.~\cite{Acosta:2021qpx} by some authors of this paper. 

The muon collider proposal has received revived interests 
in the particle 
physics community in recent years because of 
its potential of reaching very high energies
in a compact tunnel (e.g., the size of the LHC) 
at relatively low costs. It has 
been argued that the discovery potential of certain 
unknown hard processes or heavy particles at 
a 14~TeV $\mu^{+}\mu^{-}$ collider matches that at 
a 100~TeV proton-proton collider 
(e.g., FCC-hh)~\cite{Delahaye:2019omf, MCwpPhysics}, as all of the available beam
energy is carried by the interacting muons. While it is in
principle feasible to build the next electron/proton colliders
with more advanced magnets and the
construction of a new, longer {$\mathcal O$}(100)~km tunnel, 
the affordability is a concern in terms of both 
cost and time. Exploring new accelerator 
and collider technology not only provides attractive alternatives
but also will be transformative in bringing the field of 
nuclear and particle physics the furthest to new energy frontiers.

Development of muon collider technology is still at the pre-conceptual stage, with many challenges to overcome~\cite{Delahaye:2019omf}. Specifically, muons are short-lived and decay rapidly even when accelerated to TeV 
energies. Effective cooling of muon bunches to reduce its 
phase space in six dimensions~\cite{MICE:2019jkl} and 
subsequent rapid acceleration
are among the crucial elements for realizing a high energy,
high luminosity muon collider.
Beam backgrounds from muon decays also pose challenges  
to both the accelerator and the detectors. 
Radiation hazards from interacting neutrinos would need 
to be mitigated, especially at {$\mathcal O$}(10)~TeV energies. 
There is a consensus in the muon collider community that 
realizing a smaller-scale muon 
collider would be a necessary intermediate step to serve 
as a demonstrator before pursing the ultimate
{$\mathcal O$}(10)~TeV $\mu^{+}\mu^{-}$ collider. 
Such a demonstrator 
still requires tremendous R\&D effort and significant cost, so 
a compelling science program that is not
accessible by other proposed facilities is needed. For example,
the physics of a $\mu^{+}\mu^{-}$ collider with $\roots$ of 
several hundred GeV to 1 TeV may not be competitive with 
an $e^+e^-$ collider proposed with more 
established technology, such as the International
Linear Collider (ILC)~\cite{Behnke:2013xla}, the Compact Linear
Collider (CLIC)~\cite{Charles:2018vfv}, the Circular Electron Positron
Collider (CEPC)~\cite{CEPCStudyGroup:2018ghi}, and 
the FCC-ee~\cite{Abada:2019lih}. 
The international muon collider collaboration (IMCC) was
established recently and is focusing on investigating
the physics potential and design of a 3~TeV $\mu^{+}\mu^{-}$ 
collider, which is discussed in a separate white paper submitted 
to the Snowmass 2021 workshop \cite{MCwp3TeV}.

In this article, we focus on the proposal of
muon-hadron (ion) colliders at TeV energies. A MuIC and/or 
LHmuC will be simultaneously a discovery machine and a technology 
demonstrator to establish a novel muon collider. It has great
potential to attract worldwide interest and funding 
resources from both the nuclear and particle physics communities,
and thus can provide a realistic path, or staging option, toward
the realization of an ultimate $\mu^{+}\mu^{-}$ collider at $\roots=10$~TeV and beyond.

\section{Concept and Design Overview}
\label{sec:muic}

\subsection{Collider Configuration Options}

The principal thrust of our proposal is the development of 
a muon-ion collider as an upgrade to existing hadron collider 
facilities, such as RHIC/EIC and the LHC, although other facilities are not excluded but may require hadron ring development.
These muon-ion collider design scenarios are summarized 
in Table~\ref{fig:designtable} along with the expected 
maximum possible instantaneous luminosity (see more details 
on the luminosity estimation in Section~\ref{sec:lumi}).

\begin{table}[thb]
\caption{Benchmark beam energies for several $\mu p$ collider configurations studied here, the corresponding center-of-mass energy ($\sqrt{s}$), and the expected maximum possible instantaneous luminosity. }
\centering
\includegraphics[width=0.9\linewidth]{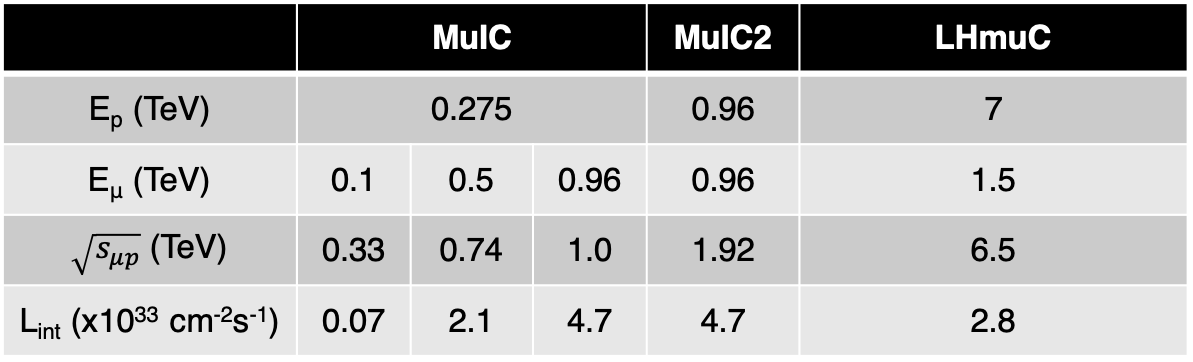}
\label{fig:designtable}
\end{table}

{\bf A Muon-Ion Collider (MuIC) at BNL} would reuse the hadron ring 
at the EIC with a maximum proton energy of 275~GeV and replace the 
electron storage ring with a high-energy muon one with a maximum 
energy of about 1~TeV ($p\,[{\rm GeV/c}]=0.3Br[{\rm T\cdot m}]$), assuming $B=11$~T 
dipole bending magnets 
(developed for the HL-LHC) and a bending radius of $r=290$~m of the 
EIC tunnel. The electron injection
and acceleration chain will be replaced by the proton-driver 
muon injection front-end including muon cooling, and a muon
acceleration ring residing either inside the EIC tunnel or 
in a separate, larger tunnel.

The MuIC's center-of-mass energy of 1.0~TeV and luminosity of
10$^{33}$--10$^{34}$~cm$^{-2}$s$^{-1}$ is comparable to the 
proposed LHeC electron-hadron collider of 1.2~TeV at CERN~\cite{Agostini:2020fmq} 
with a 50~GeV electron beam incident on one of the LHC proton rings. 
However, if the EIC hadron ring could be similarly upgraded to 
1~TeV energy, or if a 1~TeV muon storage ring were added to any 
other hadron ring facility with a 1~TeV proton energy 
(e.g. the Fermilab Tevatron), an option for doubling the 
center-of-mass energy to 2~TeV would be possible, which we 
call the {\bf MuIC2} option. Additionally, the MuIC and MuIC2 
options at BNL would offer the unique advantage of providing 
polarized beams, which is important for understanding the nucleon 
spin puzzle, as documented in the EIC white paper~\cite{Accardi:2012qut}. 

The ability to run at lower energies (and lower luminosities) 
always exists, and may in fact be the starting point in the 
development and commissioning of a new 
high-energy muon storage ring. Thus the MuIC could run at a 
lower center-of-mass energy to facilitate measurements in 
overlap with past experiments. For example, the MuIC running 
with a ${\sim}100$~GeV muon beam at BNL would have an equivalent 
center-of-mass energy to HERA. In fact a beam energy of 65~GeV, 
such as might be extracted from a ``Higgs factory'' muon collider 
facility, would provide 85\% of the HERA center-of-mass energy 
when collided with the BNL hadron beam.

{\bf A Large Hadron-Muon Collider (LHmuC) at CERN} would
achieve an even higher energy, if one of the LHC 7~TeV proton rings 
were utilized (as proposed for the LHeC) to collide 
with a new TeV-scale muon beam. If we assume that a 3~TeV $\mu^+\mu^-$ 
collider were to be constructed at CERN as 
proposed by the IMCC, one of the 1.5~TeV muon beams
could be brought into collision with one of the LHC 7~TeV hadron rings 
to achieve a 6.5~TeV muon-proton center-of-mass energy. 
The LHC, LHmuC and
3~TeV $\mu^+\mu^-$ colliders can in principle be operated 
concurrently, maximizing the scientific program potential 
at CERN. We note that this center-of-mass energy exceeds by a 
factor 2 even that possible at an electron-hadron collider facility 
where a 50~GeV electron beam is brought into collision with a 
50~TeV proton beam at a new large circular ring (FCC-eh), and 
avoids the necessity of a new 100~km circular tunnel. 
Of course a TeV-scale muon beam brought into collision 
with a ${\mathcal O}$(50)~TeV FCC proton ring would reach even higher energies. 

\begin{figure}[thb]
\centering
\includegraphics[width=0.9\linewidth]{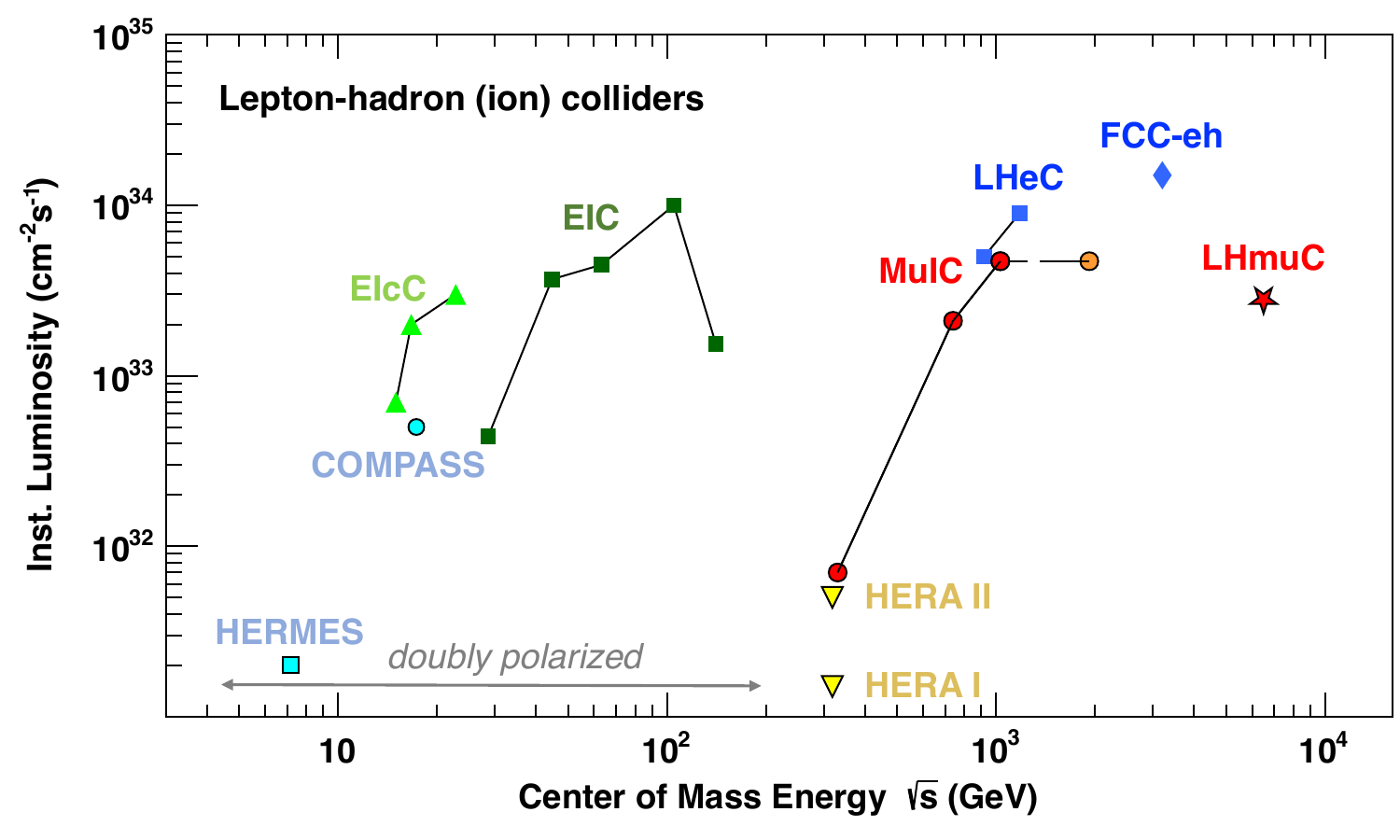}
\caption{Instantaneous luminosity and $\sqrt{s}$ for various past and
proposed future lepton-hadron colliders. Multiple operating energies 
are also shown for some facilities.}
\label{fig:discollider}
\end{figure}

The proposed muon-hadron (ion) colliders in this paper can be 
compared with other past and proposed future lepton-hadron 
facilities in Fig.~\ref{fig:discollider}, which shows the 
evolution of the instantaneous luminosity and center-of-mass 
energy of the colliders, including multiple operating energies 
for some facilities. The muon collider technology, if realized,
provides an alternative way of entering the TeV regime of
DIS physics based on existing facilities and infrastructure,
probing the structure of nucleon and nucleus down to
an unexplored Bjorken-$x$ regime of 10$^{-7}$--10$^{-8}$.

\subsection{Luminosity and Performance Matrix}
\label{sec:lumi}

A recent estimation of the instantaneous luminosity of a muon-proton collider
is discussed in Ref.~\cite{Kaya:2019ecf}, which can be generally expressed as follows:

\begin{equation}
    \mathcal{L}_{\mu p} = \frac{N^{\mu}N^{p}}{4\pi\max[\sigma^{\mu}_{x},\sigma^{p}_{x}]\max[\sigma^{\mu}_{y},\sigma^{p}_{y}]}\min[f^{\mu}_{c},f^{p}_{c}]H_{hg},
\end{equation}

\noindent Here, $N^{\mu}$ and $N^{p}$ represent the number of
particles per respective beam bunch. The transverse
RMS beam size in $x$ and $y$ for the muon and proton
beam, $\sigma^{\mu,p}_{x,y}$, is calculated as

\begin{equation}
\sigma^{\mu,p}_{x,y} = \sqrt{\varepsilon_{x,y}^{*}\beta_{x,y}^{*}m^{\mu,p}/E^{\mu,p}}
\end{equation}

where, $\varepsilon^{*}$ is the normalized transverse emittance
and $\beta^{*}$ is the amplitude function at the interaction point.
One can see that the luminosity
is determined by the beam that has a larger size.
The $f^{\mu}_{c}$, $f^{p}_{c}$ parameters are the bunch frequencies. Typically, 
the bunch frequency of proton beams is 2--3 orders of magnitude
larger than that of muon beams, so ${L}_{\mu p}$ is largely
determined by $f^{\mu}_{c}$, which is equal to the muon bunch injection 
repetition frequency ($f_{\rm rep}$) multiplied by the number of
cycles ($N_{c}$) that muons can make in a circular storage ring 
before decaying away (a muon bunch would decay away long before 
the next one is injected so there will be just one muon bunch or one train 
of muon bunches in the ring at a time). Each muon bunch will 
survive an average of about $300 B$(Tesla) cycles in a ring, since the
magnetic field defines the size of the ring for a given muon energy
and the energy defines the time dilation factor for muons.
For simplicity, the hour-glass factor, $H_{hg}$, is assumed to be unity.

\begin{table}[t!]
\centering
\caption{Proposed beam parameters and estimates of achievable luminosity
for MuIC at BNL and LHmuC at CERN for muon-proton collisions.}
\vspace{0.5cm}
\includegraphics[width=0.9\linewidth]{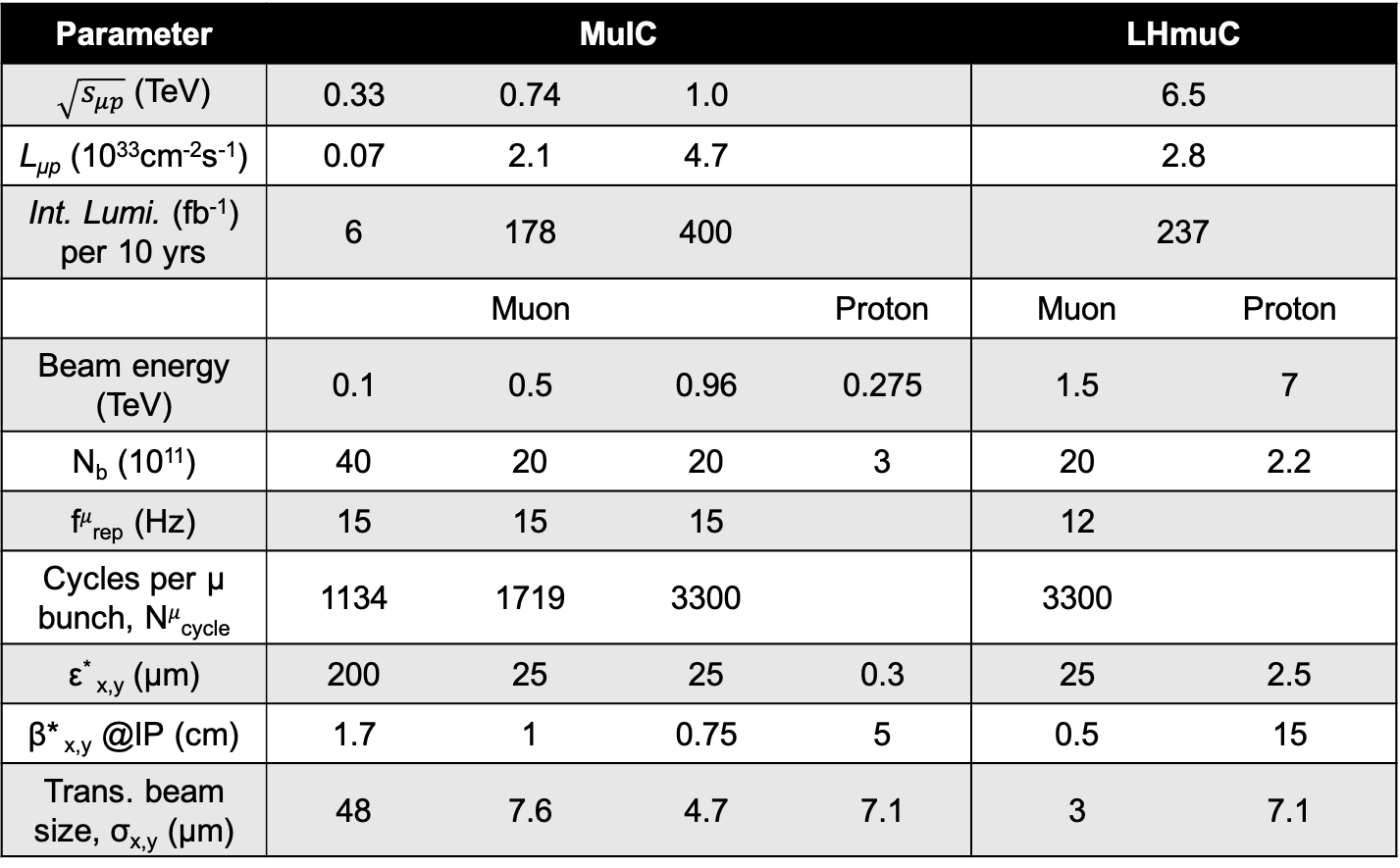}
\label{tab:table2}
\end{table}

The proposed parameters of MuIC and LHmuC are listed in 
Table~\ref{tab:table2} for muon-proton collisions. For the muon beam, 
we use the proposed parameters of the proton driver scheme
from Ref.~\cite{Palmer:2014nza,Delahaye:2019omf}. 
The muon bunch repetition frequency is taken to be 12--15~Hz.
The proton beam parameters are assumed to be those
achieved at RHIC~\cite{Aschenauer:2014cki} and the 
LHC~\cite{Kaya:2019ecf}, or foreseen to be achieved 
at EIC~\cite{eic_cdr}. Assuming the implementation of strong
hadron cooling at EIC, the normalized transverse emittance
of the proton beam is expected to be as small as 0.3$\mu$m.
Note that the proposed muon bunch intensity is an order of magnitude
larger than that of the proton bunch. Such intensity muon
beam will likely disturb the proton beam. To minimize the
beam-beam effect, we propose to split the muon bunch into a
train of 10 bunches. Also note that the EIC design 
at BNL adopts a flat transverse beam profile with the horizontal
dimension stretched to be much larger than the vertical 
dimension. 
At the interaction point, the two beams 
would intersect at a finite crossing angle of 25~mrad 
in the horizontal plane, with crab cavities compensating for
luminosity reduction effects. The purpose of such a design for EIC is to 
maximize the luminosity and, at the same time, fulfill 
other requirements such as the possibility of detecting scattered 
protons with a transverse momentum as low as 200~MeV by Roman Pots 
inside the beam pipe. For simplicity, our estimates assume 
round transverse profile of beams. Details of the beam profile
and crossing angle should be further optimized in a conceptual design.

With above parameters, we achieve a peak $\mathcal{L}_{\mu p}$ 
up to ${\approx}5\times$10$^{33}$~cm$^{-2}$s$^{-1}$ at the highest muon energy for MuIC.
If we assume a running period of 28 weeks per year and a duty cycle
of 0.5, the total delivered integrated luminosity over 
10 years are listed in Table~\ref{tab:table2}. A few hundreds of
fb$^{-1}$ can be expected at the highest muon energies.
To put the MuIC and LHmuC luminosity into context, it is 
anticipated that the EIC will deliver an integrated luminosity up to about 1.5~fb$^{-1}$/month with
$\mathcal{L}_{ep} \approx 10^{33}$~cm$^{-2}$s$^{-1}$,
and most of science cases studied at the EIC require a total
integrated luminosity of 10~fb$^{-1}$ ($\approx 30$ weeks of operations). 
Therefore, even with much less stringent requirements on the 
muon beam, e.g., a peak $\mathcal{L}_{\mu p} \approx  10^{32}$~cm$^{-2}$s$^{-1}$ operating for several years, 
the MuIC will still be a novel facility that breaks new ground 
in  high energy nuclear and particle 
physics with cutting-edge technology.

We leave for a future study an estimate of the achievable luminosity for muon-ion collisions at MuIC or LHmuC.

\subsection{Design Status}

The design of a muon-ion collider is still at the
very conceptual stage. The design and development of its
muon component, including the muon front-end and the acceleration
and collision rings, almost completely overlaps with that 
of $\mu^+\mu^-$ colliders, where the design status, parameters, 
and challenges are described in Ref.~\cite{Delahaye:2019omf}.
The Muon Accelerator Program (MAP) collaboration in the U.S. carried out
detailed studies and designs of muon colliders in 2011--2017. 

\begin{figure}[thb]
\centering
\includegraphics[width=\linewidth]{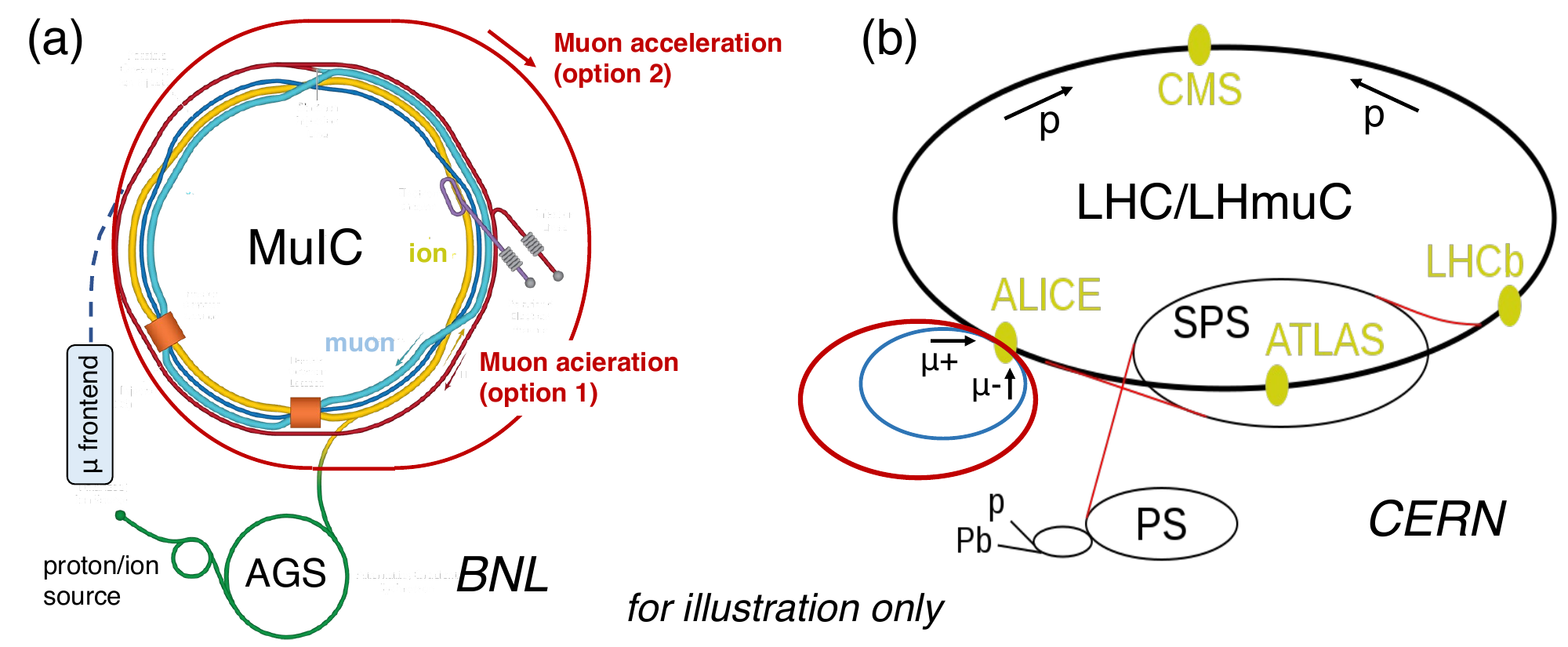}
\caption{Design sketches for (a) Muon-Ion Collider at BNL and (b) Large Hadron-Muon Collider (LHmuC) at CERN.}
\label{fig:muic-sketch}
\end{figure}

Sketches of possible MuIC and LHmuC designs are shown in Fig.~\ref{fig:muic-sketch}.
For the MuIC at BNL (Fig.~\ref{fig:muic-sketch}a, derived from the EIC design), 
essentially all electron-related components will
be replaced by those for muon beams. The proton-driver scheme is
the natural choice at BNL, although no existing proton complexes (either at BNL or CERN) meet the requirement for muon colliders of 2--4~MW in beam power
and 4--8~GeV in kinetic energy yet. A substantial upgrade is necessary. 
A dedicated muon front-end is needed to provide the
muon injection, cooling, and initial fast acceleration with recirculating linacs. Ideally, both muon
acceleration and storage rings would fit within the existing tunnel to
minimize the infrastructure cost. There are six straight sections of
the BNL tunnel that could be used for the accelerator. On the other hand,
the design by the MAP collaboration favors an acceleration ring with larger curvature for fast-ramping magnets and long straight sections for fast acceleration, and a compact circular ring
for muon storage and collisions. Therefore, it is conceivable that
a new tunnel would be added for muon acceleration, as also illustrated in
Fig.~\ref{fig:muic-sketch}a, which still fits well within the BNL campus. 
At CERN, if a 3~TeV $\mu^+\mu^-$ collider is built by the IMCC, it can
be constructed to intersect with one of LHC proton beams (e.g., at IP2 as 
in the LHeC proposal, Fig.~\ref{fig:muic-sketch}b). In this way, $\mu^+\mu^-$, $\mu p$, and $pp$ programs 
can operate concurrently at CERN, leading to a rich scientific program and
extending the lifetime of the LHC infrastructure. Once the muon collider
technology is established, an upgrade to a ${\cal O}(10+)$~TeV $\mu^+\mu^-$ collider 
can be considered using the LHC tunnel as the next stage.

\subsection{Unique Design Challenges and R\&D requirements}

We discuss some of muon-ion collider specific design challenges below: 
\begin{itemize}

\item {\bf The muon-ion interaction point} would require dedicated 
design and R\&D efforts to maximize the machine's luminosity and
meet all the science requirements. The size and intensity of muon and proton bunches are different, so how to ensure the proper crossing
of two beams without interfering each other would be a topic of
R\&D. It may also be advantageous to separate one high intensity
muon bunch into a train of lower intensity bunches to avoid disturbing the proton beam. Nevertheless, similar challenges 
are also present in the design of the electron-ion collider, where
insights will be learned.

\item {\bf Machine-detector interface (MDI):} Because of the 
asymmetric colliding configuration, the MDI requires special 
consideration somewhat different from that for $\mu^+\mu^-$ colliders. For 
example, the shielding tungsten nozzles necessary for protecting detectors against
the secondary particle background from muon decays should only be 
necessary on the incoming muon side, instead of both beam directions, 
which otherwise restrict the detector acceptance to approximately 
$|\eta|<2.4$.
Full simulations are needed to verify the level of beam-induced 
background with a single-sided nozzle configuration for muon-ion 
collisions. This will have an impact on the available detector 
acceptance, which is discussed in Section~\ref{sec:detector}.

\item {\bf Neutrino radiation protection and mitigation:}
One peculiar type of radiation background from muon colliders is
induced by collimated high intensity neutrinos from muon decays, which are mainly
aligned with the plane of the collider. This is commonly
not a concern for proton-proton or e$^{+}$e$^{-}$ colliders.
Here, the main issue is not related to the direct interactions of neutrinos
with human bodies but instead from long-term stationary exposure to secondary particles produced when neutrinos traverse dense materials (soil, buildings, etc.). 

\begin{figure}[htb]
\centering
\includegraphics[width=0.6\linewidth]{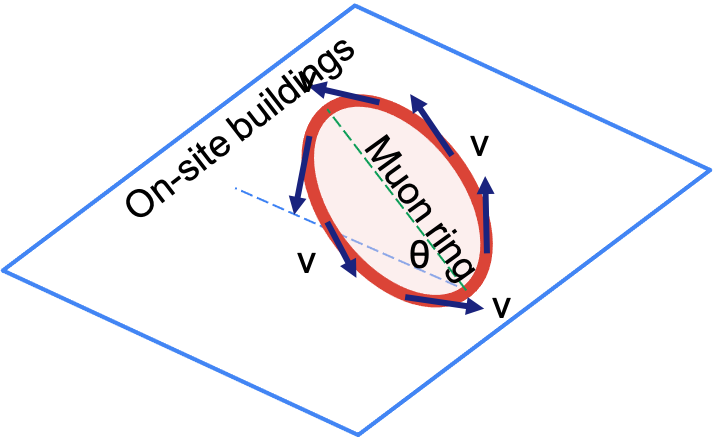}
\caption{A sketch to illustrate the proposed strategy of mitigating 
neutrino radiation backgrounds for MuIC at BNL.}
\label{fig:neutrino-sketch}
\end{figure}

Various approaches for mitigating neutrino radiation backgrounds from multi-TeV $\mu^+\mu^-$ collider are extensively discussed, for example, in Ref.~\cite{NeutrinoMitigate}.
Here, we specifically discuss the possible strategy of mitigating 
neutrino backgrounds for the MuIC at BNL. The BNL tunnel is effectively
on the surface located near the tip of Long Island, NY. As long as
we can direct radiated neutrinos to not strike on-site buildings at BNL,
their impact can be kept minimal. A sketch to illustrate our proposed
mitigation strategy is shown in Fig.~\ref{fig:neutrino-sketch}. By tilting
the muon ring (assuming six straight sections as with the existing RHIC tunnel)
by a small angle, e.g., $\theta \lesssim 1^{\circ}$, it would be sufficient
to direct most of neutrinos toward the air in one direction and toward the ground
and sea in the other direction for a given straight section, with little impact on buildings nearby. 

\item {\bf Muon beam polarization:} A unique feature of the EIC is its 
doubly-polarized beams, which is critical for understanding the nucleon 
spin puzzle and also many other physics processes which are spin or helicity
dependent. Therefore, maintaining the capability of muon beam polarization 
would be highly beneficial to the science program. The possibility of maintaining
the muon beam polarization has been discussed in Refs.~\cite{Cline:1996yi,Norum:1996mi,Neuffer:1999aw,Ankenbrandt:1999cta}.
Muon beams are produced with about 20\% longitudinal polarization. By extracting
higher energy muons at their production from pion decay, the average polarization can be
increased up to 50\% with a compromise in the luminosity. Loss of polarization
after muons go through the ionization cooling stage is estimated to be negligible.
With a series of spin rotators, it is in principle possible to manipulate and 
maintain the muon spins. However, significant efforts are needed to investigate
such a feasibility and arrive at a credible design.

\end{itemize}

\subsection{Staging Options and Desirable Demonstrators}
Staging options of MuIC have been indicated in our proposed energy
and luminosity scenarios in Tables~\ref{fig:designtable} and \ref{tab:table2},
and also in Fig.~\ref{fig:discollider}. Starting with a 100~GeV muon beam
at the MuIC would already match the center-of-mass energy and luminosity 
of HERA II. A muon beam of 100~GeV or less, similar to that proposed for a Higgs factory, can be much less demanding in muon cooling (e.g., without final cooling)
and fast-ramping magnet technology, so it can serve as a good initial
demonstrator and can be directly compared with HERA data as a calibration.
Next, the muon beam energy can be ramped up to 0.5 TeV and eventually 0.96~TeV
to reach a TeV lepton-hadron collider. If the hadron beam can also be
upgraded to about 1~TeV, the MuIC center-of-mass energy can then be further doubled
(MuIC2).

\section{Synergies with existing facilities in high energy and nuclear physics
communities}

A key merit of the muon-ion collider proposal is the strong synergy
with existing accelerator facilities in the high energy and nuclear physics
communities. Leveraging the established infrastructure, accelerator
expertise, and user community has enormous financial benefits and
is a recipe for success in the evolution of high energy physics research.
The RHIC at BNL and the LHC at CERN are currently the only operating hadron colliders in the world.
At BNL, the realization of the EIC would be nearly impossible if built
from scratch without the existence of RHIC. At CERN, the LHC was constructed
using the LEP tunnel and the existing PS and SPS accelerators. While we are not advocating against constructing more ambitious new infrastructure,
maximizing the utility
of existing ones is something the community should always try to leverage,
especially under the strong financial constraints nowadays. 

The muon collider is an attractive but yet unproven concept. The technology
needs to be first demonstrated, and a staged approach is necessary to 
achieve the ultimate $\mu^+\mu^-$ collider at 10~TeV and beyond. While it is
ideal to already identify a path toward the ultimate energy, realizing a
muon collider in any form with a strong science program would be a significant
step forward as a technology demonstrator and helps raise the priority of
muon colliders in future planning processes for high energy physics (HEP) collider programs. 
The MuIC concept at BNL would serve as such a technology demonstrator, and it has the 
potential for the U.S. to take a leadership role in future novel TeV colliders. 
The science case of a high energy, high luminosity lepton-ion collider has
already been well established with the EIC, so the MuIC can be considered
as an upgrade to the EIC.
Potential collaboration between HEP and Nuclear Physics (NP) communities (supported by different 
offices at funding agencies such as DOE and NSF) could make it more affordable to each community separately,
and helps attract broader interests worldwide. Once the technology is demonstrated,
a $\mu^+\mu^-$ collider at comparable energies 
could then be explored at the same site in a cost effective way, given strong physics motivations, or at a new site that is more suitable for much higher energies. Since the muon front-end system is expected to be
costly, perhaps elements of it could be transferred to another site if necessary and cost effective.
Similarly, the LHmuC concept will significantly broaden and enrich the CERN
program during or beyond the HL-LHC. The LHC tunnel could also be a candidate for
hosting a future 10~TeV $\mu^+\mu^-$ collider.

\begin{figure}[thb]
\centering
\includegraphics[width=0.95\linewidth]{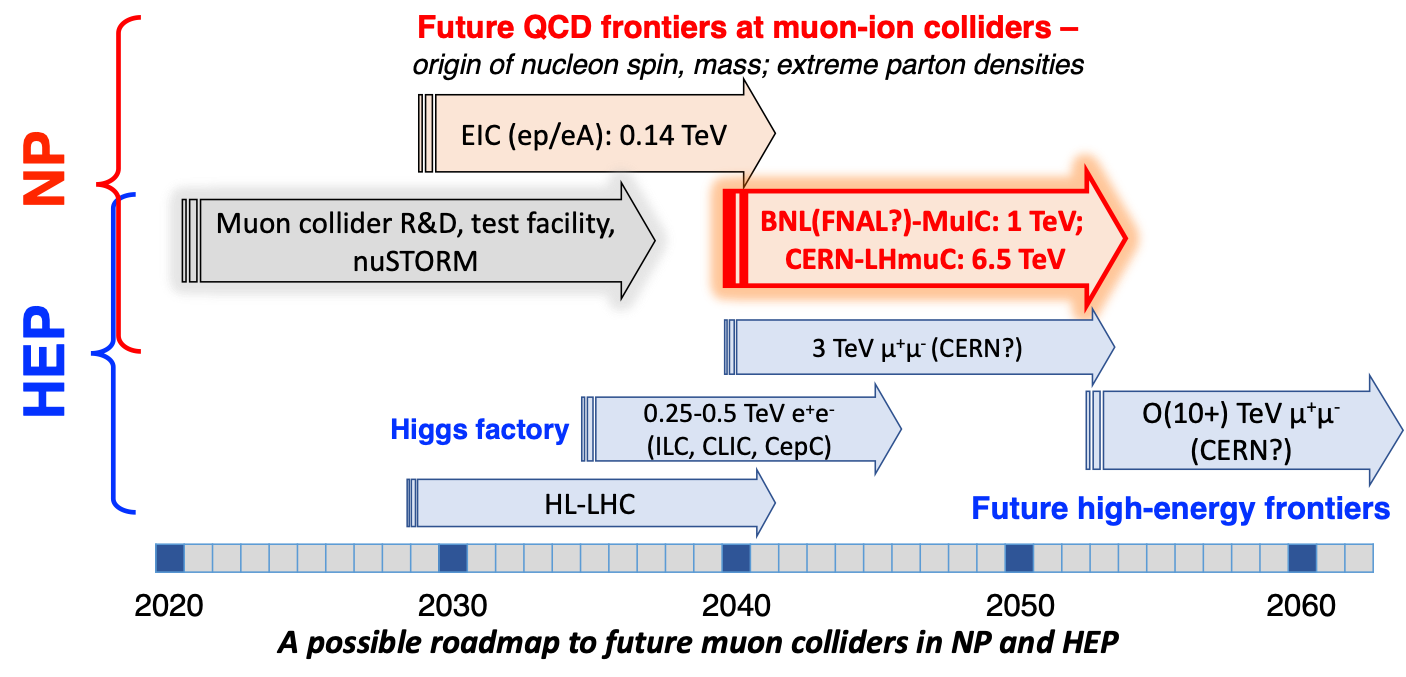}
\caption{A possible road map toward realizing future muon colliders by synergizing 
efforts of both the nuclear physics and high energy physics communities.}
\label{fig:timeline}
\end{figure}

Our view on a possible road map to a future combined facility for 
nuclear physics (NP) and high energy physics 
(HEP) based on muon collider technology is presented in Fig.~\ref{fig:timeline}. 
Based on the assessment of IMCC~\cite{Adolphsen:2022ibf}, a technically limited timeline shows about 10 years to 
establish the feasibility of a muon collider via intense R\&D 
at a testing facility in order to be ready to commit to construct a 
muon collider. We strongly advocate the HEP and NP
communities to carry out this effort jointly to open new frontiers
in their respective fields. From the NP perspective, the EIC
is the highest priority at present and there is no other program 
on the horizon planned beyond the EIC. The next NP long-range planning
process is scheduled to conclude in late 2023, so it is an excellent time to
consider the MuIC as a future option and establish a R\&D program. On the
HEP side, the landscape is more complex with many different 
proposals of future colliders. In our view, a Higgs factory with $e^{+}e^{-}$
collisions is likely the immediate next step beyond the HL-LHC, because of
its technological maturity and well-defined deliverables of high interest.
If NP and HEP jointly develop the muon collider, a 3~TeV 
$\mu^+\mu^-$ collider can be ready for construction by the IMCC at a similar
time scale to the MuIC. With the success of the first muon collider projects, 
the community would be ready to pursue a ${\cal O}(10+)$~TeV $\mu^+\mu^-$ collider 
(e.g., at CERN) in the 2050s--60s.

\section{The Science Case of TeV Muon-Ion Colliders}

The science case for a MuIC using the BNL facility was outlined in our initial concept proposal~\cite{Acosta:2021qpx}, although the scientific 
potential of muon-hadron colliders in general have been discussed previously, 
for example in Refs.~\cite{Ginzburg:1998yw,doi:10.1063/1.56424, Sultansoy:1999na, Cheung:1999wy, Canbay:2017rbg, 
Acar:2017eli, Caliskan:2018vep, Kaya:2019ecf, Ozansoy:2019rmu, Aydin:2021iky, 
Cheung:2021iev}.
A large part of the physics case of a TeV collider also overlaps
with that of the proposed LHeC, given the nearly equivalent center-of-mass 
energies. This includes structure function measurements, precision electroweak 
and QCD measurements, Higgs boson studies, and beyond Standard Model searches. 
Therefore, a more comprehensive set of topics is explored in 
Ref.~\cite{Agostini:2020fmq}. 
However, the initial state muon at the MuIC provides a complementary 
sensitivity to any lepton flavor violating processes, and the scattering
kinematics can be quite different than the LHeC, which can lead to 
different and sometimes advantageous experimental conditions 
(see Ref.~\cite{Acosta:2021qpx} and Appendix~\ref{sec:dis-kin}). 
Additionally, the MuIC at BNL also offers the possibility of the 
polarization of both beams, and a wide range of ion species for 
lepton-ion collisions, enabling a broad nuclear physics program 
and detailed studies of the spin structure of nuclei.
Thus the MuIC can extend to new regimes the science program of 
the EIC that is documented in the EIC NAS report~\cite{NAP25171} 
and Yellow Report~\cite{AbdulKhalek:2021gbh}.
Here we outline the science program of a TeV muon-hadron (ion) collider
and report on a subset of topics with quantitative studies to illustrate 
the physics potential, for several target energies and facilities.

\begin{figure}[t!]
\centering
\includegraphics[width=0.7\linewidth]{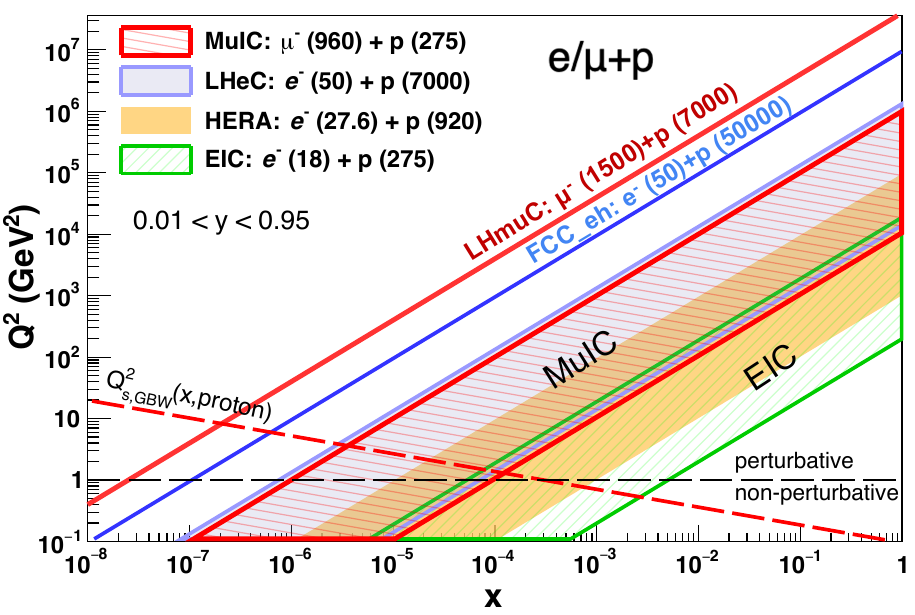}
\includegraphics[width=0.7\linewidth]{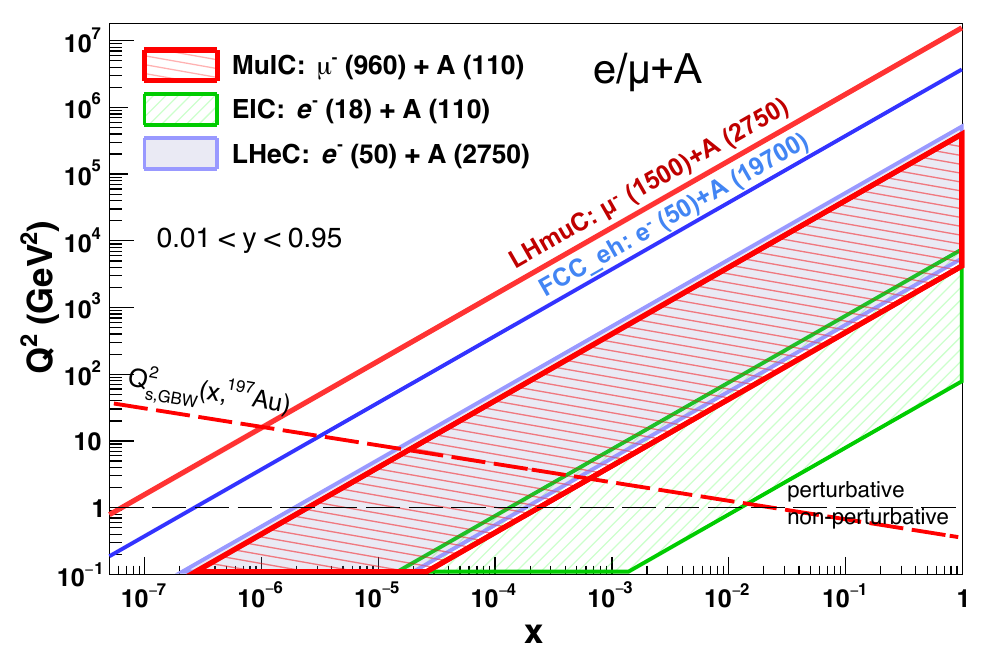}
\caption{Kinematic coverage of $Q^{2}$ and $x$ in deep 
inelastic lepton-proton (top) and lepton-nucleus (bottom) scattering 
for the muon-ion collider design options presented here and for the 
EIC at BNL, HERA at DESY, and the LHeC and FCC-eh options at CERN, 
each at their  maximum beam energies. The inelasticity ($y$) range 
is assumed to be $0.01<y<0.95$ (hatched areas). The long dashed lines
indicate the saturation 
scale as a function of $x$ in the proton and the gold ($^{197}$Au) 
nucleus from the GBW model~\cite{GolecBiernat:1998js}.}
\label{Q2x-all}
\end{figure}

The physics potential of lepton-hadron (ion) colliders is generally 
represented and compared by their kinematic coverage in $Q^{2}$ and $x$, 
as shown in Fig.~\ref{Q2x-all} for various past and future colliders.
The MuIC, and equivalently the LHeC, will significantly
extend the kinematic coverage of the EIC to much larger $Q^{2}$ and
smaller $x$ regimes, by an order of magnitude in each compared to the 
previous HERA $ep$ collider. In lepton-proton collisions (Fig.~\ref{Q2x-all}, top),
the saturation or non-linear QCD regime parametrized by the GBW model based on the HERA 
inclusive data~\cite{GolecBiernat:1998js} is clearly out of the reach for the EIC. However, 
it becomes within reach at very small $x$ values at the MuIC and LHeC, 
and is opened further by higher energy colliders such as the proposed LHmuC and FCC-eh. 
Figure~\ref{Q2x-all} also shows that the LHmuC, by colliding the existing LHC proton beam with a 1.5~TeV muon beam,
will provide even larger coverage in $Q^{2}$ and $x$ than the FCC-eh.
In lepton-nucleus collisions (Fig.~\ref{Q2x-all}, bottom), 
choosing the $^{197}$Au nucleus as a representative example, 
a factor of 6 enhancement in the saturation scale to the proton 
is expected, $Q_{\rm sat}^{2}$($x$, Au$)=A^{1/3}\, Q_{\rm sat}^{2}$($x$, proton).
While the EIC approaches the saturation regime, the MuIC and LHmuC
will bring us well into the domain to explore gluon saturation 
and nonlinear QCD phenomena. 

\begin{figure}[thb]
\centering
\includegraphics[width=0.7\linewidth]{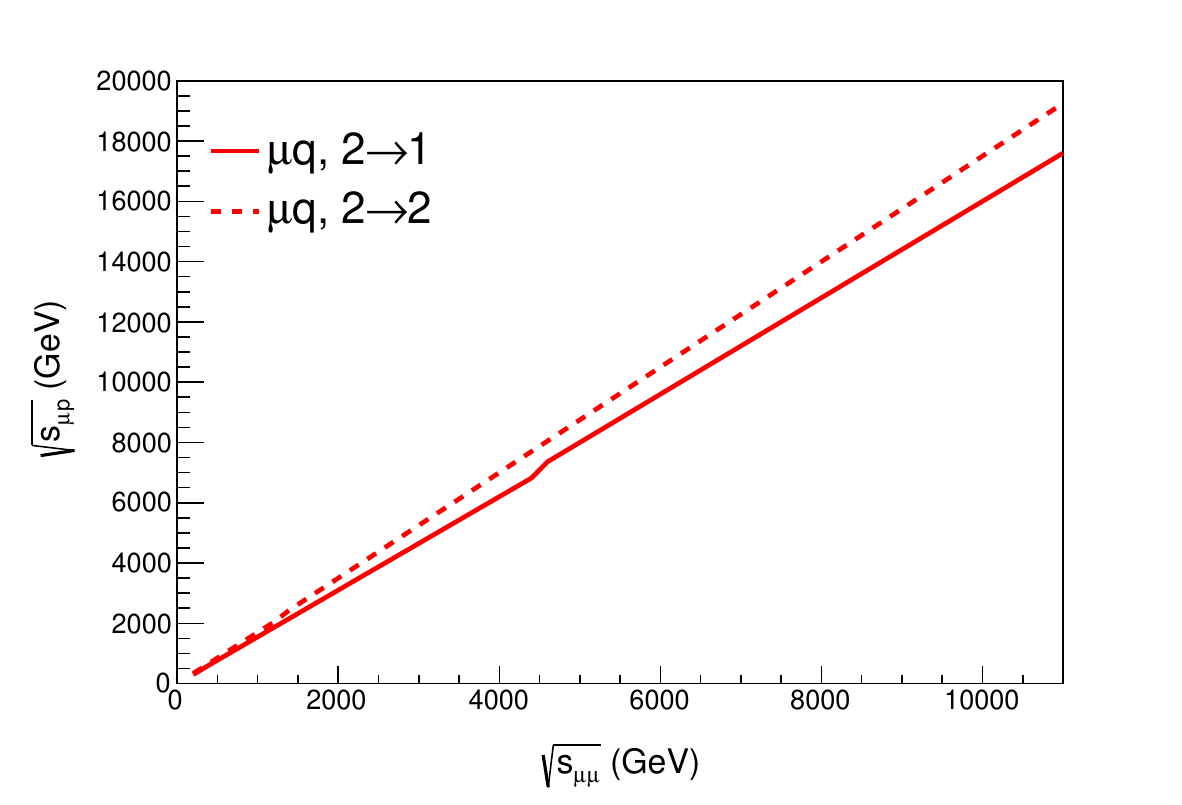}
\caption{The center-of-mass energy of $\mu p$ colliders versus that of 
$\mu^+\mu^-$ colliders, which yield equivalent cross sections. 
The solid curve corresponds to the $2\rightarrow1$ annihilation process,
while the dashed curve corresponds to the $2\rightarrow2$ scattering process.}
\label{fig:muic-vs-pp}
\end{figure}

To evaluate and compare discovery potentials of new 
physics beyond the Standard Model, Ref.~\cite{AlAli:2021let} 
argues that a $\mu^+\mu^-$ collider can be competitive
with a $pp$ collider at 5--20 times higher center-of-mass 
energy (\roots)
for discoveries via certain processes such as 
annihilation ($2\rightarrow1$) and vector
boson fusion ($2\rightarrow2$). This is because a muon carries
100\% of its available momentum in the interaction process, while
only a small fraction of the proton momentum is carried by a parton. We carry out
a similar calculation for the muon-proton collider so that
all three types of colliders can be compared. The parton luminosity
of a $\mu p$ collider is expressed as,
\begin{equation}
    \frac{dL_{i}}{d\tau}(\tau,\mu_{f})=\int_{\tau}^{1}\frac{dx}{x}f_{i}(x,\mu_{f}).
\end{equation}

\noindent
Here the $f_{i}(x,\mu_{f})$ are the parton distribution 
functions (PDFs) for parton $i$ carrying a
fraction $x$ of the longitudinal momentum, at factorization scale 
$\mu_{f}=\sqrt{\hat{s}}/2$, where $\hat{s}$ is the partonic 
center-of-mass energy and $\tau=\hat{s}/s$. The resulting
center-of-mass energy of $\mu p$ colliders versus that of 
$\mu^+\mu^-$ colliders, which yield equivalent cross sections,
is shown in Fig.~\ref{fig:muic-vs-pp}. The CT18NNLO PDF set is chosen
for this calculation. We find that a $\mu^+\mu^-$ collider is equivalent to
a $\mu p$ collider with 1.5$\times$ higher \roots\  in 
terms of its discovery potential. Therefore, the proposed LHmuC with $\roots=6.5$~TeV
matches the potential of a $\mu^+\mu^-$ collider at $\roots=4.3$~TeV,
exceeding the 3~TeV $\mu^+\mu^-$ collider proposed by the IMCC.
This exercise again highlights the unique opportunities by the TeV muon-ion (proton)
collider proposal, taking advantage of existing facilities.

\begin{figure}[thb]
\centering
\includegraphics[width=0.9\linewidth]{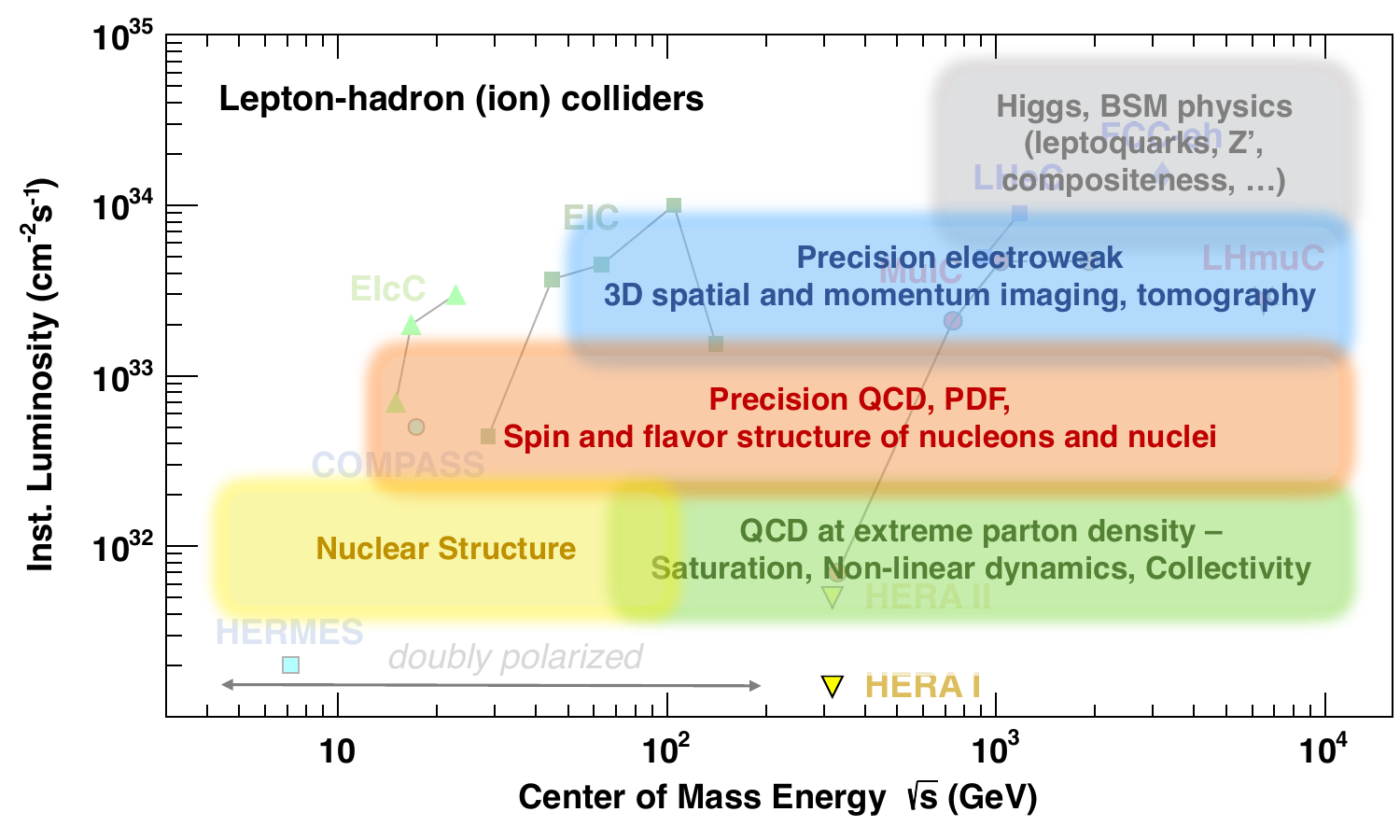}
\caption{The landscape of science at lepton-hadron (ion) colliders with
required energy and luminosity.}
\label{fig:muicpotential}
\end{figure}

The landscape of science programs at lepton-hadron (ion) colliders with
their required energy and luminosity is illustrated in Fig.~\ref{fig:muicpotential}.
Higgs boson measurements and beyond the Standard Model searches are the most demanding
in terms of the reach in energy and luminosity (${>}5 \times 10^{33}$cm$^{-2}$s$^{-1}$).
Studies of QCD physics at extreme parton densities generally
prefer high \roots\ in order to probe the smallest possible $x$ regime, but 
a luminosity at the level of 10$^{32}$~\percms\ is sufficient
because of the large cross sections. Studies of PDFs, including the spin and
flavor structure of nucleons and nuclei, require a moderate luminosity
up to about 10$^{33}$~\percms\ (where beam polarization is important 
for the spin physics program). With a peak luminosity of 
10$^{33}$--10$^{34}$~\percms, precision electroweak 
physics can be carried out as well as three-dimensional 
spatial and momentum imaging and tomography of nucleons and 
nuclei. Therefore, even if the maximal proposed luminosity cannot
be achieved, there is still a very rich physics program to explore
at muon-ion colliders.
We discuss the physics potential in detail below, including
quantitative studies on a few topics. We look forward to
engaging the broader nuclear physics community to study
the potential of muon-ion colliders.

\subsection{QCD and Nuclei}

The scientific potential of lepton-hadrons in understanding 
the physics of QCD, nucleon and nuclei has been discussed
in detail
in a series of documents on the electron-ion collider to
be constructed at BNL by the 2030s~\cite{Accardi:2012qut,AbdulKhalek:2021gbh,NAP25171}.
We briefly highlight a few selected fundamental questions
in QCD and nuclear physics that the MuIC or LHmuC 
will explore to unprecedentedly small-$x$ and large-$Q^{2}$ regimes
as a future extension of the EIC science program.

\subsubsection{Nucleon Spin and 3-D Structure}
The nucleon spin is one of its fundamental properties.
It was found that quark polarization inside a nucleon only 
contributes to about 30\% of the total spin. Therefore, the 
majority rest of the nucleon spin must be carried by the 
gluon polarization and orbital motion of quarks and gluons.
To determine the contribution of gluon polarization,
a measurement of the helicity-dependent gluon distribution function, 
$\Delta g(x)$, especially in the small $x$ region, is crucial. 
The uncertainty on the overall gluon polarization from RHIC measurements
is still rather large mainly because of the limitation in accessing 
the small $x$ region~\cite{Abdallah:2021aut}. For $x<0.01$, 
$\Delta g(x)$ is largely unconstrained. With polarized beams, 
the EIC is projected to significantly improve the precision
of gluon polarization by accessing $x$ values down to 0.001 at 
$Q^{2} \approx 10$~GeV$^{2}$~\cite{Aschenauer:2012ve}, with an
integrated luminosity of 10~fb$^{-1}$. The early phase of MuIC
is already capable of delivering 10~fb$^{-1}$ in one year data taking
at a center-of-mass energy substantially higher than EIC (e.g., 
with a 500~GeV muon beam). Assuming a polarized muon beam, the MuIC 
will extend the reach of gluon polarization in $x$ down to $10^{-5}$, 
potentially providing a more definitive answer to the gluon spin 
contribution. Furthermore, precise measurements 
of three-dimensional (3D) parton distribution functions, 
generalized parton distributions (GPDs) and
transverse-momentum-dependent (TMD) distributions, over
a much wider range of $x$ values at the MuIC could provide a 
complete picture of orbital angular momentum of quarks and 
gluons inside the nucleon.

\subsubsection{Gluon Saturation at Extreme Parton 
Densities in proton and nucleus}

The gluon density inside the nucleon increases
dramatically toward small $x$ values. 
At extreme gluon densities, the nonlinear QCD
process of gluon-gluon fusion will start
playing a key role to limit the divergence of
the gluon density. At a certain dynamic scale of
momentum transfer, known as $Q_{s}$, gluon splitting and fusion 
processes reach an equilibrium such that the gluon density 
is saturated, resulting in novel universal properties of 
hadronic matter. Examples of gluon saturation scales 
inferred from fits to HERA data, known as the GBW model,
\cite{GolecBiernat:1998js} are shown Fig.~\ref{Q2x-all}
for muon-hadron and muon-ion colliders. The large $Q_{s}$ 
scale predicted at small $x$ values, especially in large 
nuclei (enhanced by a factor of $A^{1/3}$), 
enables perturbative QCD calculations of nuclear
structure functions, as proposed in the color-glass
condensate (CGC) effective field theory~\cite{McLerran:1993ni}.
Predictions of signatures of gluon saturation in large nuclei
at the EIC are presented in Ref.~\cite{Accardi:2012qut}.
As shown in the kinematic coverage of Fig.~\ref{Q2x-all}, 
the EIC starts entering the domain of gluon saturation in 
gold nuclei at $x \approx 10^{-3}$, while the MuIC and LHmuC will
bring us well into the saturation regime. 
The MuIC will also probe the saturation
regime in the proton and other light nuclei for the first time,
which is likely not accessible by the EIC.

\begin{figure}[htb]
\centering
\includegraphics[width=0.49\linewidth]{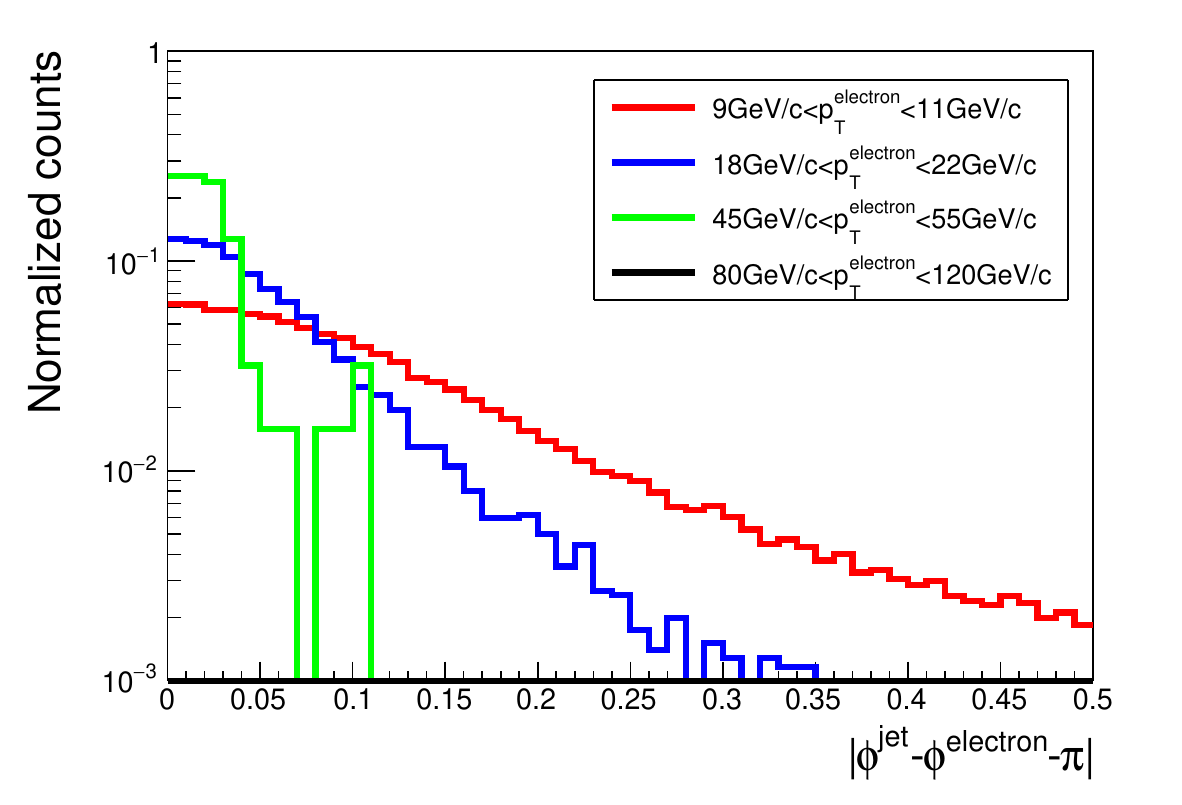}
\includegraphics[width=0.49\linewidth]{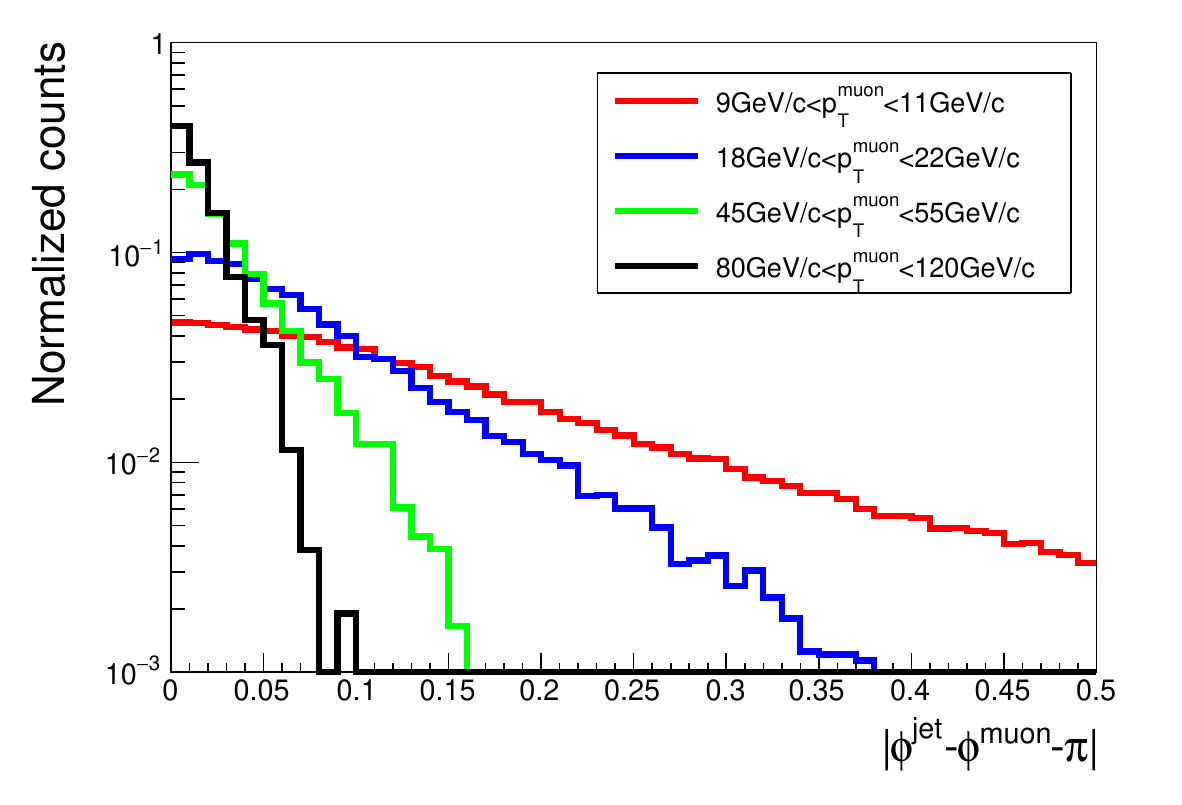}
\caption{Lepton-Jet azimuthal de-correlations at EIC (left) and MuIC (right) 
energies for various scattered lepton transverse momentum intervals, 
assuming 28 weeks of data takings in one year and 50\% duty cycle.}
\label{fig:jetdecorr}
\end{figure}

One promising observable to probe the dense gluonic medium
inside a heavy nucleus and the possible gluon saturation effect is
the decorrelation from back-to-back, in azimuth, of the jet-muon final state. As the struck parton
traverses through the cold nuclear matter, it will experience energy loss and
its direction will thus be smeared. Through this parton-medium interaction, 
properties of the dense gluon state in the nucleus can be inferred.
Compared to EIC, MuIC will
explore this decorrelation with a much wider range of lepton $p_{\rm T}$, 
as shown in Fig.~\ref{fig:jetdecorr} based on PYTHIA 8 simulations, providing
detailed information on the parton $p_{\rm T}$ dependence of the energy loss
in nuclear matter for very small $x$ regimes.

\subsubsection{QCD Collectivity in Small-System Collisions}
Discovery of QCD collective behavior in high-multiplicity
final states of small-system collisions such as pp~\cite{CMS:2010ifv} 
and pPb~\cite{CMS:2012qk} at the LHC has raised fundamental questions
on the initial state and dynamical evolution of non-perturbative
QCD systems. Since the discovery, the origin of observed
QCD collectivity has been intensely debated~\cite{Dusling:2015gta}. 
Attempts to search for such collective effects have been
extended to even smaller systems such as $ep$~\cite{ZEUS:2019jya}, $\gamma$A~\cite{ATLAS:2021jhn}, $\gamma$p~\cite{CMS-PAS-HIN-18-008}
and $e^{+}e^{-}$~\cite{Badea:2019vey} collisions. It has also been suggested
that a collective effect can also be developed in a QCD system
as small as a single parton propagating in the vacuum~\cite{Baty:2021ugw}.
The main challenge for observing the collectivity in these
very small systems is that it is difficult to create events
with high multiplicities at energies achieved by previous
colliders. In fact, a lepton-hadron or lepton-ion collider
would be an ideal, clean environment for studying such 
collectivity as Q$^{2}$ of the virtual photon can provide
a lever-arm to control 
the initial size of the system. The MuIC and LHmuC will 
substantially extend the event multiplicity reach of
previous lepton-hadron collisions to a regime that is
comparable to that in $pp$ collisions, providing unique
discovery potential.

\subsection{Standard Model Physics}

\subsubsection{Structure Function and QCD Measurements}

\begin{figure}[htb]
\centering
\includegraphics[width=0.6\linewidth]{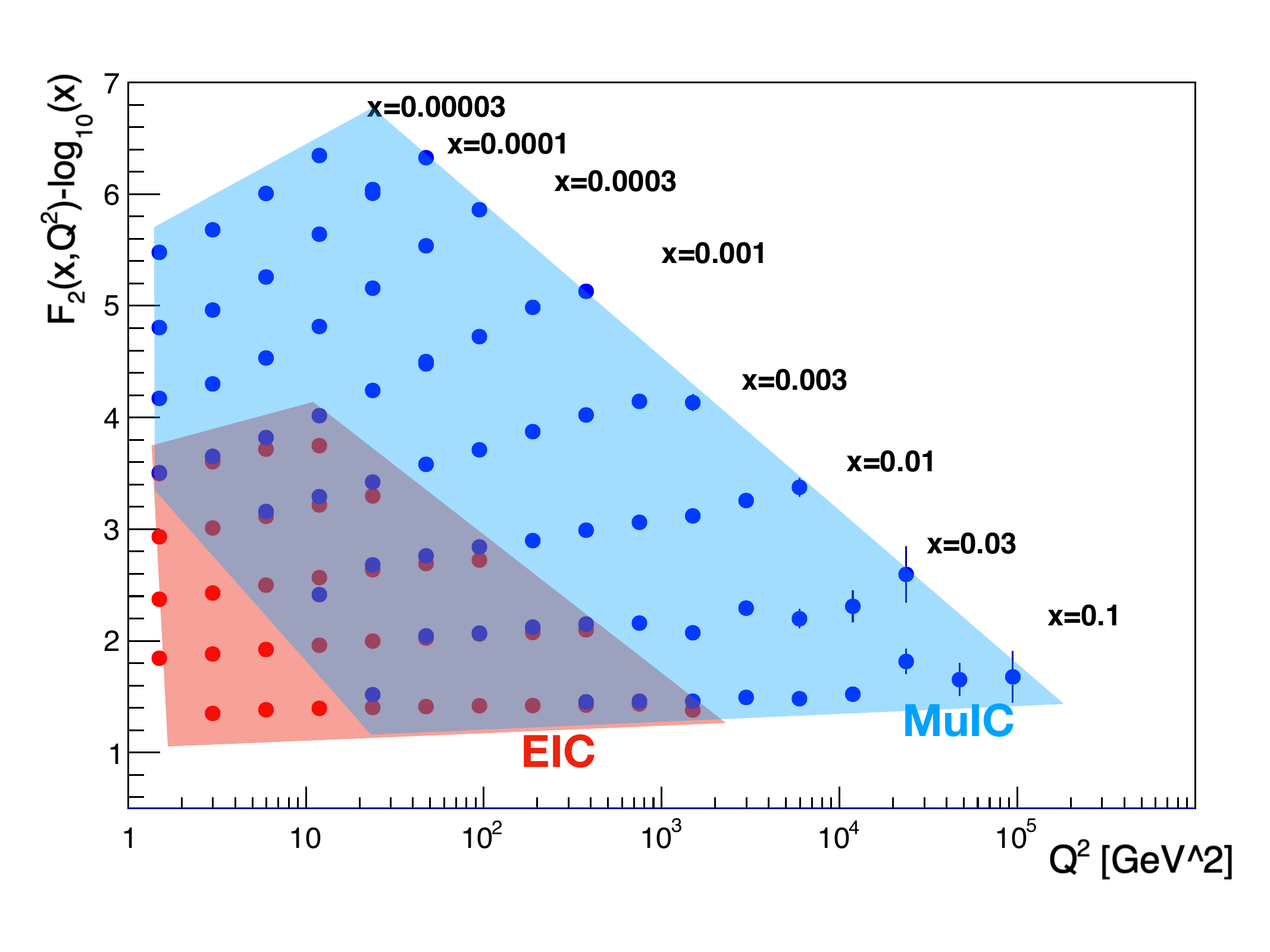}
\caption{The projected structure function ($F_{2}$) as a function of $Q^{2}$ for various $x$ ranges using PYTHIA 8 simulations assuming one-year of data taking for EIC and MuIC for 28 weeks of operation and a duty cycle of 50\%.}
\label{fig:f2}
\end{figure}

Measurements of the nucleon structure functions form the ``bread and butter'' of DIS experiments, from which information on the PDFs and spin structure (with polarized beams) of the nucleon can be obtained. This information, particularly on the flavor content of the PDFs, is quite independent from the constraints obtained from hadron collider measurements, where such information is less cleanly separated from the hard scattering process. 
As already emphasized, the MuIC can perform these measurements on protons and nuclei in new regions at low $x$ and high $Q^2$. Figure~\ref{fig:f2} shows the projected structure function ($F_{2}$) as a function of $Q^{2}$ for various $x$ ranges using pseudo-data  for EIC and MuIC assuming one-year of data taking with 28 weeks of operation and a duty cycle of 50\%. Thus, these measurements can be used to reduce the overall PDF uncertainty in calculations for cross sections at future hadron collider facilities, like FCC-hh. 

Global fits to all structure function data provide information not only on the PDFs, but simultaneously provide a precise measurement of the QCD coupling parameter $\alpha_{\rm s}(Q^2)$ through the QCD evolution equations. Moreover, as the $Q^2$ reach of the MuIC extends well into the electroweak scale (see Section~\ref{sec:highq2dis}), precise electroweak parameter measurements are also possible from the global fits. 

\begin{figure}[htb]
\centering
\includegraphics[width=0.6\linewidth]{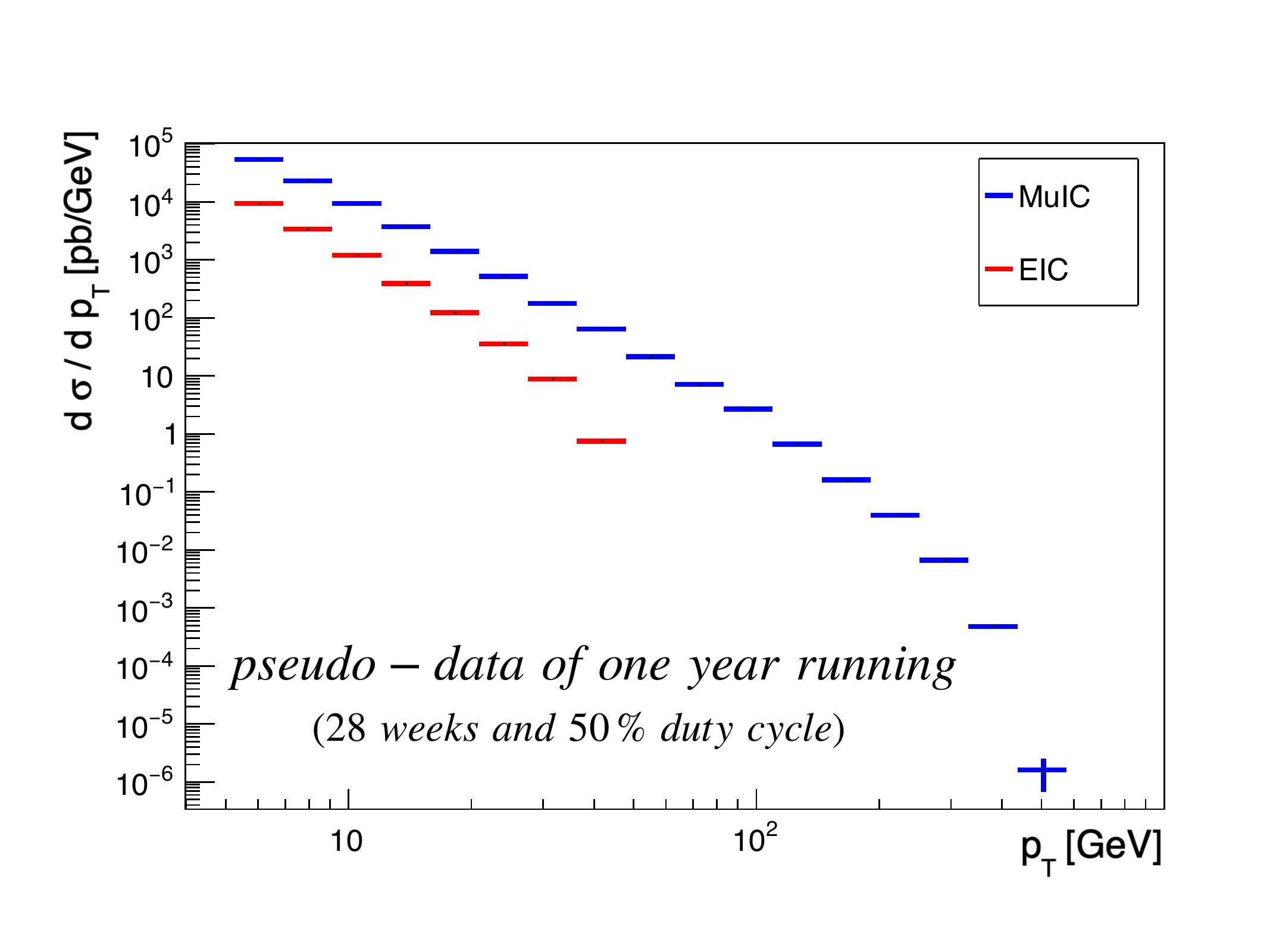}
\caption{Inclusive jet $p_{T}$ spectra in DIS events from PYTHIA 8 for EIC and MuIC at their top energies, assuming 28 weeks of data taking in one year and 50\% duty cycle. }
\label{fig:jets}
\end{figure}

In addition to the inclusive structure function measurements, 
direct measurements of inclusive jet and multijet production in DIS also allow for a precise determination of $\alpha_{\rm s}$ and its running over a wide $Q^2$ range in a single experiment. The leading-order jet transverse momentum scales approximately as $p_{\rm T} \approx (1-y) \sqrt{Q^2}$. The coverage of the MuIC for jet measurements is very similar to that of the LHeC. The only difference is that the jets tend to be more central, trending toward the muon direction, than the LHeC, where the jets are in the proton direction. 
In Fig.~\ref{fig:jets}, a study of inclusive jet $p_{T}$ spectra is performed using the PYTHIA 8 generator in muon-proton collisions at 1~TeV, and is also compared to that at the EIC. 
Jets are reconstructed by the anti-k$_{T}$ algorithm~\cite{Cacciari_2008,fastjet2012} with a cone size of 1. Projected statistical uncertainties correspond to
one year of data taking, assuming 28 weeks of operation and 50\% duty cycle. As one can see, the MuIC will reach to much higher $p_{T}$ jets than EIC.

\subsubsection{High $Q^2$ Total and Differential DIS Cross Sections}
\label{sec:highq2dis}

\begin{figure}[htb]
\centering
\includegraphics[width=0.6\linewidth]{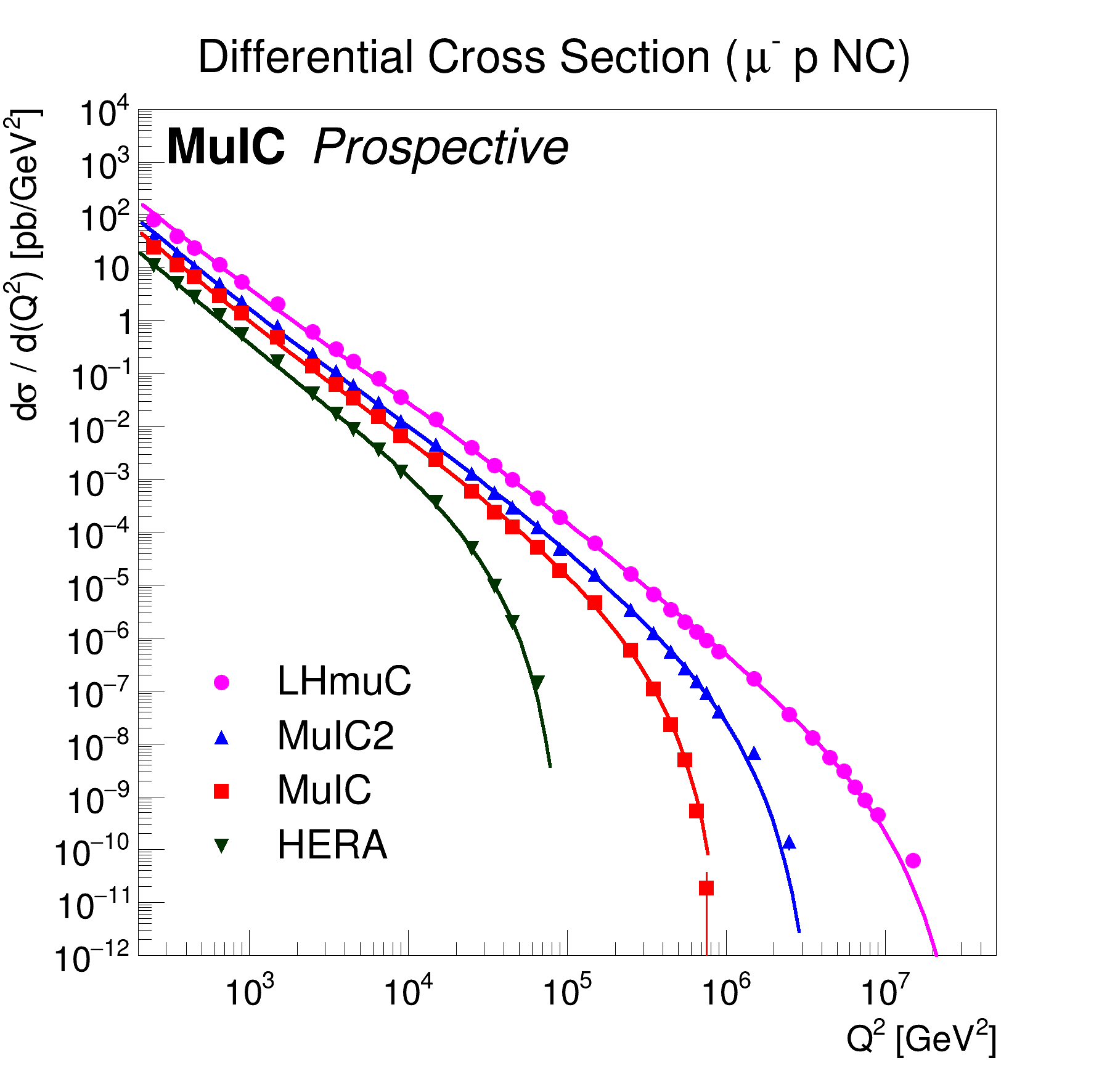}
\caption{The neutral-current differential cross section in $Q^2$ for unpolarized $\mu^- p$ deep inelastic scattering for several collider options: HERA with $\sqrt{s}=0.32$~TeV, MuIC with $\sqrt{s}=1.0$~TeV, MuIC2 with $\sqrt{s}=2.0$~TeV, and LHmuC with $\sqrt{s}=6.5$~TeV. The inelasticity variable is restricted to $0.1<y<0.9$.
}
\label{fig:DiffQ2NC-allColliders}
\end{figure}

The TeV-scale center-of-mass energy made possible by the muon-ion collider proposal described here leads to a reach in $Q^2$ well beyond that which was achieved by the HERA $e p$ collider, and thus opens sensitivity to possible new particle interactions and substructure. This is illustrated in Fig.~\ref{fig:DiffQ2NC-allColliders}, which shows the calculated neutral-current differential scattering cross section in $Q^2$ compared across different $\mu^- p$ collider options, including at the HERA center-of-mass energy. The cross sections were calculated using Pythia~8 \cite{Sjostrand:2014zea} with the NNPDF2.3 parton density set \cite{NNPDF:2017mvq} for the inelasticity range $0.1 < y < 0.9$. The HERA experiments published differential cross section measurements with polarized and unpolarized lepton beams \cite{H1:2012qti, ZEUS:2012zcp} that were sensitive to cross sections at the level of ${\cal O} (10^{-6})$ pb/GeV$^2$ with individual data samples corresponding to ${\cal O}(100)$~pb$^{-1}$, which translated to a reach in $Q^2$ of approximately $50{\,}000$~GeV$^2$ for $\sqrt{s}=318$~GeV. For a similarly-sized data sample recorded by an experiment at a muon-ion collider, the reach in $Q^2$ would be ${\approx} 200{\,}000$~GeV$^2$ for the MuIC, ${\approx} 400{\,}000$~GeV$^2$ for the MuIC2, and ${\approx} 800{\,}000$~GeV$^2$ for the LHmuC. 
However, because of the ${\sim}1/Q^4$ fall-off of the differential cross sections at high $Q^2$, the ability to probe the $Q^2$ scale corresponding to the same high $x\approx 0.5$ as HERA would require 2 (3) orders of magnitude more integrated luminosity for MuIC (MuIC2), and even 4 or more orders of magnitude for LHmuC. 
The corresponding target integrated data sample sizes would therefore be 10, 100, and 1000~fb$^{-1}$ for MuIC, MuIC2, and LHmuC, respectively.  
While the first two benchmarks are achievable within the first few years of operation at nominal luminosity, the latter is disfavored within 10 years in our estimation given in Section~\ref{sec:lumi}.  


\begin{figure}[htb]
\centering
\includegraphics[width=0.45\linewidth]{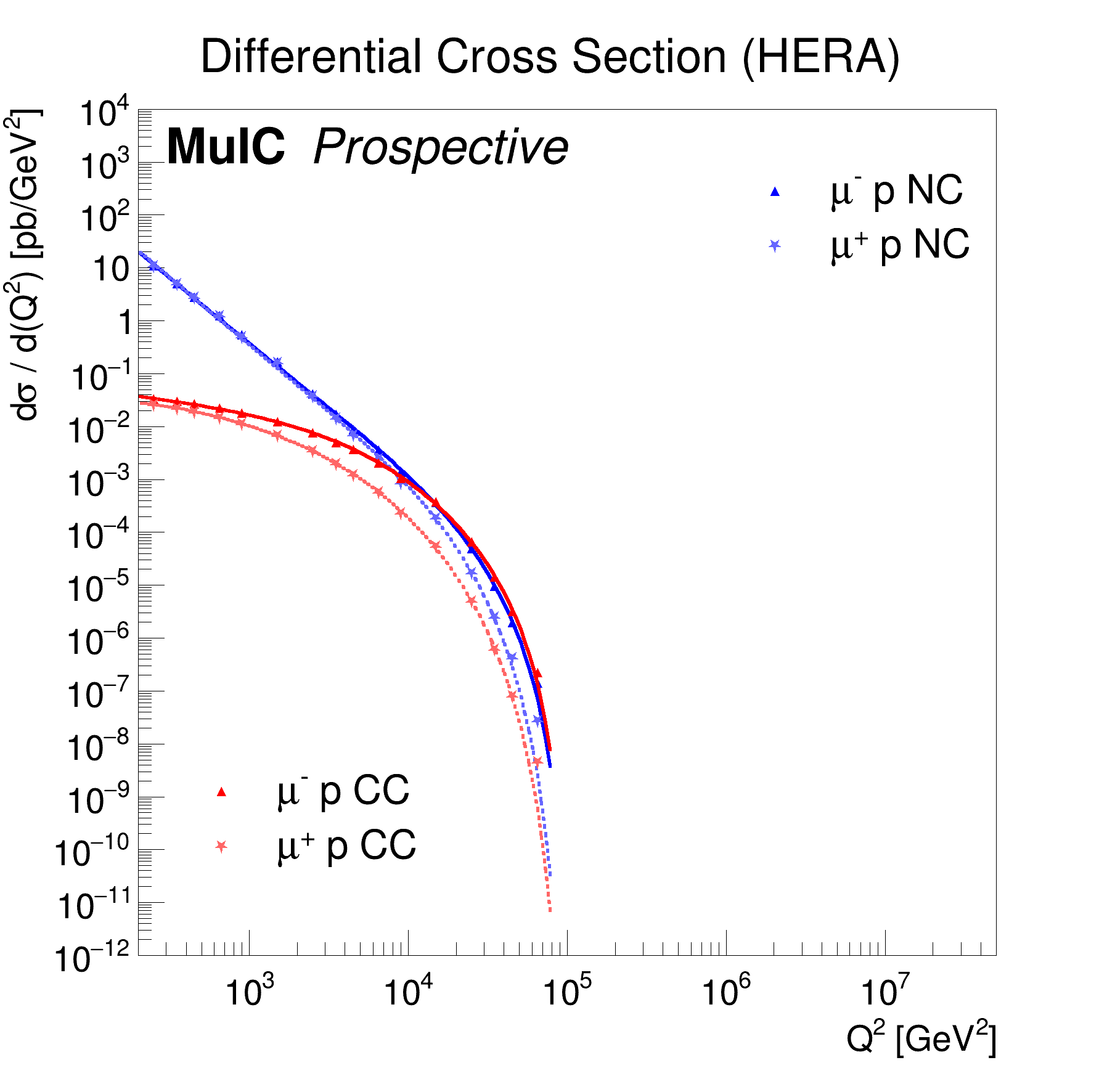}
\includegraphics[width=0.45\linewidth]{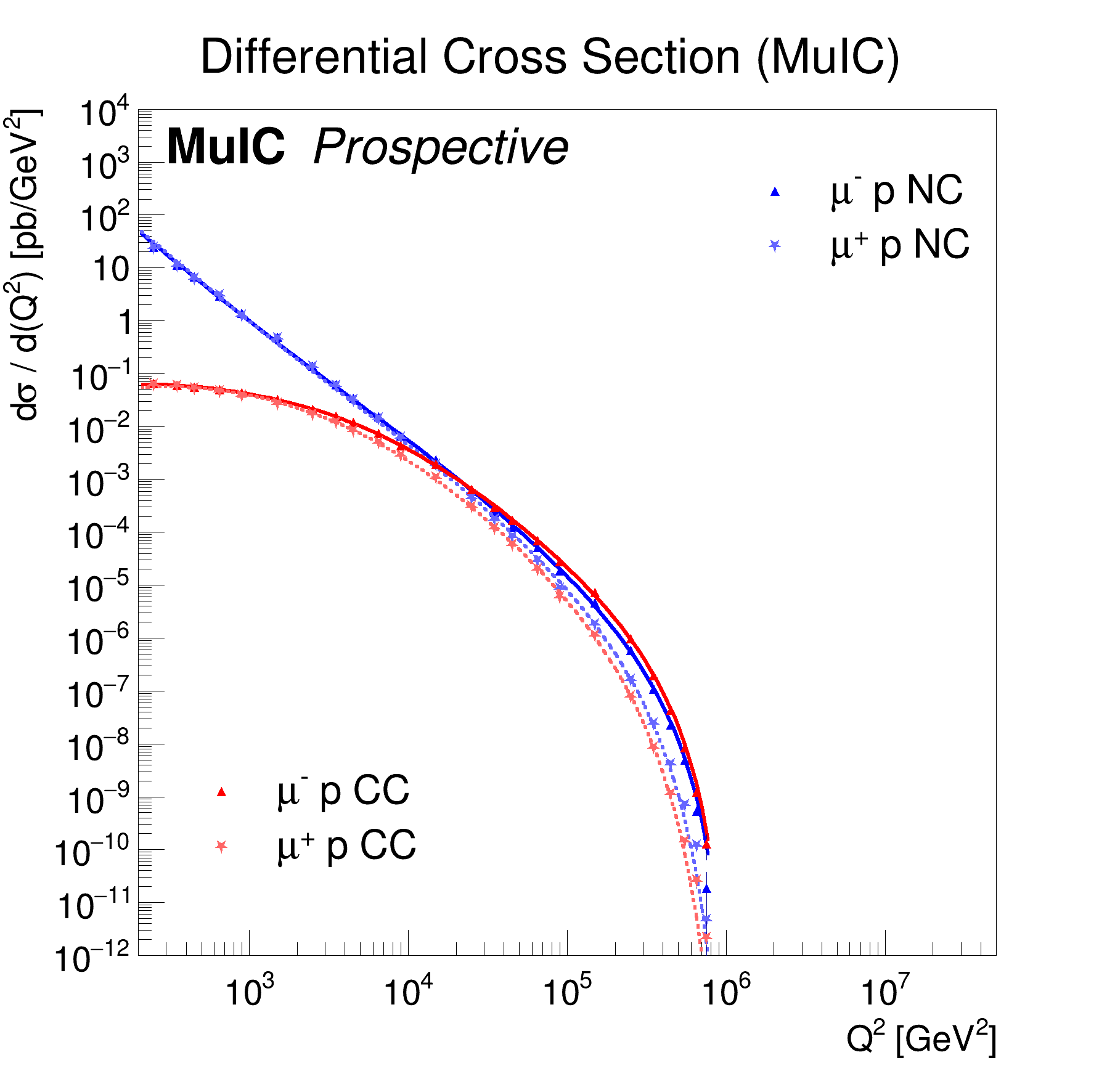}
\includegraphics[width=0.45\linewidth]{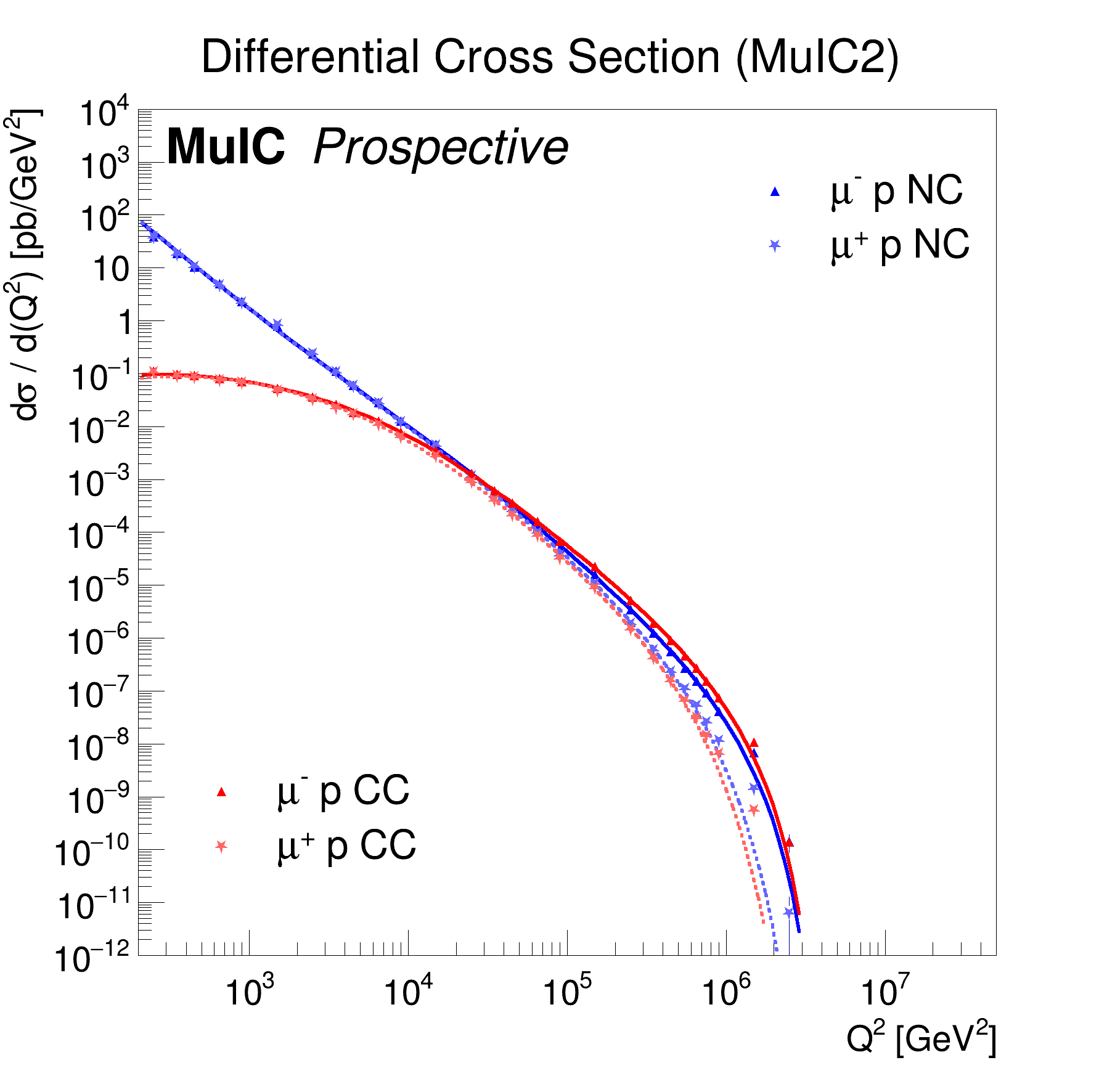}
\includegraphics[width=0.45\linewidth]{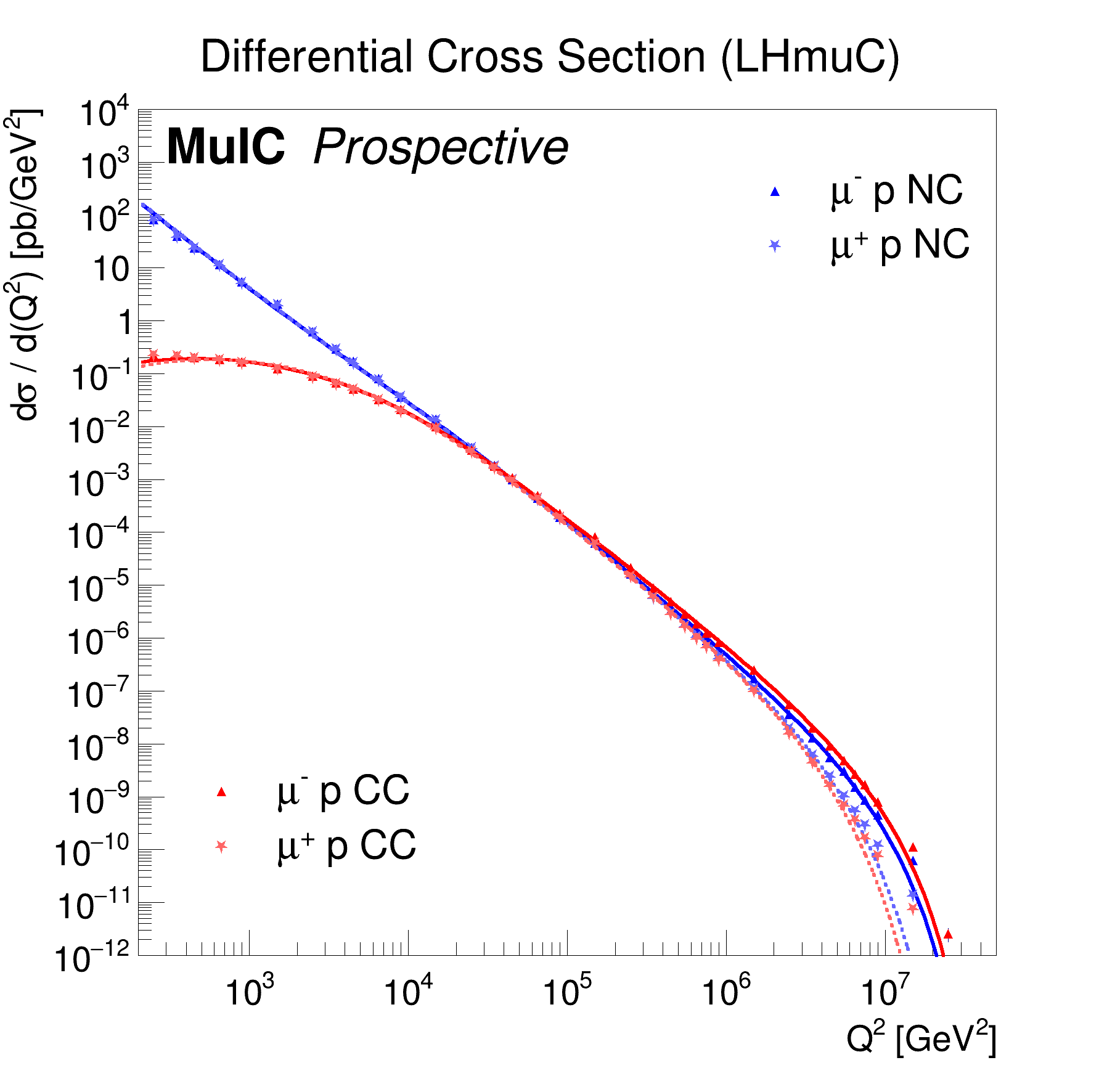}
\caption{The differential cross sections in $Q^2$ for neutral-current and charged-current deep inelastic scattering in unpolarized $\mu^- p$ and $\mu^+ p$ collisions for HERA (top-left), MuIC (top-right), MuIC2 (bottom-left), and LHmuC (bottom-right). The inelasticity variable is restricted to $0.1<y<0.9$}
\label{fig:DiffQ2-CCNC}
\end{figure}

The calculated differential cross sections for both neutral-current (NC)  and charged-current (CC) DIS and for both muon- and antimuon-proton collisions and for $0.1<y<0.9$ are shown in Fig.~\ref{fig:DiffQ2-CCNC} for center-of-mass energies corresponding to HERA, MuIC, MuIC2, and the LHMuC. The higher energy of a muon-ion collider allows for measurements well into the electroweak unification region at high $Q^2$. 
The integrated cross sections above several high $Q^2$ thresholds for neutral-current scattering  and charged-current scattering are shown in Tables~\ref{tab:NCxsec} and \ref{tab:CCxsec}, respectively, to compare the production yields at the different collider facilities. The effective total charged-current cross sections (for $Q^2>1$~GeV$^2$) are also shown in Table~\ref{tab:CCxsec}, where one can see that the cross section grows from that at HERA by a factor 3--4 for the MuIC, for example.

\begin{table}[!htb]
    \centering
    \caption{Integrated cross sections, in pb, for neutral current scattering in unpolarized $\mu^- p$ and $\mu^+ p$ collisions for various minimum thresholds on $Q^2$ in GeV$^2$ and several machine choices. The inelasticity variable is restricted to $0.1 < y < 0.9$. 
\label{tab:NCxsec}    
    }
    \begin{tabular}{l|c|c|c|c|c}
        \hline
        Machine & $Q^2>3\times 10^4$ & $Q^2>10^5$ & $Q^2>3\times 10^5$ & $Q^2>10^6$ & $Q^2>10^7$ \\ \hline 
        \multicolumn{6}{c}{$\mu^- p \to \mu^- X$} \\ \hline
        HERA  & 0.024 & -- & -- & -- & -- \\
        MuIC  & 3.7 & 0.072 & 0.0028 & -- & -- \\
        MuIC2 & 9.8 & 0.59 & 0.12 & -- & -- \\
        LHmuC & 37 & 3.4 & 1.1 & 0.060 & 0.012 \\ \hline
        \multicolumn{6}{c}{$\mu^+ p \to \mu^+ X$} \\ \hline
        HERA  & 0.0051 & -- & -- & -- & -- \\
        MuIC  & 2.1 & 0.020 & 0.0005 & -- & -- \\
        MuIC2 & 7.8 & 0.30 & 0.047 & -- & -- \\
        LHmuC & 36 & 3.0 & 0.87 & 0.032 & 0.0005 \\

        \hline
    \end{tabular}
\end{table}
        
\begin{table}[!htb]
    \centering
    \caption{Integrated cross sections, in pb, for charged current scattering in unpolarized $\mu^- p$ and $\mu^+ p$ collisions for various minimum thresholds on $Q^2$ in GeV$^2$ and several machine choices. The inelasticity variable is restricted to $0.1 < y < 0.9$. 
\label{tab:CCxsec}    
    }
    \begin{tabular}{l|c|c|c|c|c|c}
        \hline
        Machine & $Q^2>1$ & $Q^2>3\times 10^4$ & $Q^2>10^5$ & $Q^2>3\times 10^5$ & $Q^2>10^6$ & $Q^2>10^7$ \\ \hline 
        \multicolumn{7}{c}{$\mu^- p \to \nu_\mu X$} \\ \hline
        HERA  & 68 & 0.038 & -- & -- & -- & -- \\
        MuIC  & 200 & 5.2 & 0.12 & 0.0053 & -- & -- \\
        MuIC2 & 345 & 13 & 0.92 & 0.20 & -- & -- \\
        LHmuC & 860 & 43 & 4.6 & 1.6 & 0.098 & 0.020 \\ \hline
        \multicolumn{7}{c}{$\mu^+ p \to \overline{\nu}_\mu X$} \\ \hline
        HERA  & 37 & 0.00095 & -- & -- & -- & -- \\
        MuIC  & 160 & 1.4 & 0.0090 & -- & -- & -- \\
        MuIC2 & 300 & 6.5 & 0.22 & 0.029 & -- & -- \\
        LHmuC & 850 & 36 & 3.0 & 0.83 & 0.024 & -- \\

        \hline
    \end{tabular}
\end{table}

\subsubsection{Standard Model Production Cross Sections}\label{sec:sm_xsec}

Studies of the electroweak production of vector bosons and top quarks are essential ways to measure fundamental SM parameters, such as triple gauge boson couplings and CKM mixing matrix terms involving the top quark. 
The production of $W$ bosons was measured at HERA~\cite{ZEUS:2009W,H1:2009W,H1:ZEUS:2010W} 
with a total of only 23 identified events. 
The MuIC would operate at a much higher center-of-mass energy and with higher luminosity, yielding orders of magnitude more  $W$ and $Z$ bosons and top quarks.
This opens additional new opportunities for precision electroweak measurements beyond deep inelastic scattering,
adding to previous combinations based on data collected at LEP, the Tevatron, and the LHC~\cite{GFitter:2018}. The production of Higgs bosons is also very important, and is discussed separately in Section~\ref{sec:higgs_phys}.

\begin{figure}[!htb]
    \centering
    \includegraphics[width=0.32\textwidth]{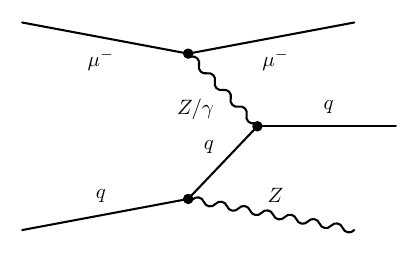}
    \includegraphics[width=0.32\textwidth]{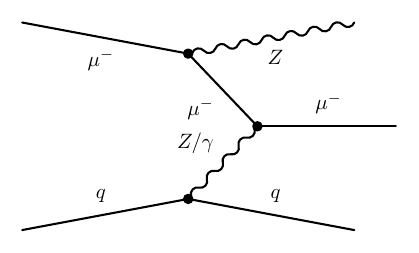}
    \includegraphics[width=0.32\textwidth]{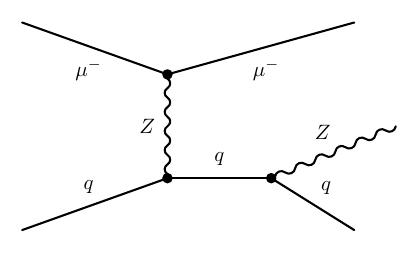}
    \caption{Diagrams of $Z$ boson production in association with $\mu^{-}$ as the scattered lepton in $\mu^{-}p$ collisions.}
    \label{fig:Zmu_diagrams}
\end{figure}

\begin{figure}[!htb]
    \centering
    \includegraphics[width=0.32\textwidth]{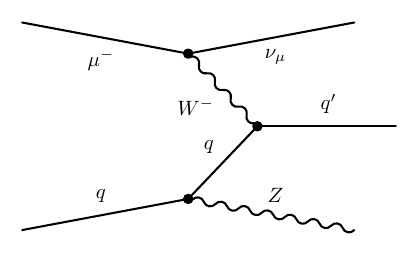}
    \includegraphics[width=0.32\textwidth]{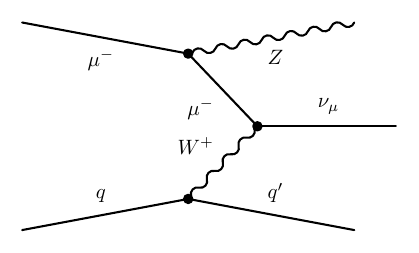}
    \includegraphics[width=0.32\textwidth]{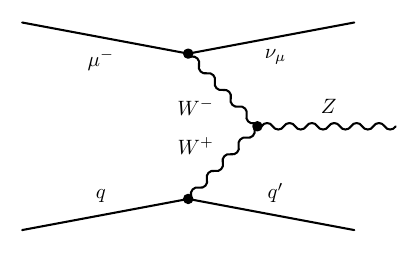}
    \includegraphics[width=0.32\textwidth]{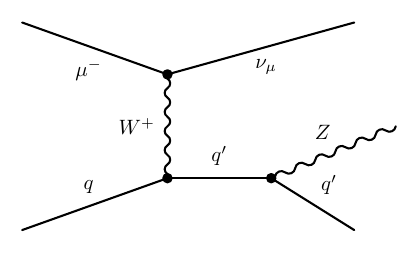}
    \includegraphics[width=0.32\textwidth]{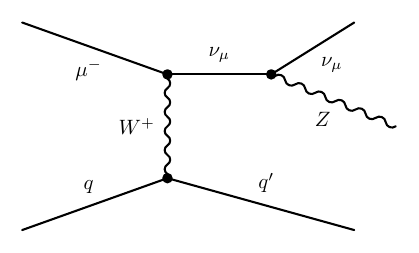}
    \caption{Diagrams of $Z$ boson production in association with $\nu_{\mu}$ as the scattered lepton in $\mu^{-}p$ collisions.}
    \label{fig:Znu_diagrams}
\end{figure}

\begin{figure}[!htb]
    \centering
    \includegraphics[width=0.32\textwidth]{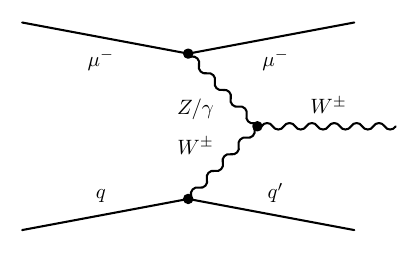}
    \includegraphics[width=0.32\textwidth]{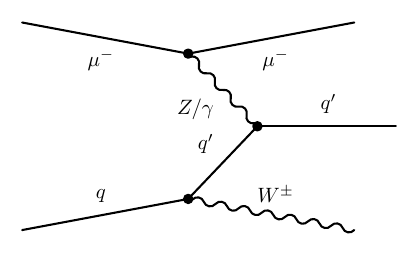}
    \includegraphics[width=0.32\textwidth]{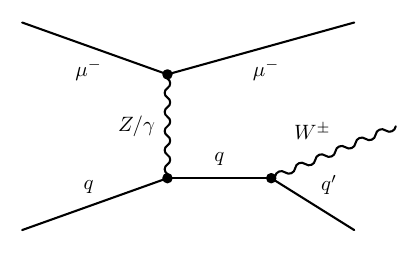}
    \caption{Diagrams of $W^{\pm}$ boson production in association with $\mu^{-}$ as the scattered lepton in $\mu^{-}p$ collisions. Both charges of the $W$ boson can be produced in these modes.}
    \label{fig:Wmu_diagrams}
\end{figure}

\begin{figure}[!htb]
    \centering
    \includegraphics[width=0.33\textwidth]{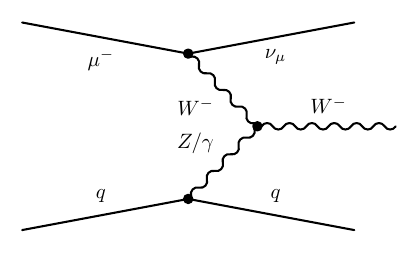}
    \includegraphics[width=0.33\textwidth]{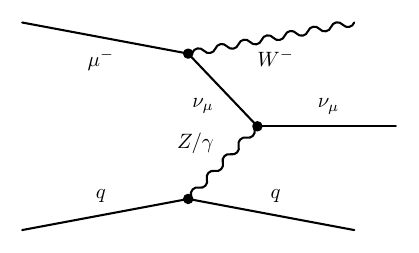}\\
    \includegraphics[width=0.33\textwidth]{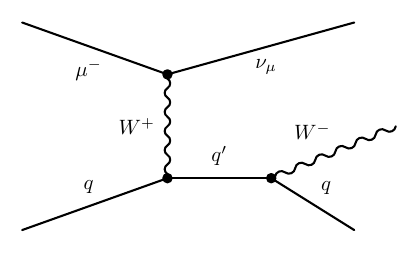}
    \includegraphics[width=0.33\textwidth]{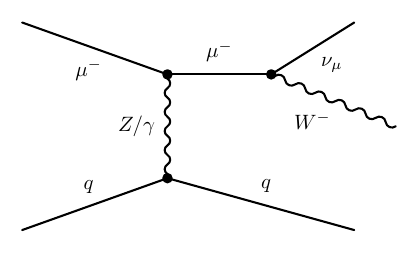}
    \caption{Diagrams of $W^{-}$ boson production in association with $\nu_{\mu}$ as the scattered lepton in $\mu^{-}p$ collisions. 
    Only negatively (positively) charged $W$ bosons are produced for initial state $\mu^-$ ($\mu^+$) leptons.}
    \label{fig:Wnu_diagrams}
\end{figure}

\begin{figure}[!htb]
    \centering
    \includegraphics[width=0.33\textwidth]{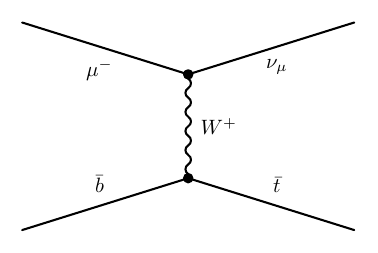}
    \includegraphics[width=0.33\textwidth]{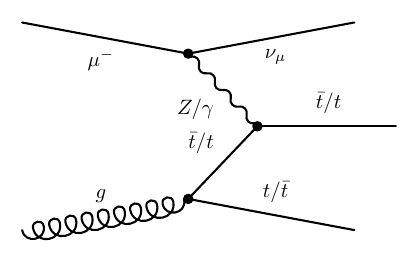}
    \caption{Diagrams of single $\bar{t}$ production and $t\bar{t}$ production in $\mu^{-}p$ collisions.}
    \label{fig:top_diagrams}
\end{figure}

The production of $W$ and $Z$ bosons in $\mu^- p$ collisions can be achieved via many diagrams, which can be categorized based on their final states into 5 processes: $Z\mu^-j$, $Z\nu_{\mu}j$, $W^+\mu^-j$, $W^-\mu^-j$, and $W^-\nu_{\mu}j$, where $j$ is the jet from the struck quark.
Figures~\ref{fig:Zmu_diagrams} and~\ref{fig:Znu_diagrams} show the leading order production diagrams of the $Z$ boson, with $\mu^{-}$ or $\nu_{\mu}$ as the scattered lepton, respectively. 
Figures~\ref{fig:Wmu_diagrams} and~\ref{fig:Wnu_diagrams} show the leading order production diagrams of the $W$ boson, with $\mu^{-}$ or $\nu_{\mu}$ as the scattered lepton, respectively.
We note that Figs.~\ref{fig:Znu_diagrams}--\ref{fig:Wnu_diagrams} include triple gauge boson terms. 
The production diagrams for top quarks are simpler at leading order, where
Fig.~\ref{fig:top_diagrams} shows the leading order production diagrams of single $\bar{t}$ quark and $t\bar{t}$ pair in $\mu^{-}p$ collisions. The corresponding diagrams for $\mu^+ p$ collisions can be obtained through the appropriate application of charge conjugation.

Cross sections for these SM particle productions are calculated with MadGraph~\cite{Madgraph}, version 3.3.1,  using the PDF set NNPDF31\_nlo\_pdfas~\cite{NNPDF:2017mvq} for the proton.
These cross sections are compiled in Tables~\ref{tab:muNeg_p_highE_Znu_xsec} to \ref{tab:muNeg_p_highE_ttbar_lepcut_xsec}.
The production processes for $Z\nu_{\mu}$, $W^-\nu_{\mu}$, and single $\bar{t}$ quark involve a $W$ boson exchange on the muon side, and the outgoing $\nu_{\mu}$ has a well-defined finite $p_{\rm T}$.
Processes for $Z\mu^{-}$, $W^{\pm}\mu^{-}$, and $t\bar{t}$ pair production, on the other hand, 
involve a $Z/\gamma$ exchange on the muon side.
Their cross sections increase significantly when the virtual photon mass is close to 0, in which case the $p_{\rm T}$ of the scattered muon also approaches 0.

In total, the inclusive $W^\pm$ production cross section in $\mu^{-}p$ collisions for the MuIC at BNL is 19.4~pb, yielding $2.1\times 10^4$ leptonic $W\to \ell \nu$ decays into each lepton flavor for 10~fb$^{-1}$ of integrated luminosity. This increases by an order of magnitude for the LHmuC configuration.

\begin{table*}[!htb]
    \centering
    \caption{Cross sections for the $Z\nu_{\mu}$ process in $\mu^{-}p$ collisions for different beam energy configurations. The $\mu^{-}$ beam energy is unpolarized in all cases.}
    \begin{tabular}{l|ccc}
        \hline
        $E_{\mu}\times E_{p}$ (TeV$^{2}$) & $\sigma$ (pb)  & Scale unc.&
        PDF$\oplus\alpha_{s}$ unc.    \\
        \hline
        $0.96 \times 0.275$   &
        0.34               &
        ${}^{+5.3\%}_{-4.6\%}$   &
        ${}^{+0.9\%}_{-0.9\%}$    \\
        $0.96 \times 0.96$    &
        1.16                &
        ${}^{+2.6\%}_{-2.4\%}$   &
        ${}^{+0.8\%}_{-0.8\%}$    \\
        $1.5 \times 7$        &
        6.49                &
        ${}^{+1.5\%}_{-1.9\%}$    &
        ${}^{+0.7\%}_{-0.7\%}$    \\
        $1.5 \times 13.5$     &
        9.51                &
        ${}^{+2.7\%}_{-3.1\%}$   &
        ${}^{+0.7\%}_{-0.7\%}$    \\
        $1.5 \times 20$       &
        11.9&
        ${}^{+3.4\%}_{-3.9\%}$   &
        ${}^{+0.7\%}_{-0.7\%}$  \\
        $1.5 \times 50$       &
        19.3                 &
        ${}^{+5.0\%}_{-5.6\%}$    &
        ${}^{+0.7\%}_{-0.7\%}$   \\
        \hline
    \end{tabular}
    \label{tab:muNeg_p_highE_Znu_xsec}
\end{table*}

\begin{table*}[!htb]
    \centering
    \caption{Cross sections for the $W^{-}\nu_{\mu}$ process in $\mu^{-}p$ collisions for different beam energy configurations. The $\mu^{-}$ beam energy is unpolarized in all cases.}
    \begin{tabular}{l|ccc}
        \hline
        $E_{\mu}\times E_{p}$ (TeV$^{2}$) & $\sigma$ (pb)  & Scale unc. &
        PDF$\oplus\alpha_{s}$ unc.   \\
        \hline
        $0.96 \times 0.275$   &
        1.80               &
        ${}^{+2.8\%}_{-5.6\%}$   &
        ${}^{+1.4\%}_{-1.4\%}$    \\
        $0.96 \times 0.96$    &
        7.47                &
        ${}^{+7.9\%}_{-11\%}$   &
        ${}^{+1.4\%}_{-1.4\%}$    \\
        $1.5 \times 7$        &
        52.8                &
        ${}^{+15\%}_{-17\%}$    &
        ${}^{+1.3\%}_{-1.3\%}$    \\
        $1.5 \times 13.5$     &
        79.8                &
        ${}^{+16\%}_{-18\%}$   &
        ${}^{+1.2\%}_{-1.2\%}$    \\
        $1.5 \times 20$       &
        100                &
        ${}^{+17\%}_{-19\%}$   &
        ${}^{+1.2\%}_{-1.2\%}$  \\
        $1.5 \times 50$       &
        167                 &
        ${}^{+19\%}_{-20\%}$    &
        ${}^{+1.2\%}_{-1.2\%}$   \\
        \hline
    \end{tabular}
    \label{tab:muNeg_p_highE_Wnu_xsec}
\end{table*}

\begin{table*}[!htb]
    \centering
    \caption{Cross sections for the $Z\mu^{-}$ process in $\mu^{-}p$ collisions for different beam energy configurations and with different cutoffs on the scattered muon $p_{\rm T}$. The listed cross sections are in pb, with scale and PDF$\oplus\alpha_{s}$ uncertainties. The $\mu^{-}$ beam energy is unpolarized in all cases.}
    \begin{tabular}{l|c|c|c|c}
        \hline
        $E_{\mu}\times E_{p}$ (TeV$^{2}$) & 
        Inclusive & 
        $p^{\ell}_{\rm T} >$ 1 GeV & 
        $p^{\ell}_{\rm T} >$ 2 GeV & 
        $p^{\ell}_{\rm T} >$ 5 GeV \\
        \hline
        $0.96 \times 0.275$   &
        3.33 ${}^{+0\%}_{-0.4\%}$ ${}^{+0.7\%}_{-0.7\%}$ &   
        0.73 ${}^{+0\%}_{-1.2\%}$ ${}^{+0.8\%}_{-0.8\%}$ &
        0.60 ${}^{+0\%}_{-1.3\%}$ ${}^{+0.8\%}_{-0.8\%}$ &
        0.44 ${}^{+0\%}_{-1.3\%}$ ${}^{+0.8\%}_{-0.8\%}$ \\
        $0.96 \times 0.96$    &
        7.44 ${}^{+2.7\%}_{-3.7\%}$ ${}^{+0.7\%}_{-0.7\%}$ &
        1.57 ${}^{+2.5\%}_{-4.5\%}$ ${}^{+0.7\%}_{-0.7\%}$ &
        1.31 ${}^{+2.7\%}_{-5.0\%}$ ${}^{+0.7\%}_{-0.7\%}$ &
        0.97 ${}^{+2.5\%}_{-4.7\%}$ ${}^{+0.7\%}_{-0.7\%}$ \\
        $1.5 \times 7$        &
        25.8 ${}^{+8.6\%}_{-9.5\%}$ ${}^{+0.8\%}_{-0.8\%}$ &
        5.24 ${}^{+9.6\%}_{-11\%}$ ${}^{+0.8\%}_{-0.8\%}$ &
        4.34 ${}^{+9.8\%}_{-11\%}$ ${}^{+0.8\%}_{-0.8\%}$ &
        3.26 ${}^{+10\%}_{-12\%}$ ${}^{+0.7\%}_{-0.7\%}$ \\
        $1.5 \times 13.5$     &
        34.5 ${}^{+10\%}_{-11\%}$ ${}^{+0.9\%}_{-0.9\%}$ &
        7.00 ${}^{+12\%}_{-12\%}$ ${}^{+0.8\%}_{-0.8\%}$ &
        5.81 ${}^{+11\%}_{-13\%}$ ${}^{+0.8\%}_{-0.8\%}$ &
        4.34 ${}^{+12\%}_{-13.0\%}$ ${}^{+0.8\%}_{-0.8\%}$ \\
        $1.5 \times 20$       &
        41.1 ${}^{+11\%}_{-12\%}$ ${}^{+1.0\%}_{-1.0\%}$ &
        8.31 ${}^{+12\%}_{-13\%}$ ${}^{+0.9\%}_{-0.9\%}$ &
        6.87 ${}^{+13\%}_{-14\%}$ ${}^{+0.8\%}_{-0.8\%}$ &
        5.14 ${}^{+13\%}_{-14\%}$ ${}^{+0.8\%}_{-0.8\%}$ \\
        $1.5 \times 50$       &
        59.9 ${}^{+13\%}_{-14\%}$ ${}^{+1.3\%}_{-1.3\%}$ &
        12.1 ${}^{+14\%}_{-15\%}$ ${}^{+1.1\%}_{-1.1\%}$ &
        10.1 ${}^{+15\%}_{-15\%}$ ${}^{+1.1\%}_{-1.0\%}$ &
        7.50 ${}^{+15\%}_{-16\%}$ ${}^{+1.0\%}_{-1.0\%}$ \\
        \hline
    \end{tabular}
    \label{tab:muNeg_p_highE_Zmu_lepcut_xsec}
\end{table*}

\begin{table*}[!htb]
    \centering
    \caption{Cross sections for the  $W^{+}\mu^{-}$ process in $\mu^{-}p$ collisions for different beam energy configurations and with different cutoffs on the scattered muon $p_{\rm T}$. The listed cross sections are in pb, with scale uncertainties and PDF$\oplus\alpha_{s}$ uncertainties. The $\mu^{-}$ beam energy is unpolarized in all cases.}
    \begin{tabular}{l|c|c|c|c}
        \hline
        $E_{\mu}\times E_{p}$ (TeV$^{2}$) &
        Inclusive &
        $p^{\ell}_{\rm T} >$ 1 GeV & 
        $p^{\ell}_{\rm T} >$ 2 GeV & 
        $p^{\ell}_{\rm T} >$ 5 GeV \\
        \hline
        $0.96 \times 0.275$   &
        8.93 ${}^{+1.0\%}_{-1.2\%}$ ${}^{+0.7\%}_{-0.7\%}$ &
        2.29 ${}^{+2.4\%}_{-2.5\%}$ ${}^{+0.8\%}_{-0.8\%}$ &
        1.86 ${}^{+2.6\%}_{-2.7\%}$ ${}^{+0.8\%}_{-0.8\%}$ &
        1.32 ${}^{+3.2\%}_{-3.1\%}$ ${}^{+0.8\%}_{-0.8\%}$ \\
        $0.96 \times 0.96$    &
        22.4 ${}^{+1.2\%}_{-1.7\%}$ ${}^{+0.7\%}_{-0.7\%}$ &
        6.19 ${}^{+0\%}_{-0.4\%}$ ${}^{+0.7\%}_{-0.7\%}$ &
        5.13 ${}^{+0\%}_{-0.3\%}$ ${}^{+0.7\%}_{-0.7\%}$ &
        3.77 ${}^{+0.4\%}_{-0.7\%}$ ${}^{+0.7\%}_{-0.7\%}$ \\
        $1.5 \times 7$        &
        90.1 ${}^{+6.0\%}_{-6.7\%}$ ${}^{+1.0\%}_{-1.0\%}$ &
        27.4 ${}^{+4.6\%}_{-5.3\%}$ ${}^{+0.8\%}_{-0.8\%}$ &
        23.1 ${}^{+4.3\%}_{-5.0\%}$ ${}^{+0.8\%}_{-0.8\%}$ &
        17.6 ${}^{+4.0\%}_{-4.6\%}$ ${}^{+0.8\%}_{-0.8\%}$ \\
        $1.5 \times 13.5$     &
        124  ${}^{+7.4\%}_{-8.0\%}$ ${}^{+1.1\%}_{-1.1\%}$ &
        38.7 ${}^{+5.9\%}_{-6.5\%}$ ${}^{+0.9\%}_{-0.9\%}$ &
        32.6 ${}^{+5.6\%}_{-6.3\%}$ ${}^{+0.9\%}_{-0.9\%}$ &
        25.0 ${}^{+5.2\%}_{-5.9\%}$ ${}^{+0.8\%}_{-0.8\%}$ \\
        $1.5 \times 20$       &
        150  ${}^{+8.1\%}_{-8.8\%}$ ${}^{+1.1\%}_{-1.1\%}$ &
        47.0 ${}^{+6.6\%}_{-7.3\%}$ ${}^{+0.9\%}_{-0.9\%}$ &
        40.0 ${}^{+6.4\%}_{-7.0\%}$ ${}^{+0.9\%}_{-0.9\%}$ &
        30.6 ${}^{+5.9\%}_{-6.5\%}$ ${}^{+0.9\%}_{-0.9\%}$ \\
        $1.5 \times 50$       &
        225  ${}^{+9.9\%}_{-10\%}$ ${}^{+1.3\%}_{-1.3\%}$ &
        72.8 ${}^{+8.4\%}_{-8.9\%}$ ${}^{+1.0\%}_{-1.0\%}$ &
        61.7 ${}^{+8.2\%}_{-8.7\%}$ ${}^{+1.0\%}_{-1.0\%}$ &
        47.8 ${}^{+7.7\%}_{-8.2\%}$ ${}^{+1.0\%}_{-1.0\%}$ \\
        \hline
    \end{tabular}
    \label{tab:muNeg_p_highE_Wpmu_lepcut_xsec}
\end{table*}

\begin{table*}[!htb]
    \centering
    \caption{Cross sections for the $W^{-}\mu^{-}$ process in $\mu^{-}p$ collisions for different beam energy configurations and with different cutoffs on the scattered muon $p_{\rm T}$. The listed cross sections are in pb, with scale and PDF$\oplus\alpha_{s}$ uncertainties. The $\mu^{-}$ beam energy is unpolarized in all cases.}
    \begin{tabular}{l|c|c|c|c}
        \hline
        $E_{\mu}\times E_{p}$ (TeV$^{2}$) &
        Inclusive &
        $p^{\ell}_{\rm T} >$ 1 GeV & 
        $p^{\ell}_{\rm T} >$ 2 GeV & 
        $p^{\ell}_{\rm T} >$ 5 GeV \\
        \hline
        $0.96 \times 0.275$   &
        8.69 ${}^{+0.7\%}_{-1.0\%}$ ${}^{+0.9\%}_{-0.9\%}$ &
        2.10 ${}^{+1.6\%}_{-2.0\%}$ ${}^{+0.9\%}_{-0.9\%}$ &
        1.71 ${}^{+1.8\%}_{-2.1\%}$ ${}^{+0.9\%}_{-0.9\%}$ &
        1.23 ${}^{+2.4\%}_{-2.4\%}$ ${}^{+0.9\%}_{-0.9\%}$ \\
        $0.96 \times 0.96$    &
        21.2 ${}^{+1.7\%}_{-2.3\%}$ ${}^{+0.8\%}_{-0.8\%}$ &
        5.76 ${}^{+0.7\%}_{-1.4\%}$ ${}^{+0.8\%}_{-0.8\%}$ &
        4.79 ${}^{+0.6\%}_{-1.2\%}$ ${}^{+0.8\%}_{-0.8\%}$ &
        3.57 ${}^{+0.2\%}_{-0.7\%}$ ${}^{+0.8\%}_{-0.8\%}$ \\
        $1.5 \times 7$ &
        86.7 ${}^{+6.7\%}_{-7.4\%}$ ${}^{+1.0\%}_{-1.0\%}$ &
        26.8 ${}^{+5.5\%}_{-6.3\%}$ ${}^{+0.9\%}_{-0.9\%}$ &
        22.8 ${}^{+5.4\%}_{-6.1\%}$ ${}^{+0.9\%}_{-0.9\%}$ &
        17.8 ${}^{+5.0\%}_{-5.7\%}$ ${}^{+0.8\%}_{-0.8\%}$ \\
        $1.5 \times 13.5$     &
        121  ${}^{+7.9\%}_{-8.6\%}$ ${}^{+1.1\%}_{-1.1\%}$ &
        38.3 ${}^{+6.8\%}_{-7.6\%}$ ${}^{+1.0\%}_{-1.0\%}$ &
        32.6 ${}^{+6.6\%}_{-7.4\%}$ ${}^{+0.9\%}_{-0.9\%}$ &
        25.6 ${}^{+6.2\%}_{-6.9\%}$ ${}^{+0.9\%}_{-0.9\%}$ \\
        $1.5 \times 20$       &
        145  ${}^{+8.6\%}_{-9.3\%}$ ${}^{+1.2\%}_{-1.2\%}$ &
        47.0 ${}^{+7.4\%}_{-8.2\%}$ ${}^{+1.0\%}_{-1.0\%}$ &
        40.1 ${}^{+7.4\%}_{-8.1\%}$ ${}^{+1.0\%}_{-1.0\%}$ &
        31.6 ${}^{+7.0\%}_{-7.7\%}$ ${}^{+0.9\%}_{-0.9\%}$ \\
        $1.5 \times 50$       &
        221  ${}^{+11\%}_{-11\%}$ ${}^{+1.4\%}_{-1.4\%}$ &
        73.6 ${}^{+9.3\%}_{-9.9\%}$ ${}^{+1.1\%}_{-1.1\%}$ &
        63.3 ${}^{+9.0\%}_{-9.7\%}$ ${}^{+1.1\%}_{-1.1\%}$ &
        50.3 ${}^{+8.6\%}_{-9.3\%}$ ${}^{+1.2\%}_{-1.1\%}$ \\
        \hline
    \end{tabular}
    \label{tab:muNeg_p_highE_Wnmu_lepcut_xsec}
\end{table*}

\begin{table*}[!htb]
    \centering
    \caption{Cross sections for the $\bar{t}\,\nu_{\mu}$ process in $\mu^{-}p$ collisions for different beam energy configurations. The $\mu^{-}$ beam energy is unpolarized in all cases.}
    \begin{tabular}{l|ccc}
        \hline
        $E_{\mu}\times E_{p}$ (TeV$^{2}$) & $\sigma$ (pb)  & Scale unc. &
        PDF$\oplus\alpha_{s}$ unc.    \\
        \hline
        $0.96 \times 0.275$   &
        0.95               &
        ${}^{+5.7\%}_{-7.9\%}$   &
        ${}^{+2.5\%}_{-2.5\%}$    \\
        $0.96 \times 0.96$    &
        5.44                &
        ${}^{+8.1\%}_{-10\%}$   &
        ${}^{+1.8\%}_{-1.8\%}$    \\
        $1.5 \times 7$        &
        48.1                &
        ${}^{+12\%}_{-14\%}$    &
        ${}^{+1.5\%}_{-1.5\%}$    \\
        $1.5 \times 13.5$     &
        75.0                &
        ${}^{+13\%}_{-15\%}$   &
        ${}^{+1.4\%}_{-1.4\%}$    \\
        $1.5 \times 20$       &
        96.1                &
        ${}^{+14\%}_{-15\%}$   &
        ${}^{+1.4\%}_{-1.4\%}$  \\
        $1.5 \times 50$       &
        164.1                 &
        ${}^{+15\%}_{-16\%}$    &
        ${}^{+1.4\%}_{-1.4\%}$   \\
        \hline
    \end{tabular}
    \label{tab:muNeg_p_highE_tbar_xsec}
\end{table*}

\begin{table*}[!htb]
    \centering
    \caption{Cross sections for the $t\bar{t}\,\mu^{-}$ process in $\mu^{-}p$ collisions for different beam energy configurations and with different cutoffs on the scattered muon $p_{\rm T}$. The listed cross sections are in pb, with scale and PDF$\oplus\alpha_{s}$ uncertainties. The $\mu^{-}$ beam energy is unpolarized in all cases.}
    \begin{tabular}{l|c|c|c|c}
        \hline
        $E_{\mu}\times E_{p}$ (TeV$^{2}$) & 
        Inclusive &
        $p^{\ell}_{\rm T} >$ 1 GeV & 
        $p^{\ell}_{\rm T} >$ 2 GeV & 
        $p^{\ell}_{\rm T} >$ 5 GeV \\
        \hline
        $0.96 \times 0.275$   &
        0.014  ${}^{+26\%}_{-19\%}$ ${}^{+3.2\%}_{-3.2\%}$ &
        0.0056 ${}^{+26\%}_{-19\%}$ ${}^{+3.6\%}_{-3.6\%}$ &
        0.0048 ${}^{+26\%}_{-19\%}$ ${}^{+3.6\%}_{-3.6\%}$ &
        0.0038 ${}^{+26\%}_{-19\%}$ ${}^{+3.8\%}_{-3.8\%}$ \\
        $0.96 \times 0.96$    &
        0.221 ${}^{+19\%}_{-15\%}$ ${}^{+1.7\%}_{-1.7\%}$ &
        0.099 ${}^{+20\%}_{-15\%}$ ${}^{+1.7\%}_{-1.7\%}$ &
        0.087 ${}^{+20\%}_{-15\%}$ ${}^{+1.8\%}_{-1.8\%}$ &
        0.072 ${}^{+20\%}_{-15\%}$ ${}^{+1.8\%}_{-1.8\%}$ \\
        $1.5 \times 7$        &
        4.62 ${}^{+11\%}_{-9.2\%}$ ${}^{+1.1\%}_{-1.1\%}$ &
        2.28 ${}^{+11\%}_{-9.5\%}$ ${}^{+1.1\%}_{-1.1\%}$ &
        2.05 ${}^{+11\%}_{-9.5\%}$ ${}^{+1.1\%}_{-1.1\%}$ &
        1.76 ${}^{+11\%}_{-9.6\%}$ ${}^{+1.1\%}_{-1.1\%}$ \\
        $1.5 \times 13.5$     &
        8.24 ${}^{+11\%}_{-9.4\%}$ ${}^{+1.0\%}_{-1.0\%}$ &
        4.12 ${}^{+11\%}_{-9.1\%}$ ${}^{+1.0\%}_{-1.0\%}$ &
        3.72 ${}^{+10\%}_{-9.0\%}$ ${}^{+1.0\%}_{-1.0\%}$ &
        3.20 ${}^{+10\%}_{-8.9\%}$ ${}^{+1.0\%}_{-1.0\%}$ \\
        $1.5 \times 20$       &
        11.3 ${}^{+12\%}_{-10\%}$ ${}^{+1.0\%}_{-1.0\%}$ &
        5.70 ${}^{+12\%}_{-10\%}$ ${}^{+1.0\%}_{-1.0\%}$ &
        5.16 ${}^{+11\%}_{-9.9\%}$ ${}^{+1.0\%}_{-1.0\%}$ &
        4.43 ${}^{+11\%}_{-9.9\%}$ ${}^{+1.0\%}_{-1.0\%}$ \\
        $1.5 \times 50$       &
        22.2 ${}^{+14\%}_{-12\%}$ ${}^{+1.0\%}_{-1.0\%}$ &
        11.3 ${}^{+14\%}_{-12\%}$ ${}^{+1.0\%}_{-1.0\%}$ &
        10.3 ${}^{+14\%}_{-12\%}$ ${}^{+1.0\%}_{-1.0\%}$ &
        8.90 ${}^{+14\%}_{-12\%}$ ${}^{+1.0\%}_{-1.0\%}$ \\
        \hline
    \end{tabular}
    \label{tab:muNeg_p_highE_ttbar_lepcut_xsec}
\end{table*}

\begin{figure}[!htb]
    \centering
    \includegraphics[width=0.45\textwidth]{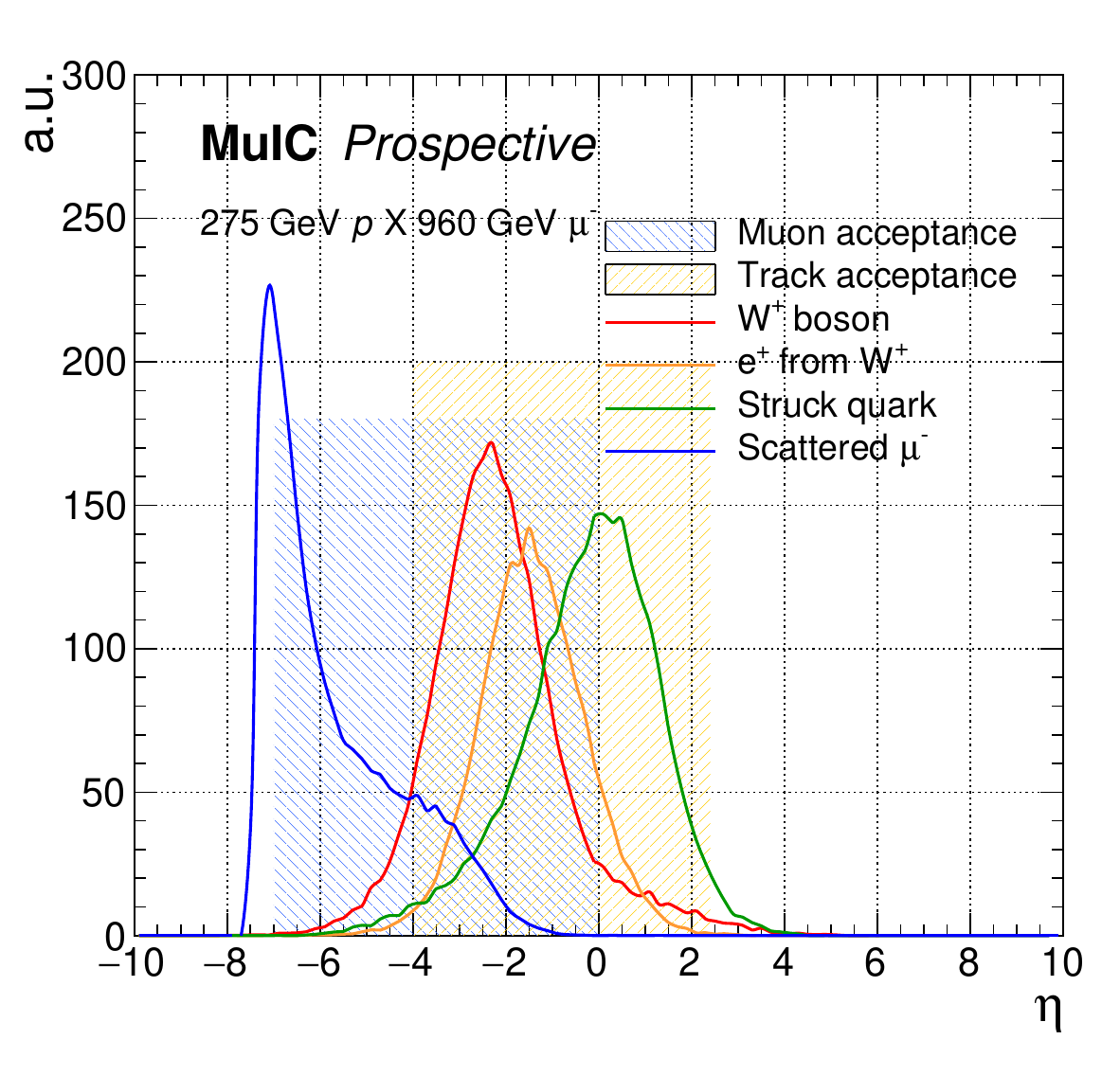}
    \includegraphics[width=0.45\textwidth]{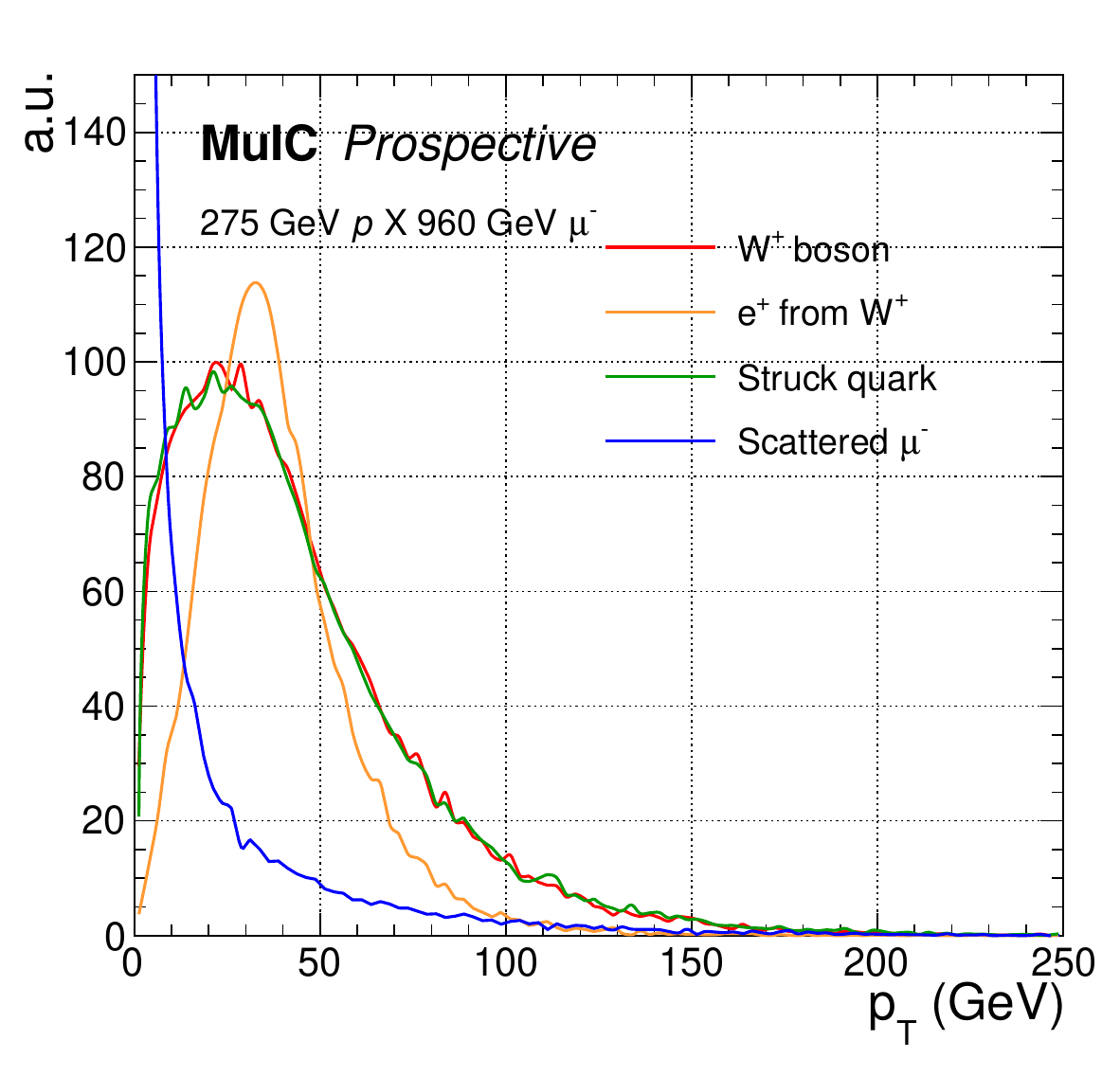}
    \caption{Generator-level kinematic distributions of $\eta$ (left) and $p_{\rm T}$ (right) for different outgoing particles in $W^{+}\mu^{-} \to e^{+}\nu_{e}\mu^{-}$ events at $E_{\mu^{-}} = 960$ GeV and $E_{p} = 275$ GeV collisions. A fiducial cut on the outgoing scattered muon $p^{\ell}_{\rm T} > 1$~GeV is applied to this sample.
    }
    \label{fig:W_gen_dists}
\end{figure}

A fiducial selection on the scattered muon is important to tag  $Z/\gamma$ exchange diagrams from those with $W$ exchange from the incoming muon, and to separate to some degree the pure photon exchange diagram. 
Therefore, for these processes, Tables~\ref{tab:muNeg_p_highE_Zmu_lepcut_xsec}, \ref{tab:muNeg_p_highE_Wpmu_lepcut_xsec}, \ref{tab:muNeg_p_highE_Wnmu_lepcut_xsec}, and \ref{tab:muNeg_p_highE_ttbar_lepcut_xsec} list the inclusive cross sections without any fiducial selection and with selections requiring $p^{\ell}_{\rm T} > $ 1, 2, and 5 GeV, as may be required to scatter into the acceptance of a detector.
As an example, the kinematic profile of $W^{+}\mu^{-}$ events are shown in Fig.~\ref{fig:W_gen_dists}. This process (illustrated in Fig.~\ref{fig:Wmu_diagrams}) can happen with a photon or $Z$ boson exchange from the muon leg. The pseudorapidity distribution of the scattered muon shows a peaking structure at large negative 
$\eta$,\footnote{We assume the convention of DIS colliders that the hadron beam defines the $+z$ direction.} 
which comes from the diagrams with photon exchange where the virtual photon mass is close to 0. The cutoff towards $\eta=-8$ is because of the cutoff on lepton $p_{\rm T}$. The small bump on the right tail of this $\eta$ distribution shows that diagrams with $Z$ boson exchange start to be the main contribution in this region.  The scattered muon, struck parton, and $W$ boson decay products all fall within a central tracking acceptance of $-4 < \eta < 2.4$. The scattered muon with $p^{\ell}_{\rm T} > $ 1 GeV would fall within a muon spectrometer acceptance of $\eta\gtrsim -7$, as would be required for low $Q^2$ DIS measurements as well (see Appendix~\ref{sec:dis-kin}).

\subsubsection{Higgs Boson Production}
\label{sec:higgs_phys}

The prospects for Higgs boson measurements at a lepton-hadron collider are exciting as the experimental environment can be cleaner than at hadron colliders, which suffer from pileup effects and large QCD cross sections. However, the effects of beam-induced backgrounds will need to be assessed experimentally for $\mu p$ collisions. The environment at $e^+e^-$ (or $\mu^+\mu^-$) ``Higgs factories'' would be cleaner still and may offer larger luminosity. Nevertheless, the possibility to measure some decay modes that are extremely difficult at hadron colliders, such as bottom and charm decays and possibly even gluon or light quark decays with sufficient integrated luminosity, should not be discounted.

\begin{figure}[!htb]
    \centering
    \includegraphics[width=0.45\textwidth]{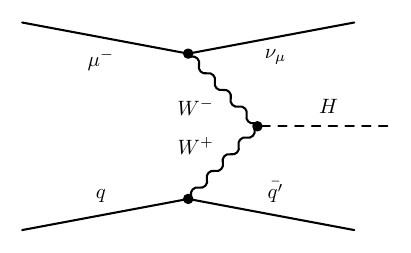}
    \includegraphics[width=0.45\textwidth]{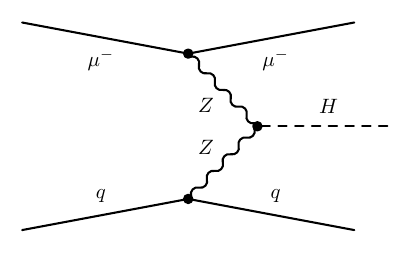}
    \caption{Vector boson fusion Higgs production mode via charged current (left) and neutral current (right) exchanges.}
    \label{fig:VBF_diagrams}
\end{figure}

\begin{figure}
    \centering
    \includegraphics[width=0.45\textwidth]{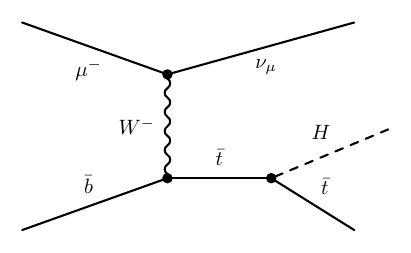}
    \includegraphics[width=0.45\textwidth]{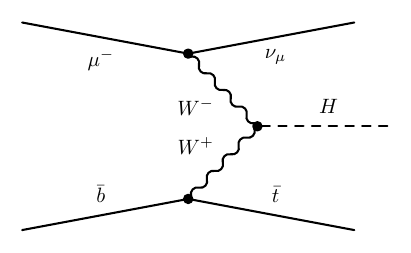}
    \caption{Higgs production associated with a top quark, in the Higgs-strahlung mode (left) and VBF mode (right).}
    \label{fig:tH_diagrams}
\end{figure}

The main production mechanism of Higgs bosons in $\mu^- p$ collisions, as is also the case in multi-TeV $\mu^+\mu^-$ collisions, is vector boson fusion (VBF) via charged-current or neutral-current exchanges. 
Figure~\ref{fig:VBF_diagrams} shows the leading order (LO) diagrams for such production.
The VBF mode here does not consider a top quark in the final state.
The Higgs boson can also be produced in association with a top quark, in the VBF or Higgs-strahlung processes, as shown in Fig.~\ref{fig:tH_diagrams}.
These two modes together are referred to as the $t\,H$ mode in the following.

The Higgs boson production cross sections at the MuIC have been calculated with MadGraph \cite{Madgraph}, version 3.3.1, using the PDF set PDF4LHC15\_nlo\_mc\_pdfas~\cite{Butterworth:2059563} for the proton. These cross sections are shown in Table~\ref{tab:muNeg_p_xsec} for $\mu^{-}$ beams and Table~\ref{tab:muPos_p_xsec} for $\mu^{+}$ beams using the nominal beam energies of $E_{\mu} = 960$~GeV and $E_{p} = 275$~GeV and for several different muon beam polarization configurations. The total cross section as well as the production process cross sections are reported. For unpolarized beams, the total Higgs boson cross section is 77~fb for $\mu^- p$ collisions at the MuIC. This is comparable to the cross section of 130~fb at the proposed LHeC with a 60~GeV electron beam energy \cite{Agostini:2020fmq}. In fact, as will be noted below, the Higgs decay products are more central in collisions at the MuIC than at the LHeC, and should lead to a relatively larger cross section in the experiment acceptance. The production cross section can be enhanced with appropriate longitudinal polarization of the muon beam.

Scenarios with higher beam energies are also studied. Table~\ref{tab:muNeg_p_highE_xsec} shows the CC and NC Higgs boson production cross sections, along with uncertainties, in several scenarios with different $\mu^{-}$ and proton energies. 
The listed uncertainties arise from the renormalization and factorization scale uncertainties ($\mu_{R}$ and $\mu_{F}$, which are varied by a factor of 2 around the $Z$ boson mass),  and from PDF set variations including also $\alpha_{s}$ variations.
As shown in the table, the leading uncertainty is the scale uncertainty.
The PDF uncertainty is always around 1\% (smaller than the scale uncertainty), while the $\alpha_{s}$ uncertainty is negligible.

\begin{table}[!htb]
    \centering
    \caption{Cross sections, in fb, for 125 GeV Higgs boson production in  $\mu^{-}p$ scattering. The $\mu^{-}$ beam energy is 960 GeV and the proton beam energy is 275 GeV. P is the polarization of the muon beam.}
    \begin{tabular}{l|cccccccc}
         \hline
                     & P = $-40\%$ & P = $-20\%$ & P = $-10\%$ & P = 0 \% & 
                       P = 10\%  & P = 20\%  & P = 40\%  & P = 100\% \\
         \hline
         $\sigma_{CC}$ & 91.1      & 78.2      & 71.7      & 65.1     &
                       58.8      & 52.1      & 39.0      & 0 \\
         $\sigma_{NC}$ & 12.6      & 12.1      & 11.9      & 11.6     &
                       11.4      & 11.1      & 10.5      & 8.9 \\
         $\sigma_{tH}$ & 0.0224    & 0.0187    & 0.0174    & 0.0158   &
                       0.0139    & 0.0128    & 0.0096    & 0   \\
         \hline
         total       & 103.7     & 90.3      & 83.6      & 76.7     &
                       70.2      & 63.2      & 49.5      & 8.9 \\
         \hline
    \end{tabular}
    \label{tab:muNeg_p_xsec}
\end{table}

\begin{table}[!htb]
    \centering
    \caption{Cross sections, in fb, for 125 GeV Higgs boson production in $\mu^{+}p$ scattering. The $\mu^{+}$ beam energy is 960 GeV and the proton beam energy is 275 GeV. P is the polarization of the muon beam.}
    \begin{tabular}{l|cccccccc}
         \hline
                     & P = 40\%  & P = 20\%  & P = 10\%  & P = 0 \% & 
                       P = $-10\%$ & P = $-20\%$ & P = $-40\%$ & P = $-100\%$ \\
         \hline
         $\sigma_{CC}$ & 45.0      & 38.2      & 35.6      & 32.1     &
                       28.9      & 25.6      & 19.2      & 0 \\
         $\sigma_{NC}$ & 12.4      & 12.0      & 11.7      & 11.6     &
                       11.3      & 11.0      & 10.6      & 9.1 \\
         $\sigma_{tH}$ & 0.0220    & 0.0190    & 0.0173    & 0.0157   &
                       0.0142    & 0.0127    & 0.0093    & 0   \\
         \hline
         total       & 57.4      & 50.2      & 47.3      & 43.7     &
                       40.2      & 36.6      & 29.8      & 9.1 \\
         \hline
    \end{tabular}
    \label{tab:muPos_p_xsec}
\end{table}

The $t\,H$ process is subject to significant destructive interference between the Higgs-strahlung and the VBF-tH diagrams. Its cross section is unobservable at low collision energies but may become visible at high energies. The cross section for the $t\,H$ process is summarized in Table~\ref{tab:muNeg_p_highE_tH_xsec}.

\begin{table}[!htb]
    \centering
    \caption{The Higgs boson cross section for different beam energy scenarios, separately for CC and NC exchange. The $\mu^{-}$ beam energy is unpolarized in all cases. Uncertainties arising from the scale and from the PDF and $\alpha_{s}$ variations are also listed, as discussed in the text.}
    \begin{tabular}{l|ccc|ccc}
        \hline
        $E_{\mu}\times E_{p}$ (TeV$^{2}$) & $\sigma_{CC}$ (fb)  & Scale unc. &
        PDF$\oplus\alpha_{s}$ unc. & $\sigma_{NC}$ (fb) & Scale unc.        & PDF$\oplus\alpha_{s}$ unc. \\
        \hline
        $0.96 \times 0.275$          &
        64.5                       &
        ${}^{+6.5\%}_{-5.5\%}$   &
        ${}^{+1.3\%}_{-1.3\%}$ &
        11.6                       &
        ${}^{+6.1\%}_{-5.2\%}$   &
        ${}^{+1.2\%}_{-1.2\%}$   \\
        $0.96 \times 0.96$           &
        235                        &
        ${}^{+3.4\%}_{-3.1\%}$   &
        ${}^{+1.2\%}_{-1.2\%}$  &
        47.9                       &
        ${}^{+2.5\%}_{-2.4\%}$   &
        ${}^{+1.0\%}_{-1.0\%}$  \\
        $1.5 \times 7$               &
        1337                       &
        ${}^{+1.5\%}_{-2.0\%}$   &
        ${}^{+1.1\%}_{-1.1\%}$    &
        327                        &
        ${}^{+2.9\%}_{-3.6\%}$   &
        ${}^{+1.1\%}_{-1.1\%}$   \\
        $1.5 \times 13.5$           &
        1955                       &
        ${}^{+2.9\%}_{-3.4\%}$    &
        ${}^{+1.2\%}_{-1.2\%}$   &
        496                        &
        ${}^{+4.3\%}_{-5.1\%}$   &
       ${}^{+1.2\%}_{-1.2\%}$   \\
        $1.5 \times 20$             &
        2422                       &
        ${}^{+3.6\%}_{-4.2\%}$    &
        ${}^{+1.2\%}_{-1.2\%}$  &
        628                        &
        ${}^{+5.2\%}_{-6.0\%}$   &
        ${}^{+1.2\%}_{-1.2\%}$   \\
        $1.5 \times 50$              &
        3883                       &
        ${}^{+5.5\%}_{-6.1\%}$    &
        ${}^{+1.4\%}_{-1.4\%}$   &
        1053                       &
        ${}^{+7.0\%}_{-7.9\%}$   &
        ${}^{+1.4\%}_{-1.4\%}$   \\
        \hline
    \end{tabular}
    \label{tab:muNeg_p_highE_xsec}
\end{table}

\begin{table}[!htb]
    \centering
    \caption{tH cross section in different beam energy scenarios. The $\mu^{-}$ beam energy is unpolarized in all cases. Uncertainties arising from the scale and PDF set are also listed, as discussed in the text.}
    \begin{tabular}{l|ccc}
        \hline
        $E_{\mu}\times E_{p}$ (TeV$^{2}$) & $\sigma_{CC}$ (fb)  & Scale unc. &
        PDF$\oplus\alpha_{s}$ unc. \\
        \hline
        $0.96 \times 0.275$   &
        0.019               &
        ${}^{+0.8\%}_{-2.7\%}$   &
        ${}^{+14\%}_{-14\%}$   \\
        $0.96 \times 0.96$    &
        0.54                &
        ${}^{+3.8\%}_{-5.8\%}$   &
        ${}^{+6.3\%}_{-6.3\%}$   \\
        $1.5 \times 7$        &
        24.1                &
        ${}^{+7.7\%}_{-9.6\%}$    &
        ${}^{+3.5\%}_{-3.5\%}$  \\
        $1.5 \times 13.5$     &
        48.7                &
        ${}^{+8.7\%}_{-10\%}$   &
        ${}^{+3.2\%}_{-3.2\%}$  \\
        $1.5 \times 20$       &
        71.8                &
        ${}^{+9.2\%}_{-11\%}$   &
        ${}^{+3.0\%}_{-3.0\%}$  \\
        $1.5 \times 50$       &
        161                 &
        ${}^{+10\%}_{-12\%}$    &
        ${}^{+2.6\%}_{-2.6\%}$  \\
        \hline
    \end{tabular}
    \label{tab:muNeg_p_highE_tH_xsec}
\end{table}

\begin{figure}[!htb]
    \centering
    \includegraphics[width=0.45\textwidth]{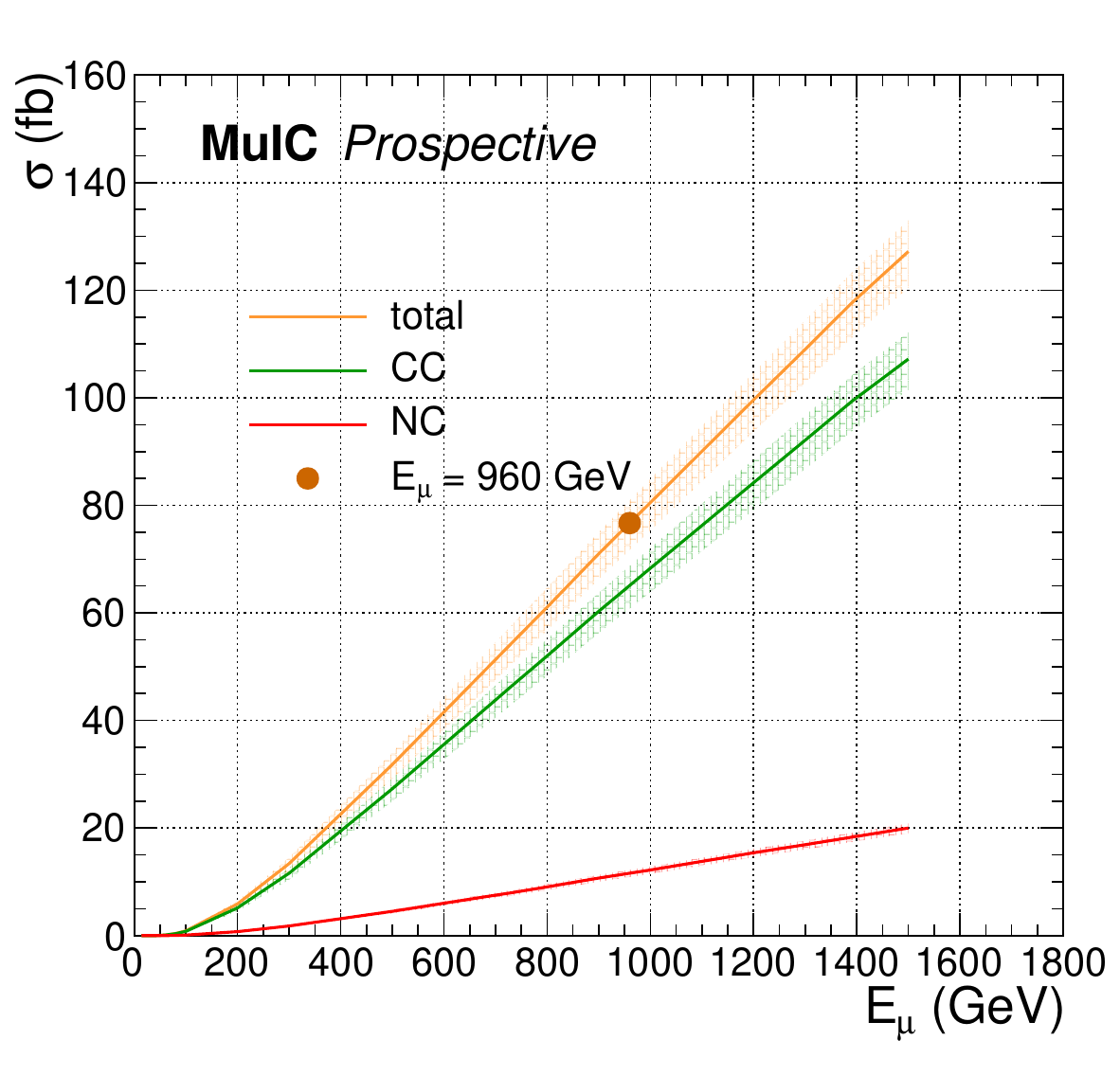}
    \includegraphics[width=0.45\textwidth]{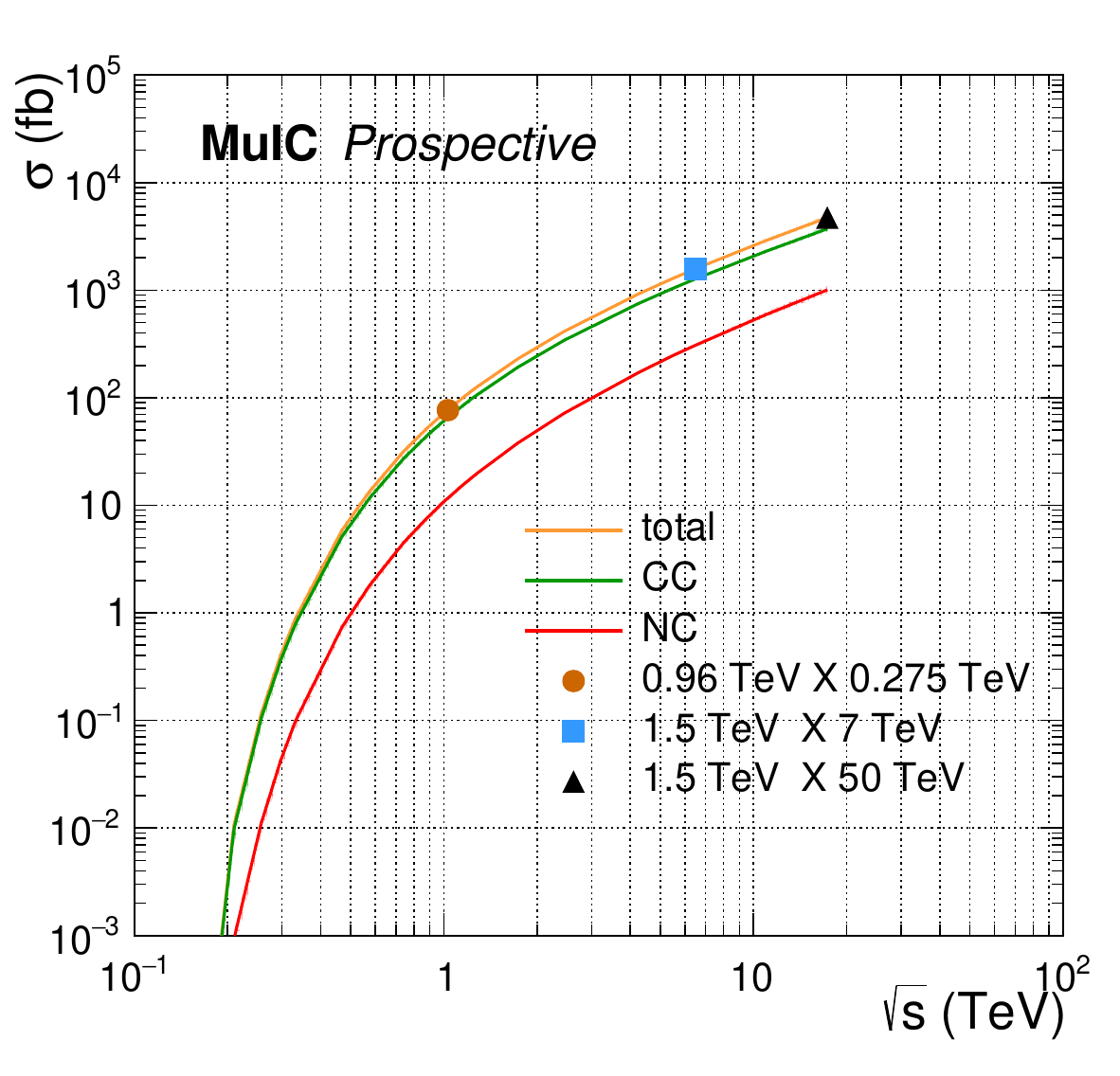}
    \caption{Left: The Higgs boson cross sections as a function of the $\mu^{-}$ beam energy, with the proton energy fixed at 275~GeV. Right: The Higgs boson cross section scan as a function of the $\mu^{-} p$ center-of-mass energy $\sqrt{s}$. Shaded bands shows the scale uncertainties. Nominal beam energy choices for the machine configurations studied here are indicated by the symbols. }
    \label{fig:muNeg_xsec}
\end{figure}

A scan of the Higgs production cross section in $\mu^{-}p$ collisions is also performed for different beam energies, as shown in Fig.~\ref{fig:muNeg_xsec}.  The left plot shows the cross section as a function of the $\mu^{-}$ beam energy with the proton beam held fixed at the EIC energy of 275~GeV. The right plot shows the cross section as a function of the $\mu^- p$ center-of-mass energy. The muon beam is assumed unpolarized, and only LO diagrams are considered. The cross section exceeds 1~pb for $\sqrt{s} \gtrsim 5$~TeV. In comparison, we note that the cross section for Higgs production at a $\mu^+\mu^-$ collider operating at $\sqrt{s}=3$~TeV center is 500~fb \cite{Costantini:2020stv}. This is actually about 3 times less than the cross section possible when colliding one of the 1.5~TeV muon beams of that collider with a 7~TeV LHC proton beam. 

The SM Higgs boson decays dominantly into a $b\bar{b}$ pair\footnote{For single $b$ quarks, charge conjugation is implied hereinafter, unless stated otherwise.} with a branching ratio of 58\%. The observation of this decay mode has been reported by both the ATLAS and CMS Collaborations~\cite{atlas2018hbb, cms2018hbb}. The inclusive NC+CC production yield of $H\to b\bar{b}$ in unpolarized 960~GeV $\mu^- \times$ 275~GeV $p$ collisions at the MuIC for 400~fb$^{-1}$ of integrated luminosity (see Section~\ref{sec:lumi}) is $17\,800$. This number grows by about a factor 4 for 2~TeV collisions (MuIC2) at the same luminosity, and by a factor 13 for 6.5~TeV collisions (LHmuC) at a lower integrated luminosity of 237 fb$^{-1}$. 
This opens new opportunities to measure the $Hbb$ coupling strength, and possibly others, at the MuIC.

\begin{figure}[!htb]
    \centering
    \includegraphics[width=0.45\textwidth]{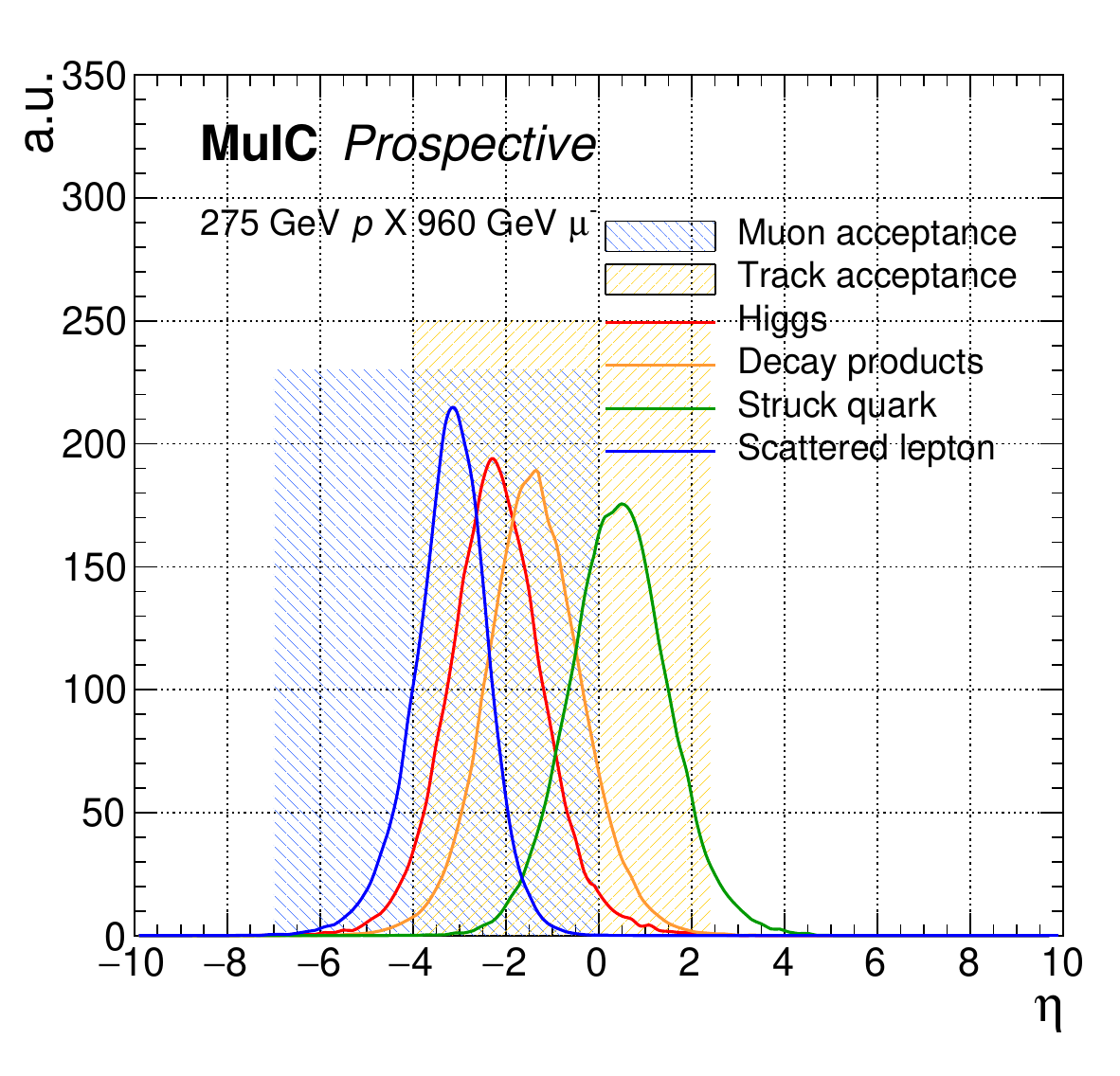}
    \includegraphics[width=0.45\textwidth]{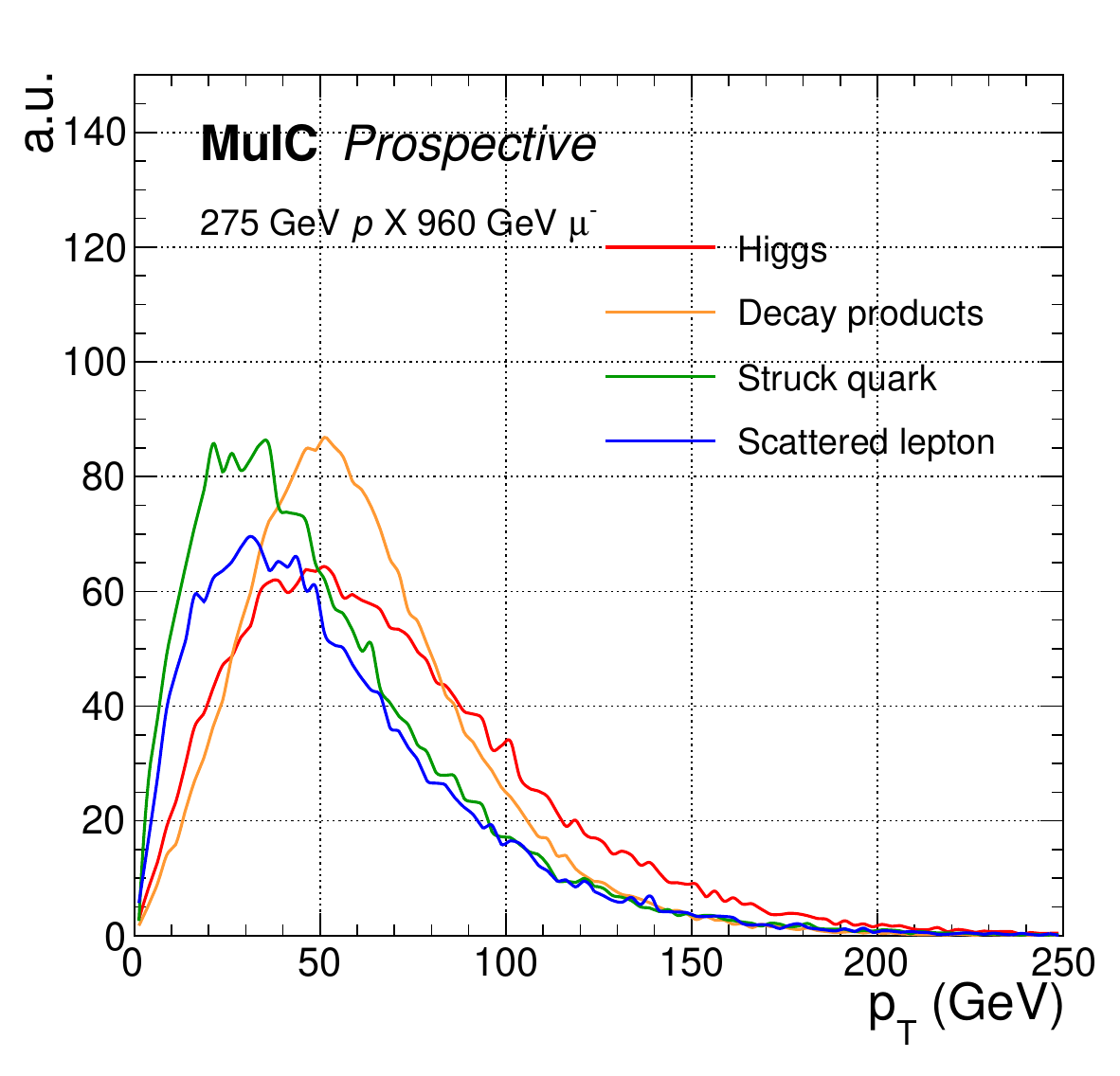}
    \caption{Generator-level kinematic distributions of $\eta$ (left) and $p_{\rm T}$ (right) for different final state objects in $H \to b\bar{b}$ events in $\mu^- p$ collisions with $E_{\mu} = 960$ GeV and $E_{p} = 275$ GeV. Nominal tracking and muon acceptance regions as would be required for DIS measurements are indicated by the hatched areas.}
    \label{fig:H_gen_dists}
\end{figure}

The kinematics of Higgs boson production and decay at the MuIC have been studied, without losing generality, using simulated $H \to b\bar{b}$ samples. The hard process is generated with MadGraph5~\cite{Madgraph} while the shower activities are simulated with Pythia~8.3 \cite{Sjostrand:2014zea}, with $E_{p} = 275$~GeV and $E_{\mu} = 960$~GeV for the NC and CC collisions.
Figure~\ref{fig:H_gen_dists} shows the generator-level distributions of $\eta$  and $p_{\rm T}$ of the Higgs boson, the Higgs decay products, the struck quark, and the scattered lepton. The Higgs decay products and the scattered quark are quite central. A tracker acceptance spanning the range $-4 < \eta < 2.4$ covers nearly all of the final state particles, and a muon acceptance extended down to $\eta = -7$ in the muon beam direction as would be required for low $Q^2$ DIS measurements at the MuIC (see Appendix~\ref{sec:dis-kin}) easily covers the NC final state muon.  This is in contrast to the distributions at the LHeC (see Ref.~\cite{Agostini:2020fmq}), where the Higgs decay products peak in the forward (proton) direction around $\eta \approx 2$ and extend as far forward as $\eta \approx 6$. The struck quark is even more forward peaking at $\eta \approx 5$ and extending further. Thus the Higgs decays are more central in a MuIC detector and should lead to a larger acceptance.

To estimate the sensitivity for the $Hbb$ coupling measurement at the MuIC, the background processes  $Z
\mu^-, Z \to b\bar{b}$ (Z\_NC), $Z\nu_{\mu}, Z \to b\bar{b}$ (Z\_CC), and gluon-initiated $b\bar{b}$ production (DIS\_bb) are also considered. The detector simulation is performed on both the signal and background samples using Delphes~\cite{delphes}, with the detector parameterization described in Appendix~\ref{sec:delphes}. The Higgs and $Z$ boson processes are simulated in their full inclusive phase space. The DIS process, which is by far the dominant background, is generated in a fiducial phase space of 50~GeV~$< m_{b\bar{b}} <$~170~GeV, $p_{\rm T}(b) >$~20~GeV, and $|\eta(b)|<$~6.
For this preliminary estimate, other background processes such as the light-flavor fake backgrounds and multi-jet backgrounds are not considered.

Object and event selections are applied to enhance the $H \to b\bar{b}$ signal over background. Jets are selected with $p_{\rm T} > $~25~GeV and $-5 < \eta < 2.4$, while the b-tagging is only available within the tracker acceptance $-4 < \eta < 2.4$. Muons are selected with $p_{\rm T} >$ 5~GeV in the central muon system $-4 < \eta < 0$, or without $p_{\rm T}$ requirement in the far-backward region $-7 < \eta < -4$. B-jets are rejected if they are within a $\Delta R = 0.4$ cone around any selected muon. The baseline event selection requires 2 selected b-tagged jets in the event. To target the H\_CC process and reduce the large DIS background, an additional set of selections are applied: no selected muon in the event, a light flavor jet (assumed from the struck quark) in addition to the 2 selected b-jets in the event, $p_{\rm T}(H) > 20$~GeV for the dijet Higgs candidate, and missing transverse energy $E_{\rm T}^{\text{miss}} > 30$~GeV. 

\begin{figure}[!htb]
    \centering
    \includegraphics[width=0.45\textwidth]{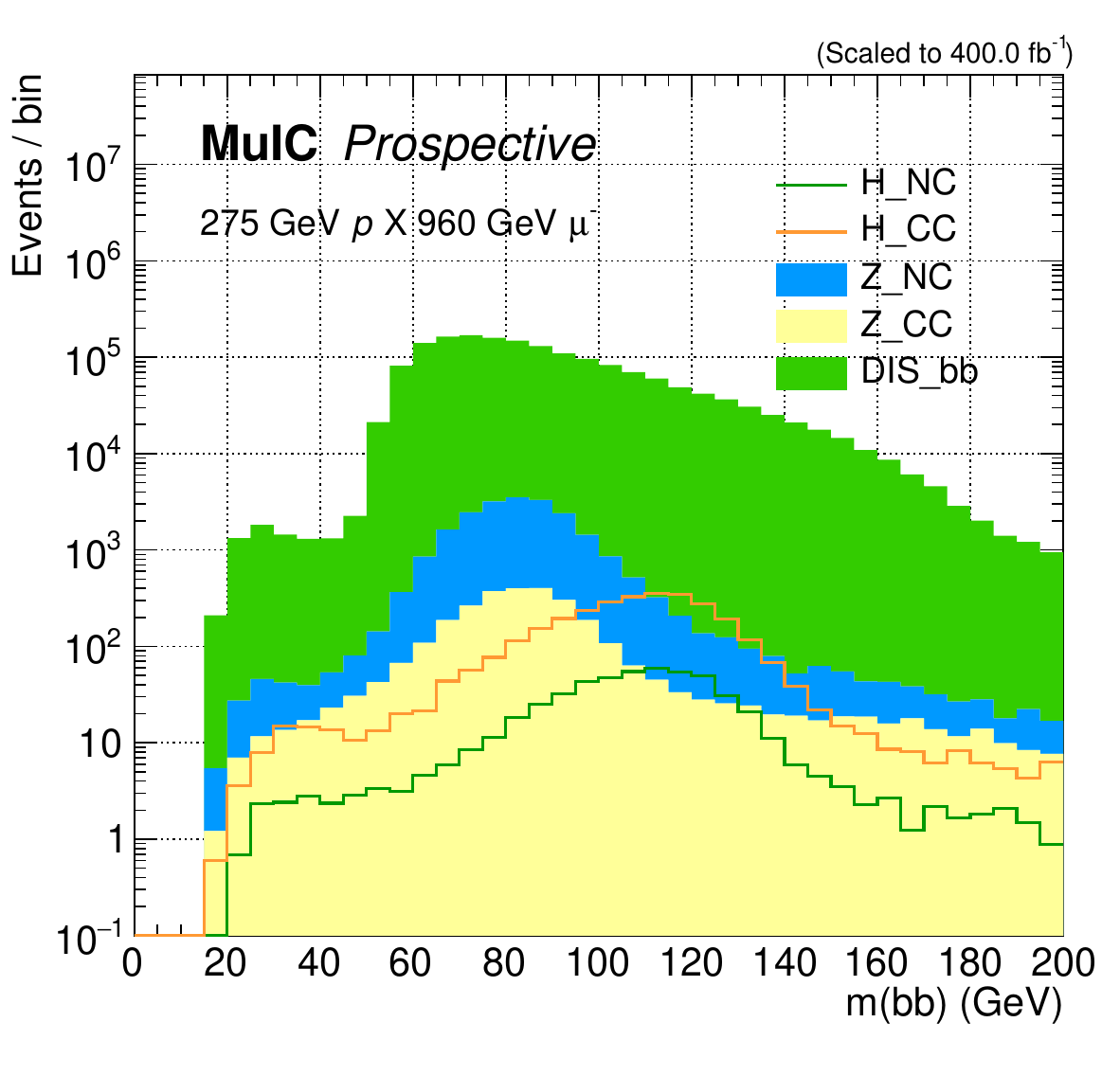}
    \includegraphics[width=0.45\textwidth]{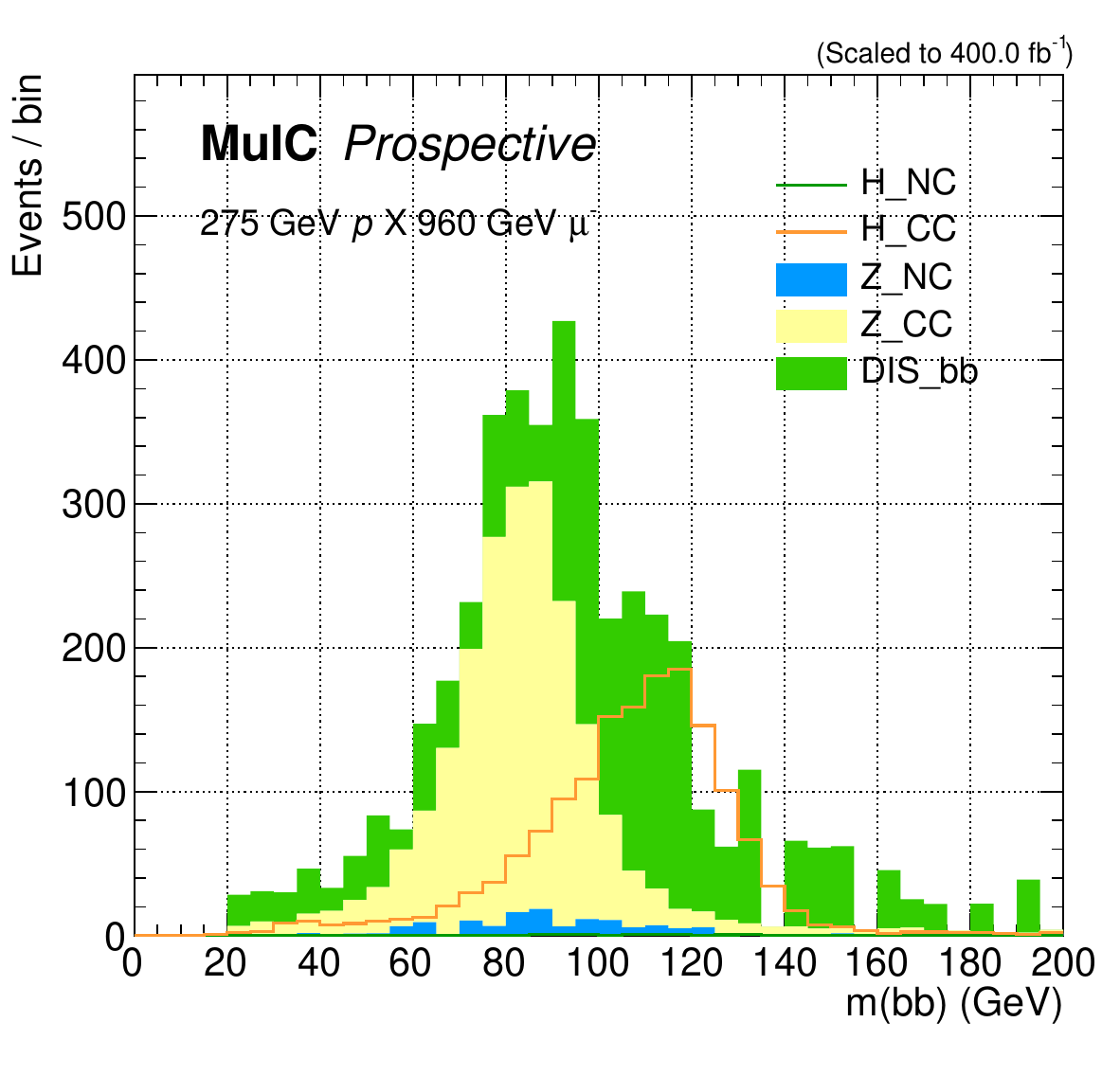}
    \caption{Dijet invariant mass distribution of $H \to b\bar{b}$ events and background processes in $\mu^- p$ collisions with $E_{\mu} = 960$ GeV and $E_{p} = 275$ GeV. 
    All events are generated with MadGraph5~\cite{Madgraph}, showered with PYTHIA8.3~\cite{Sjostrand:2014zea}, and simulated with Delphes~\cite{delphes} for detector responses. 
    Events are subject to simple cut-based selection criteria, with the left plot showing distributions before event selection and the right plot after event selection.
    Note that the DIS background is generated with a fiducial cut of 50~GeV $< m_{b\bar{b}} <$ 170~GeV. Low and high tails of the DIS background in the left plot are from resolution effects.}
    \label{fig:Hbb_mass}
\end{figure}

Figure~\ref{fig:Hbb_mass} shows the distribution of the dijet invariant mass with only the baseline selection (left) and with the full selection criteria (right). 
The full selection achieves a very good signal efficiency and reduces most of the DIS and Z\_NC backgrounds. Improvements are expected with more optimized selections and more sophisticated analysis techniques.
In the range of 100~GeV to 130~GeV, the expected signal yield is around 900, with the $S/B$ close to 1. This corresponds to a statistical uncertainty of about 3\%, which is comparable to the envisioned statistical uncertainty of the $H\to b\bar{b}$ analysis in $pp$ collisions with 3000~fb$^{-1}$ of HL-LHC data~\cite{Cepeda:2650162}.

This result is also consistent with that expected at the LHeC~\cite{Agostini:2020fmq}, which has about 1.7 times larger Higgs cross sections, but less acceptance than for the MuIC. The kinematic profile and expected sensitivity for the $H\to b\bar{b}$ measurement is comparable between the LHeC and MuIC. For the case of LHmuC, where we expect 10 times more Higgs events, it is possible to achieve percent or sub-percent level precision on the $Hbb$ coupling measurement.

The $H\to c\bar{c}$ branching ratio is about 20 times smaller than the $H \to b\bar{b}$ decay, while the $Z\to c\bar{c}$ and DIS backgrounds are essentially the same as for the $b\bar{b}$ case. The flavor tagging algorithm envisioned at the LHeC~\cite{Agostini:2020fmq} expects about 28\% efficiency for the $H\to b\bar{b}$ process and about 11\% efficiency for the $H\to c\bar{c}$ process in the $H\to c\bar{c}$ analysis. A similar level of performance can be assumed at the MuIC. This leads to a statistical uncertainty of the $H\to c\bar{c}$ signal at the same order of the signal strength. With the luminosities assumed in Section~\ref{sec:lumi}, the MuIC would not produce enough events for a precision measurement of the $Hcc$ coupling, while the LHMuC would produce 10 times more Higgs events and offer an exciting opportunity for such a measurement.

Some other Higgs channels are also of interest. With 400 fb$^{-1}$ of integrated luminosity at MuIC, a total about 1900 $H\to \tau^-\tau^+$ events and about 2500 $H\to gg$ events are expected. The experimental sensitivity to these channels largely depends on the tagging algorithms and the contamination of DIS backgrounds, which  are not elaborated in this paper.

\subsection{Beyond Standard Model Physics}

\subsubsection{Leptoquark Processes and Lepton Flavor Violation}

Leptoquarks (LQ) are hypothetical bosons that are predicted by many theories beyond the Standard Model (SM), such as grand unified theories, technicolor, composite models with quarks and lepton sub-structure,
and R-parity violating supersymmetry. They are color-triplets that carry quantum numbers such as spin and
fractional electric charge, and uniquely carry both baryon and lepton numbers allowing them to mediate quark-lepton transitions.
Historically, LQ searches have only considered models that couple to a lepton and quark of the same generation, motivated by limits placed on flavor changing neutral currents, proton decay, and other rare processes. However, there is growing observational evidence that supports inter-generational mixing in LQ decays. Most notably, a measurement of the ratio $R_K$ by the LHCb collaboration~\cite{lhcblu} hints at a possible breaking of lepton universality, and results of the muon g-2 collaboration~\cite{g-2} see increased tension with the SM in the measured muon magnetic moment. The exchange of a LQ boson can explain either anomaly, assuming that the LQ can couple to quarks and leptons of different generation.

The MuIC will provide unique testing grounds for LQ searches with mixed couplings to second and third generation leptons and quarks. Of particular interests are the t-channel and s-channel production of LQs that couple to a muon and a b-quark or to a muon and a top quark, as well as LQs that can couple both to $\tau$ lepton and a quark, and a muon and a quark. To study these processes, the software toolbox described in~\cite{LQBox} is used to generate, at leading order, muon-proton collisions with specific scalar LQ models that can couple of different quarks and leptons generations (i.e. the R2 and S3 models), and two proton beam energy configurations, 275 GeV and 1 TeV. The Yukawa coupling at the LQ-lepton-quark vertex is set to 0.1 in these simulations.

\begin{figure}[!htb]
    \centering
    \includegraphics[width=0.65\textwidth]{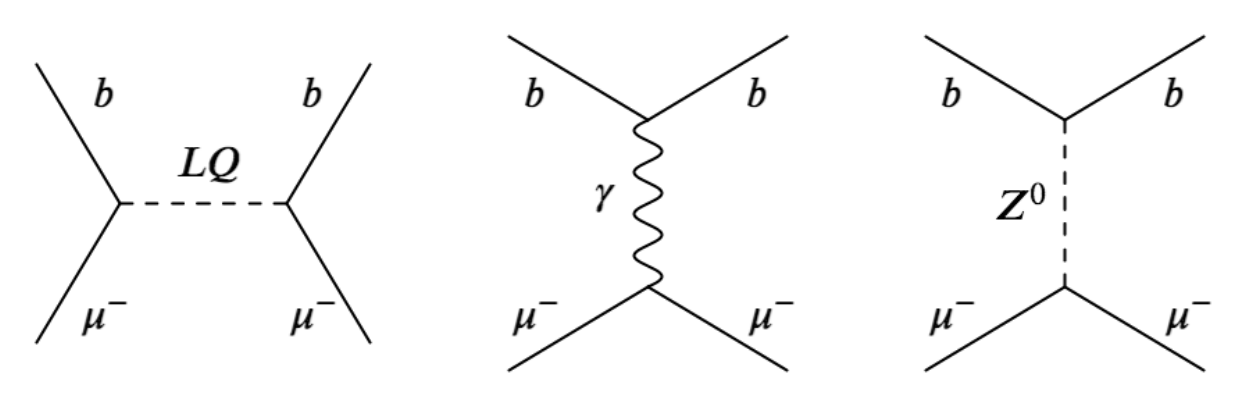}
    \caption{Diagrams of the production of a muon and a b-quark in the final state of a muon-proton collision, via the s-channel production of a S3 LQ, and via SM electroweak t-channel processes.}
    \label{fig:Feynman_mub}
\end{figure}

\begin{figure}[!htb]
    \centering
    \includegraphics[width=0.45\textwidth]{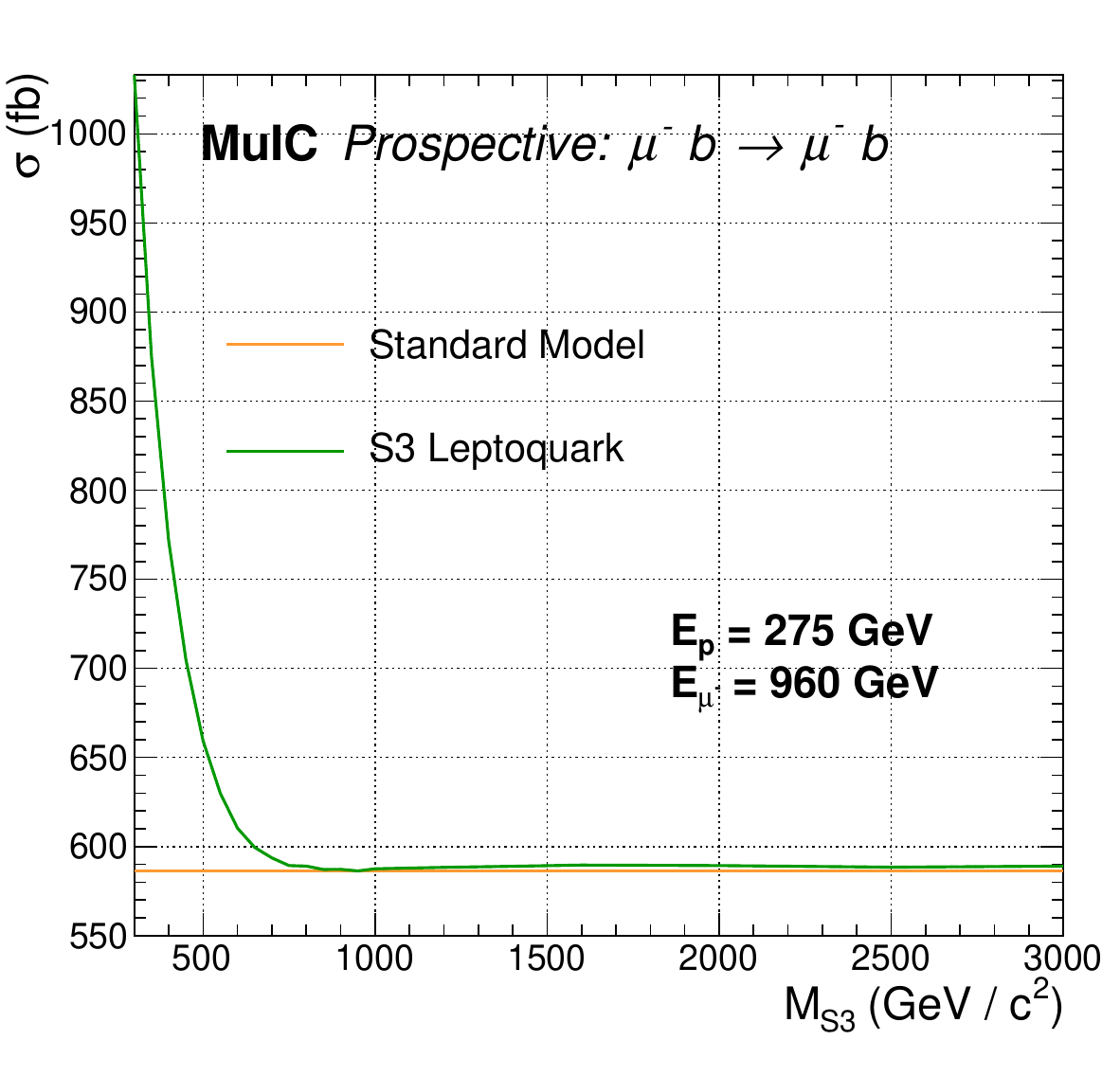}
    \includegraphics[width=0.45\textwidth]{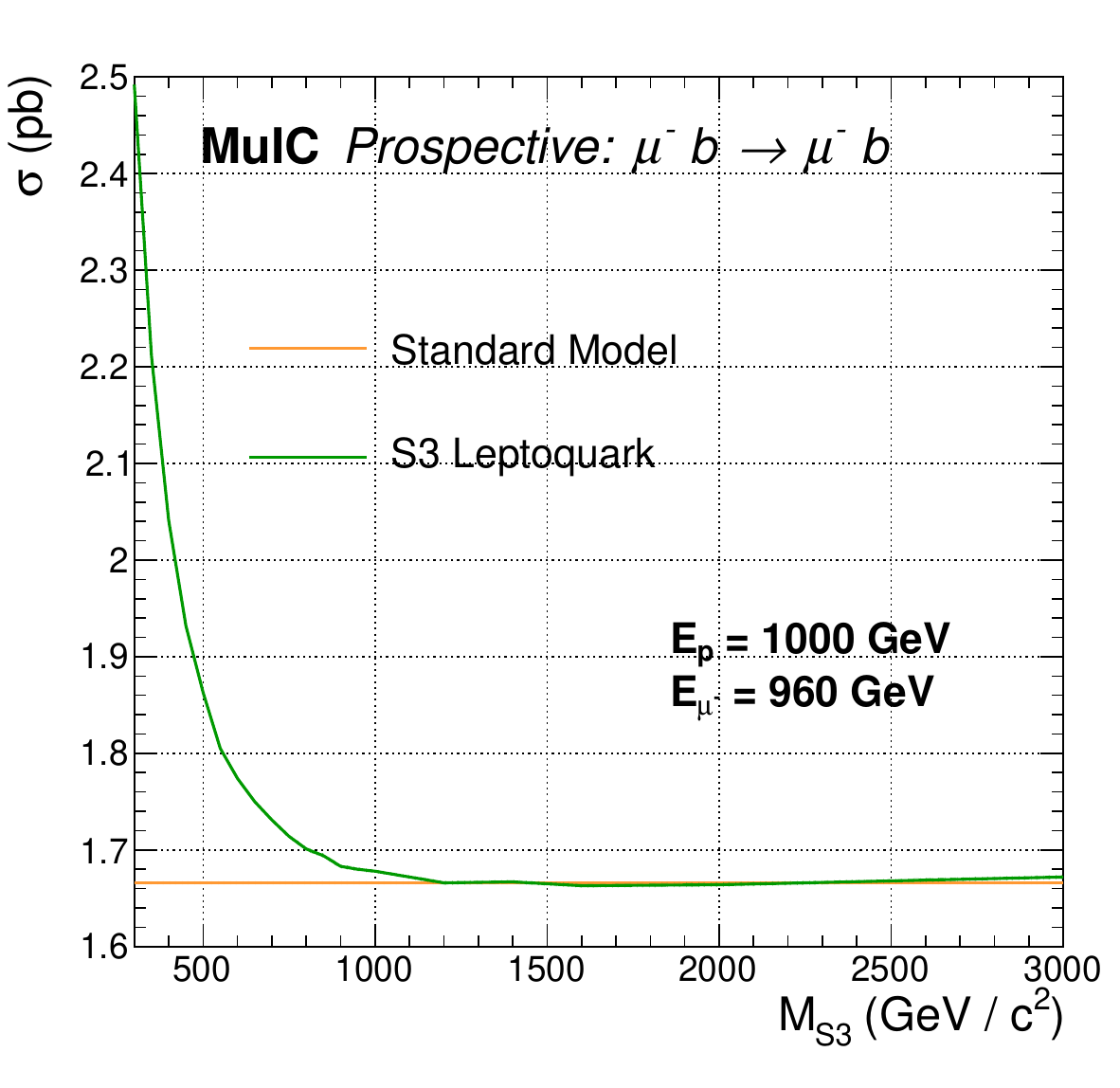}
    \caption{Cross-section of the production of $\mu$p $\rightarrow$ $\mu$b, including s-channel production and decay of a S3 LQ, as a function of LQ mass for a proton beam energy of 275 GeV (left) and 1 TeV (right).}
    \label{fig:xsec_mub_1}
\end{figure}

\begin{figure}[!htb]
    \centering
    \includegraphics[width=0.45\textwidth]{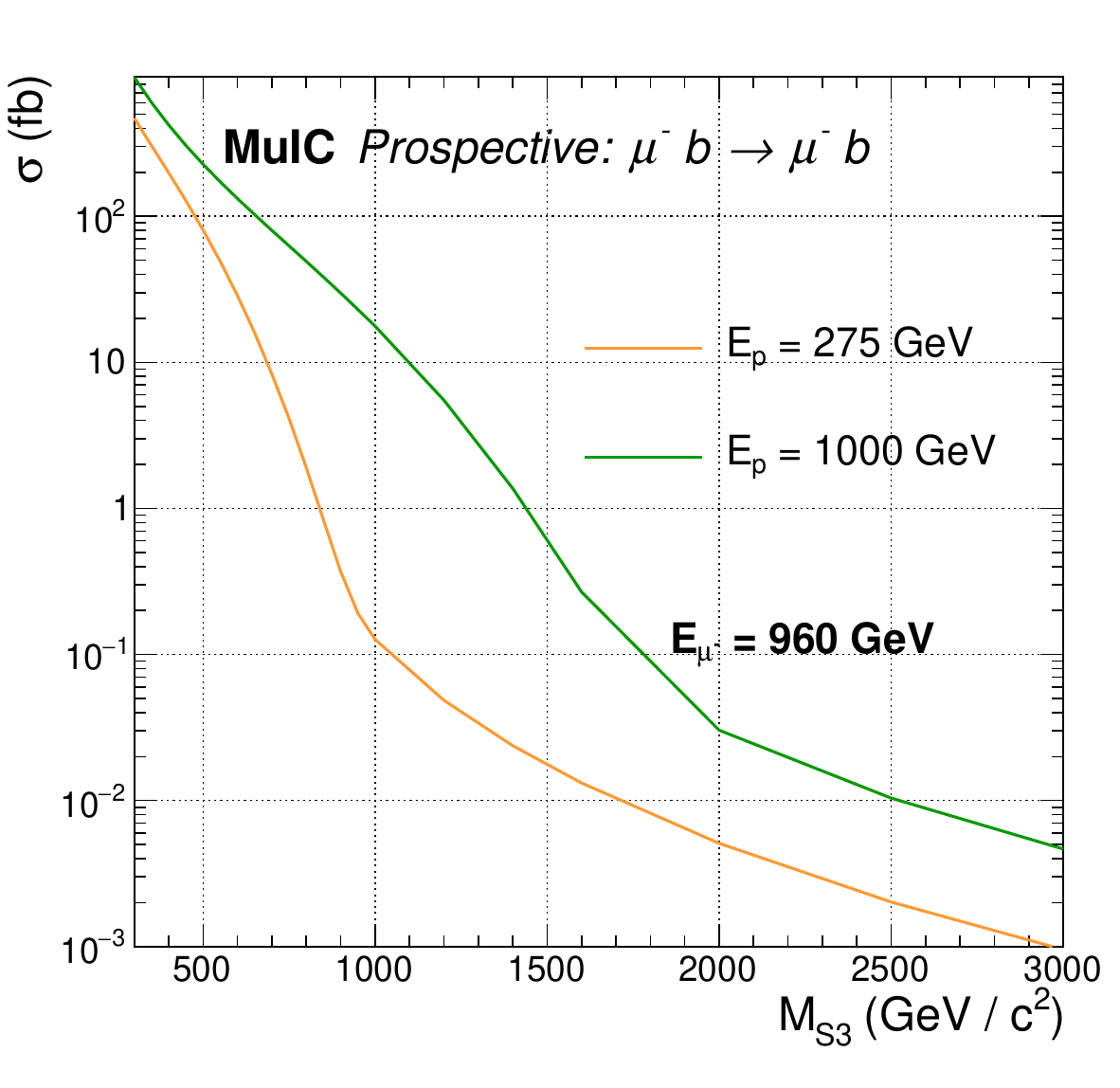}
    \caption{SM backgrounds-subtracted cross-section of $\mu$p $\rightarrow$ $\mu$b via s-channel production of a S3 LQ, as a function of LQ mass.}
    \label{fig:xsec_mub_2}
\end{figure}

\begin{figure}[!htb]
    \centering
    \includegraphics[width=0.45\textwidth]{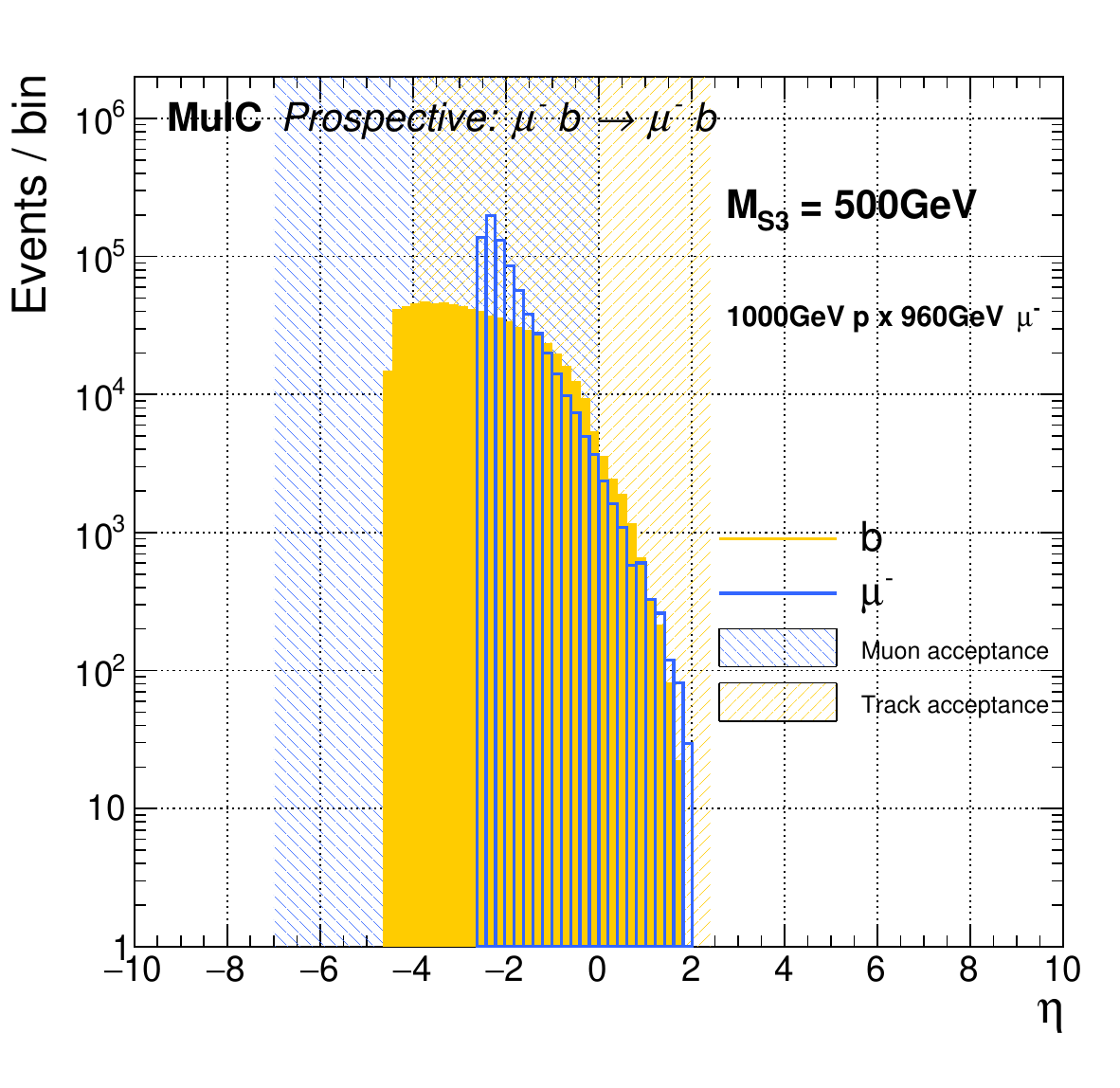}
    \includegraphics[width=0.45\textwidth]{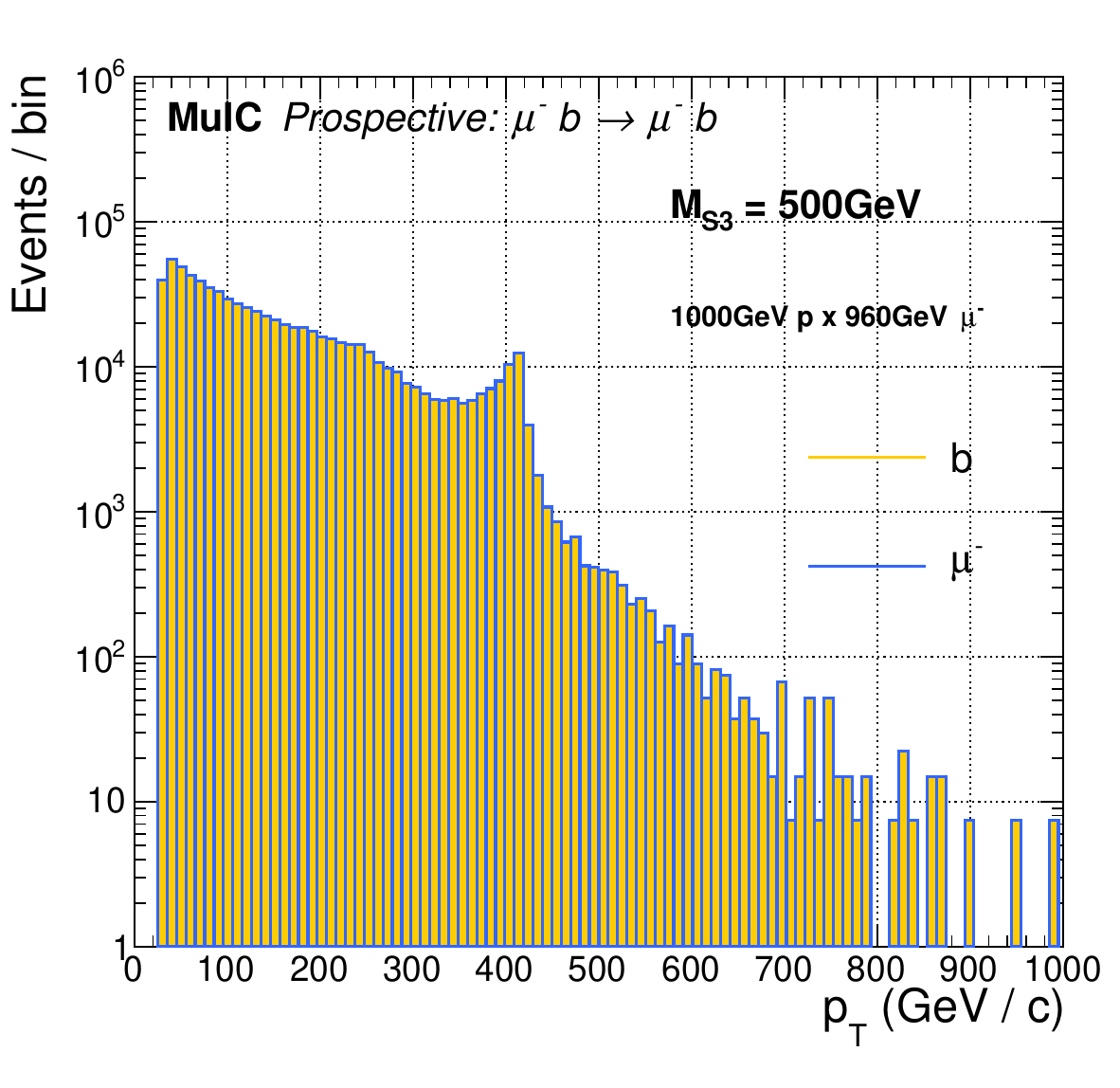}
    \caption{Generator-level kinematic distributions, $\eta$ (left) and $p_{\rm T}$ (right), of the muon and b-quark produced via LQ S3 s-channel exchange, for a LQ mass of 500 GeV, 1 TeV proton and 960 GeV muon collisions. An integrated luminosity of 10 fb$^{-1}$ is assumed.}
    \label{fig:xsec_mub_3}
\end{figure}

The cross-section for the production of $\mu$-b via s-channel exchange of a S3 LQ as well as t-channel SM electroweak $\mu$-b production (see Fig.~\ref{fig:Feynman_mub}) is shown in Fig.~\ref{fig:xsec_mub_1} as a function of LQ mass and for 275 GeV and 1 TeV proton beam energies, respectively. Fig.~\ref{fig:xsec_mub_2} shows the LQ production cross-section with SM backgrounds subtracted, and Fig.~\ref{fig:xsec_mub_3} illustrates the kinematic distributions ($p_{\rm T}$ and $\eta$), at generator level, of the outgoing muon and b-quark. Tagging the outgoing muon with a far-backward muon spectrometer, as well as a selection based on the kinematic properties and correlations among the final state muon and b-jet will be instrumental in optimizing the sensitivity for this channel.

The kinematic selection used for these LQ studies consists of a 20 GeV threshold on the $p_{\rm T}$ of the outgoing quarks, a 10 GeV threshold on the $p_{\rm T}$ of the outgoing leptons, and a minimum separation in $\eta-\phi$ space of 0.4 between the outgoing lepton and quark.

\begin{figure}[!htb]
    \centering
    \includegraphics[width=0.45\textwidth]{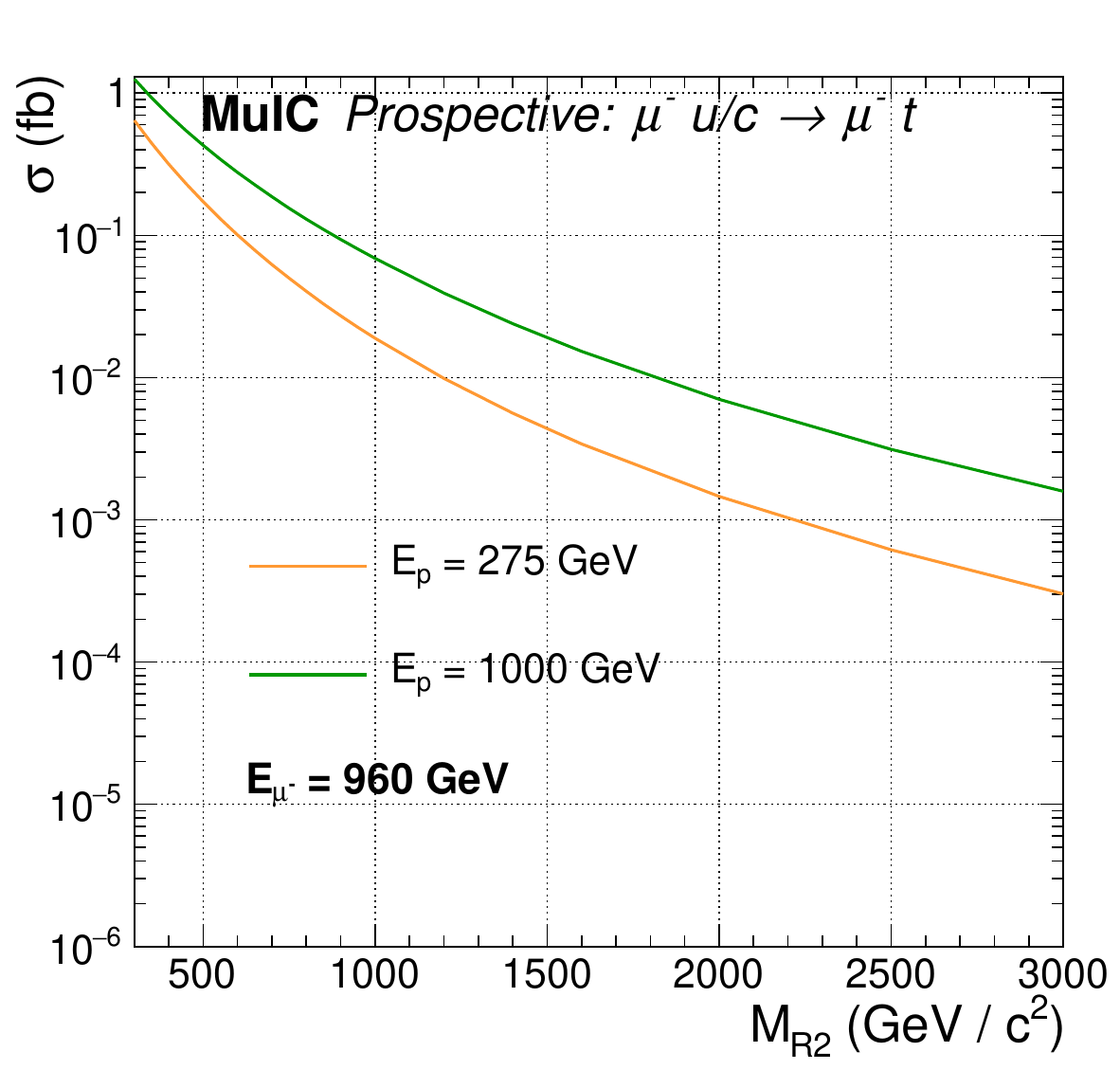}
    \includegraphics[width=0.22\textwidth]{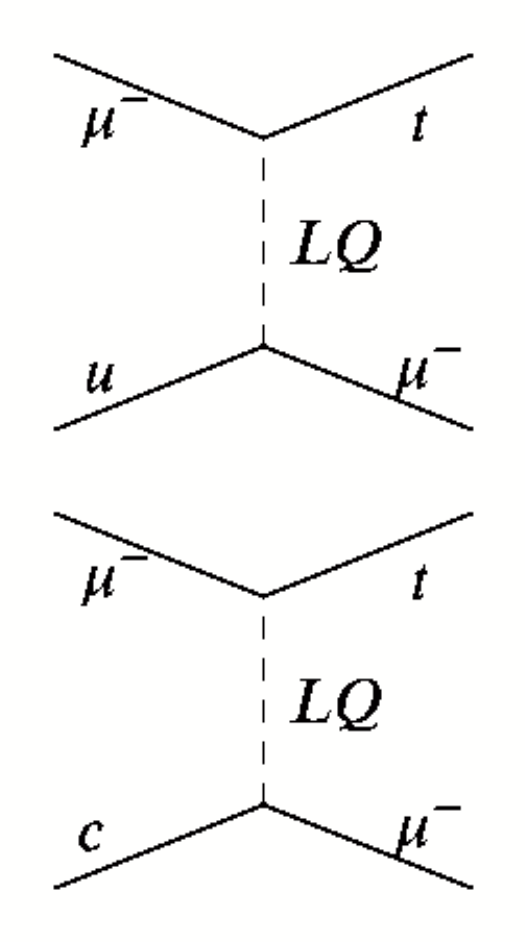}
    \caption{(Left) Cross-section for the production of muon and a top quark via a t-channel R2 LQ, as a function of LQ mass, and for two proton beam energies. (Right) R2 LQ t-channel exchange with final states consisting of a muon and a top quark.}
    \label{fig:xsec_mutop_1}
\end{figure}

\begin{figure}[!htb]
    \centering
    \includegraphics[width=0.45\textwidth]{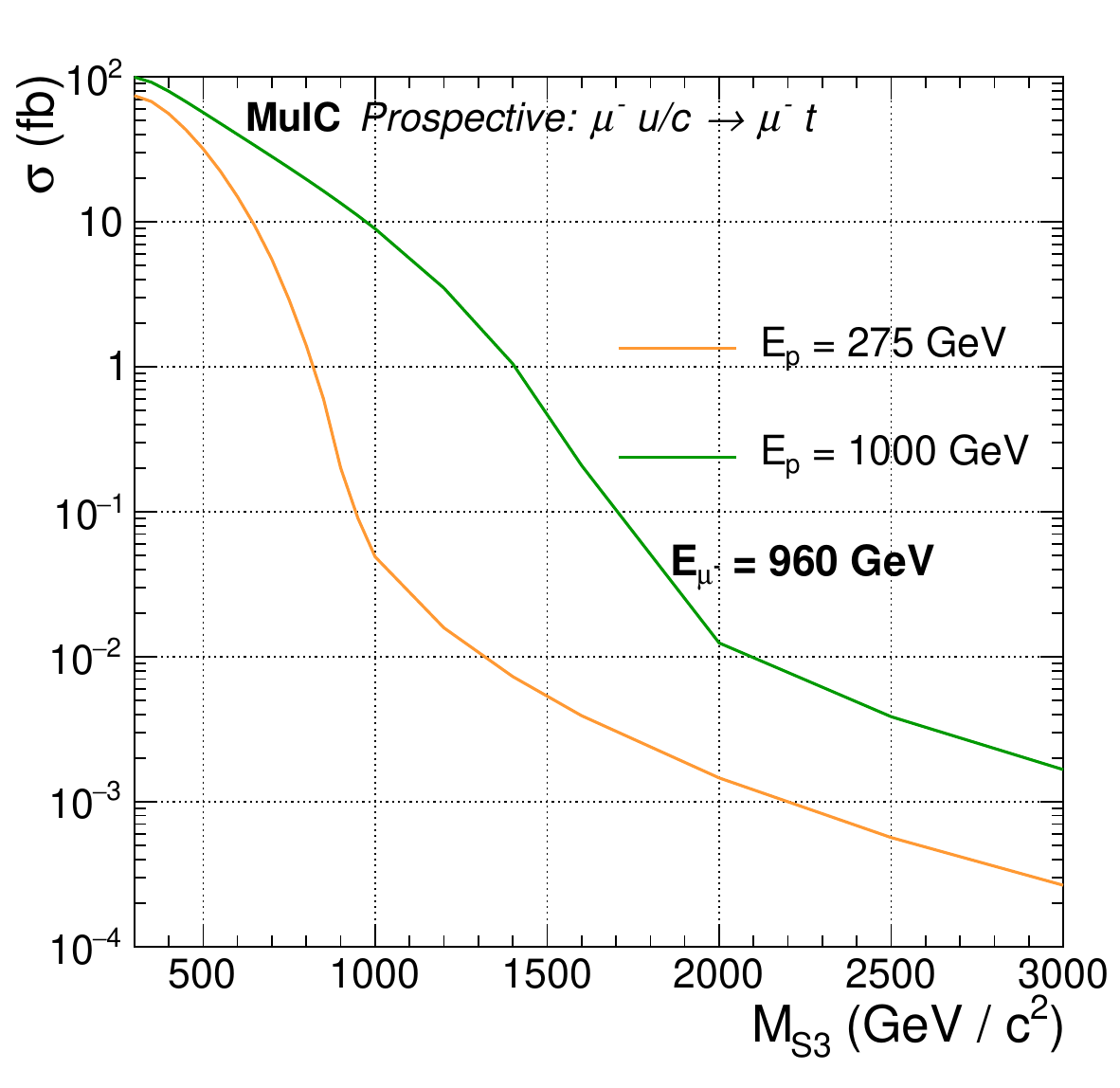}
    \includegraphics[width=0.22\textwidth]{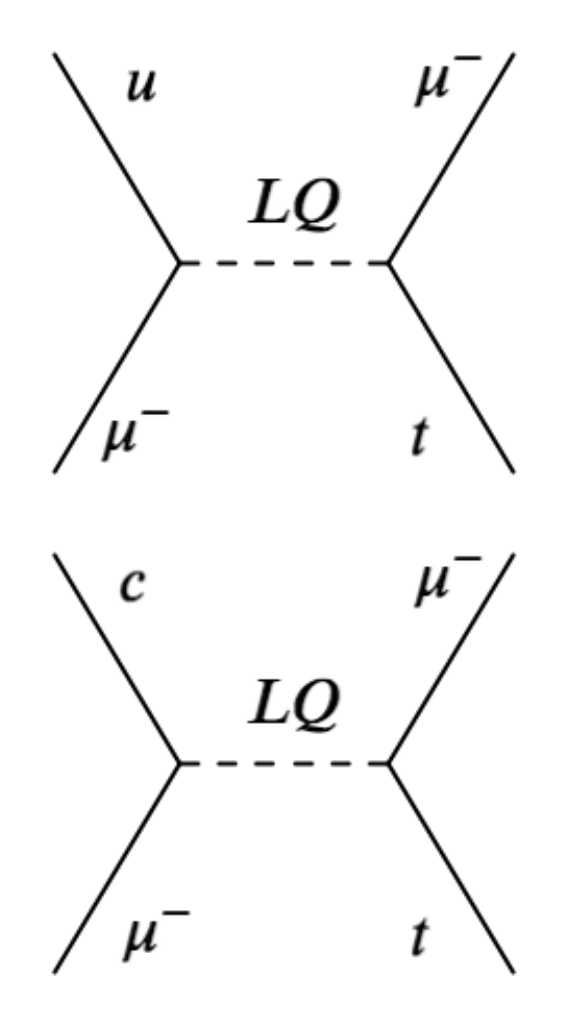}
    \caption{(Left) Cross-section for the production of muon and a top quark via a s-channel S3 LQ, as a function of LQ mass, and for two proton beam energies. (Right) S3 LQ s-channel exchange with final states consisting of a muon and a top quark.}
    \label{fig:xsec_mutop_2}
\end{figure}

The production of other third generation quarks and leptons is also investigated. Fig.~\ref{fig:xsec_mutop_1} shows the diagram and cross-section for the t-channel production of a R2 LQ, coupling to the initial muon and either an up or charm quark from the initial proton, and decaying to a muon and a top quark. Fig.~\ref{fig:xsec_mutop_2} shows similar couplings for the s-channel production cross-section of a S3 LQ, decaying to a muon and a top quark. 

\begin{figure}[!htb]
    \centering
    \includegraphics[width=0.45\textwidth]{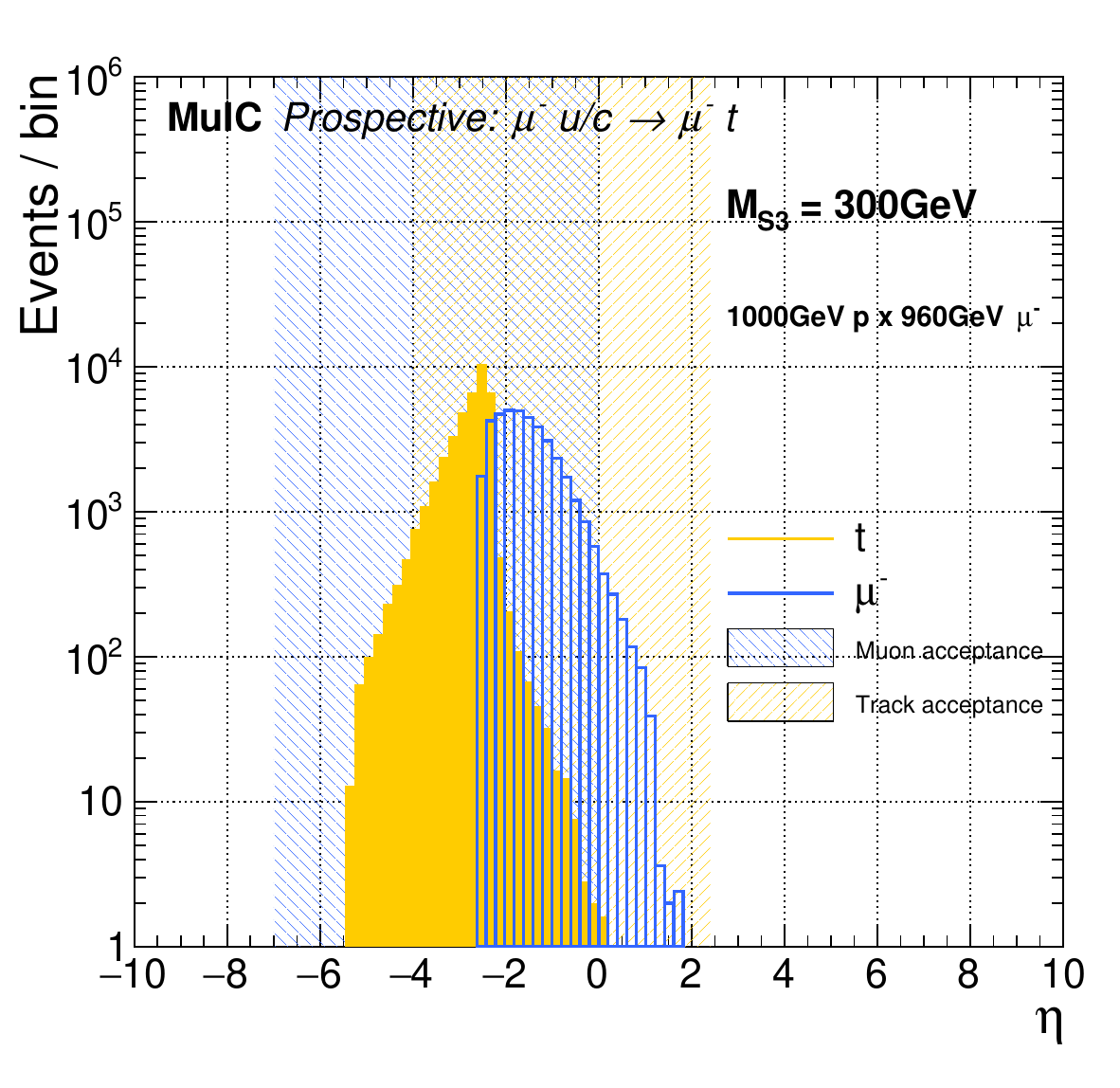}
    \includegraphics[width=0.45\textwidth]{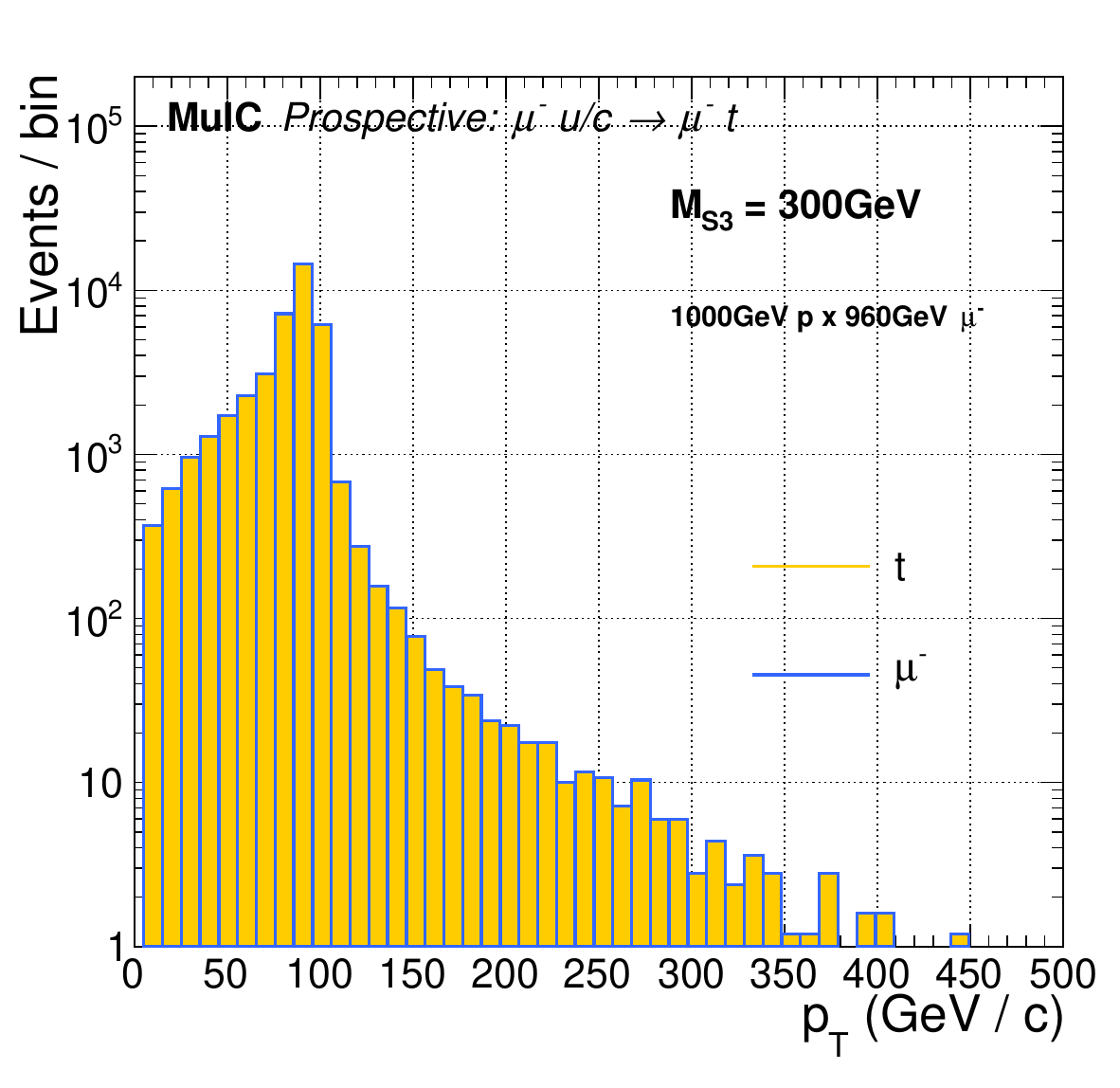}
    \caption{Generator-level kinematic distributions, $\eta$ (left) and $p_{\rm T}$ (right), of the muon and  top quark produced via LQ S3 s-channel exchange, for a LQ mass of 300 GeV, 1 TeV proton and 960 GeV muon collisions. An integrated luminosity of 10 fb$^{-1}$ is assumed.}
    \label{fig:xsec_mutop_3}
\end{figure}

The $\eta$ and $p_{\rm T}$ distributions---at generator level---for the final state muon and top quark are shown in Fig~\ref{fig:xsec_mutop_3}. Since there are no competing leading order SM processes with a top quark and a muon produced in the final state, the requirements of a muon tagged in the far-backward spectrometer, the kinematic reconstruction of the top quark from its decay products, and constraints on the reconstructed missing transverse energy in the event should be effective means of reducing other SM backgrounds (e.g. single or pair production of top quarks in association with $\nu_\mu$ or from production of $W$ bosons in association with muons).

\begin{figure}[!htb]
    \centering
    \includegraphics[width=0.35\textwidth]{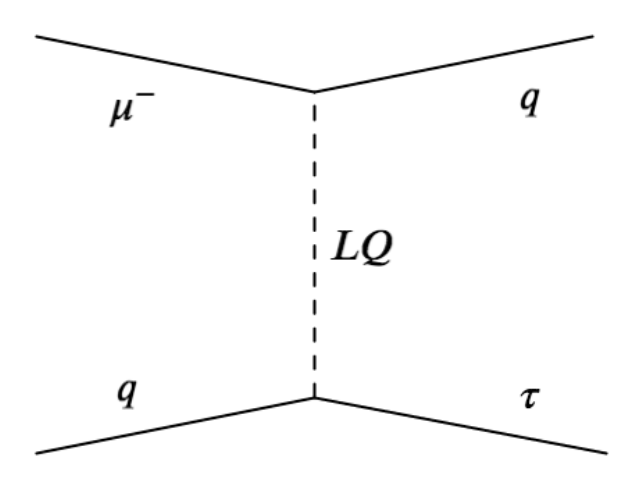}
    \caption{Diagram of the production of a $\tau$ lepton and a jet in the final state of a muon-proton collision, via the t-channel production of a LQ (either a R2 or a S3).}
    \label{fig:Feynman_mutau}
\end{figure}

\begin{figure}[!htb]
    \centering
    \includegraphics[width=0.45\textwidth]{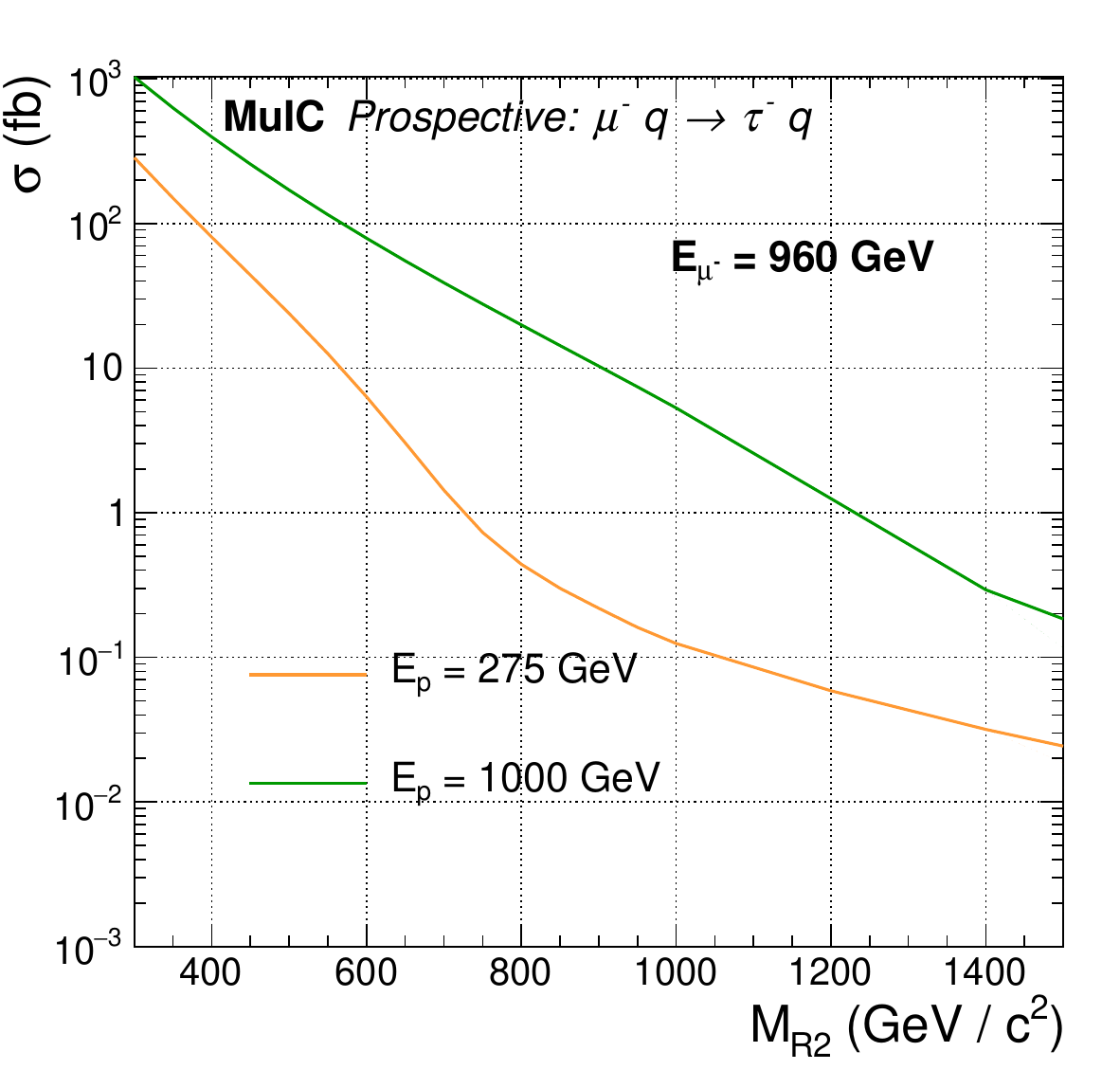}
    \includegraphics[width=0.45\textwidth]{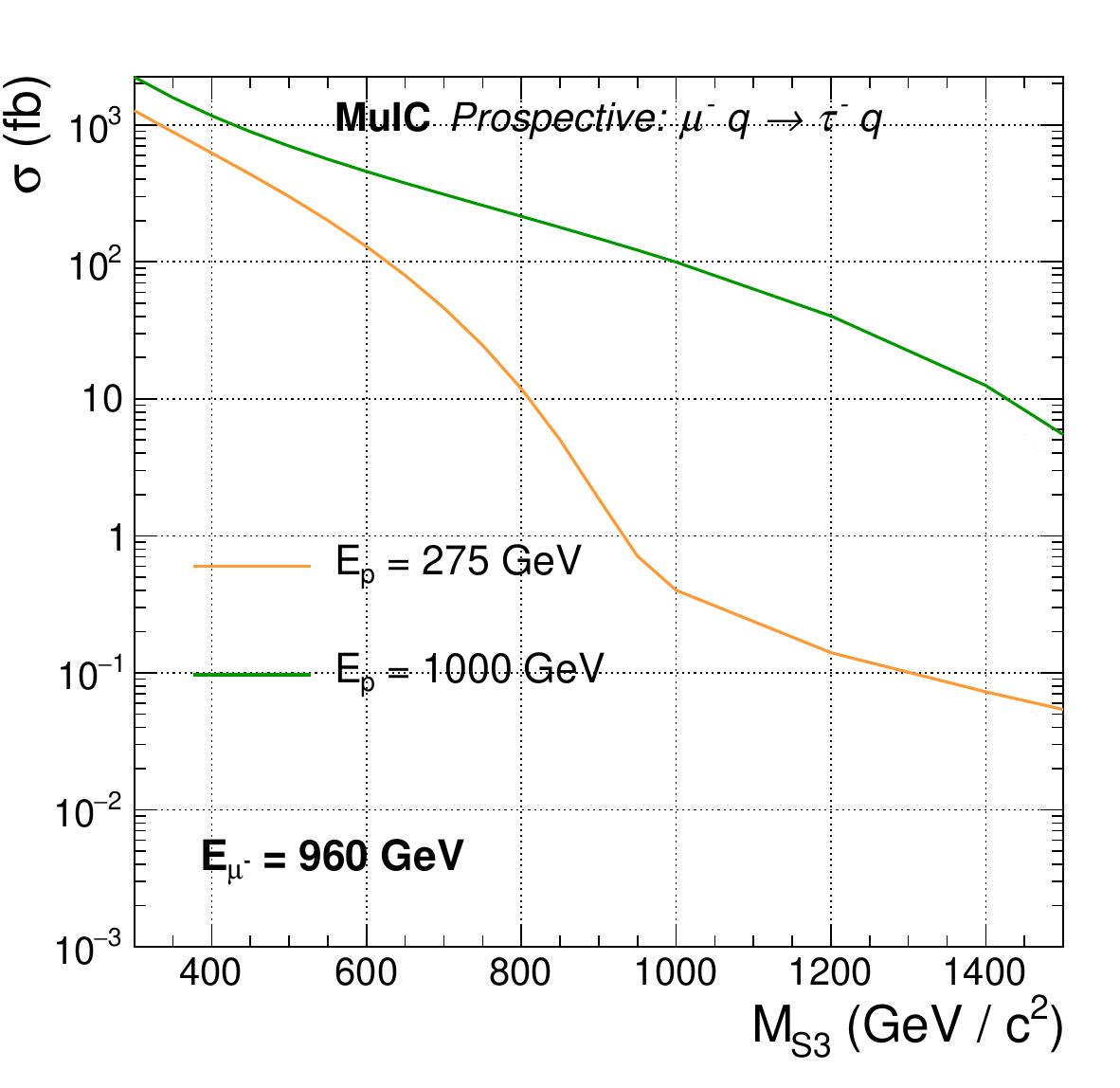}
    \caption{Cross-section for the production of $\tau$ lepton and a quark via a t-channel R2 LQ (Left) or a t-channel S3 LQ (Right), as a function of LQ mass, and for two proton beam energies.}
    \label{fig:xsec_mutau_1}
\end{figure}

\begin{figure}[!htb]
    \centering
    \includegraphics[width=0.45\textwidth]{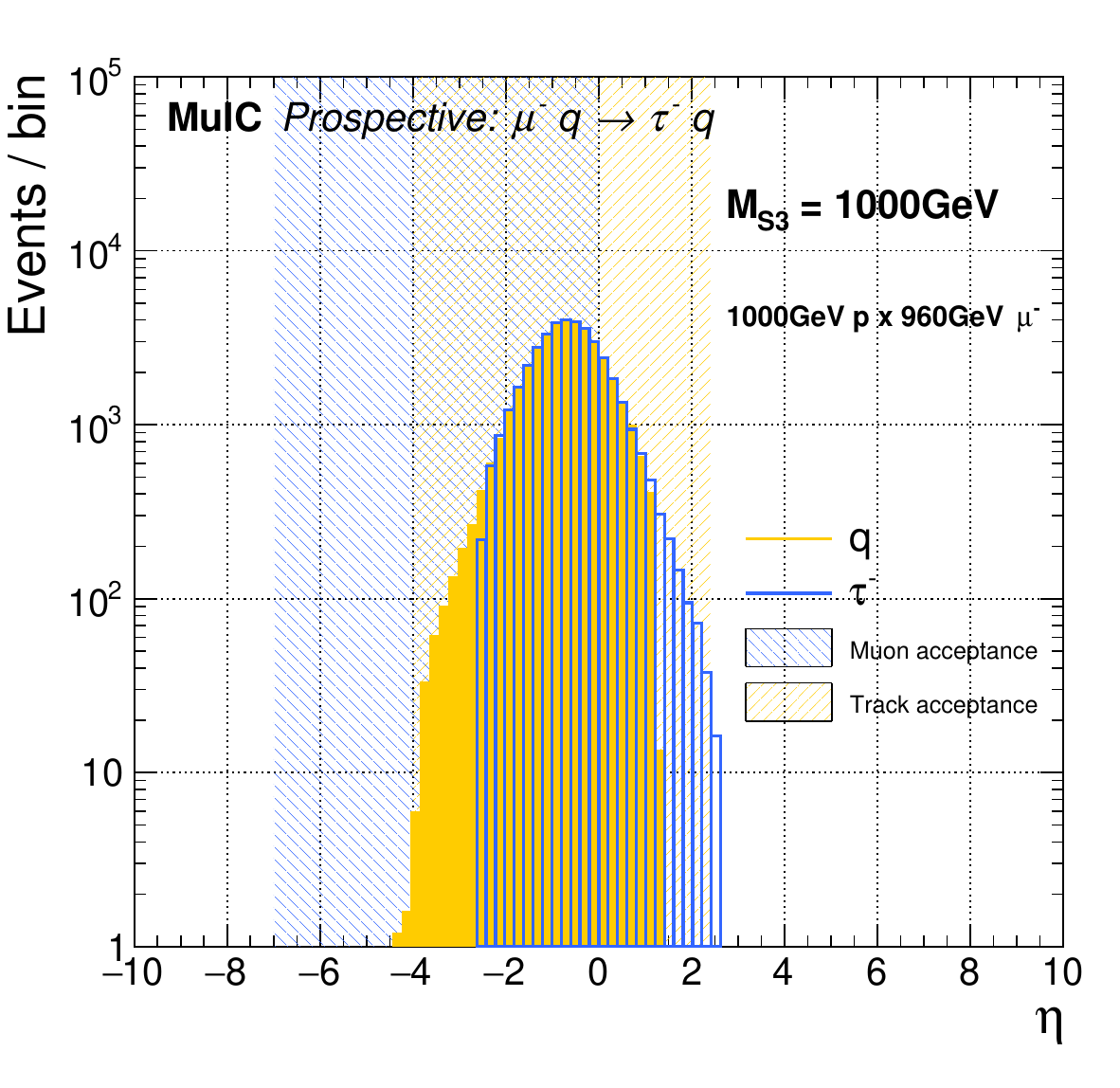}
    \includegraphics[width=0.45\textwidth]{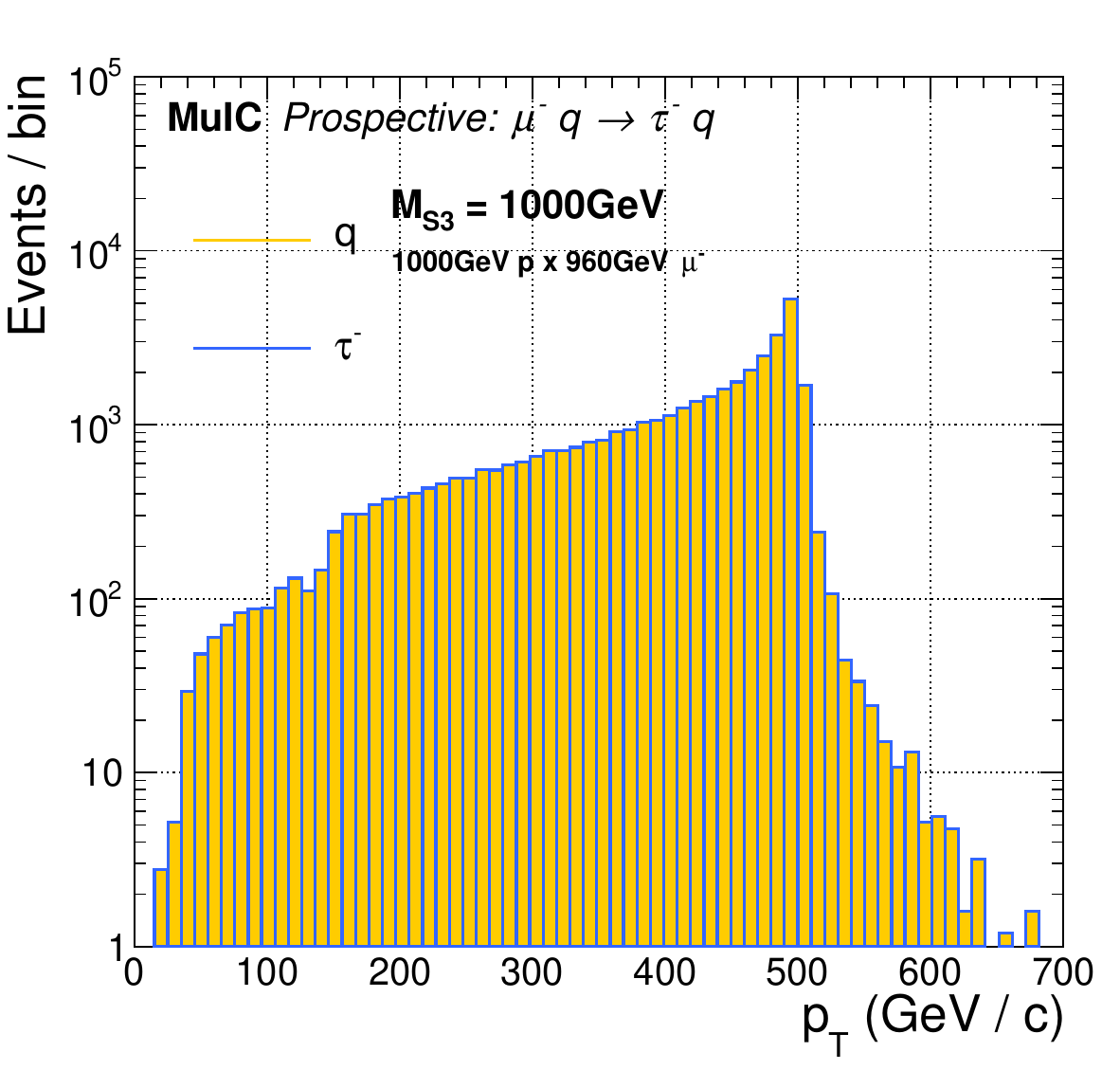}
    \caption{Generator-level kinematic distributions, $\eta$ (left) and $p_{\rm T}$ (right), of the $\tau$ lepton and the quark produced via LQ S3 s-channel exchange, for a LQ mass of 1 TeV, in 1 TeV proton and 960 GeV muon collisions. An integrated luminosity of 10 fb$^{-1}$ is assumed.}
    \label{fig:xsec_mutau_2}
\end{figure}

Finally, a striking observation of LFV would be the observation of $\mu$ p $\rightarrow$ $\tau$ jet events, which can occur via the t-channel exchange of a S3 or a R2 LQ, as shown in Fig.~\ref{fig:Feynman_mutau} (such processes can also be mediated by the exchange of a heavy $Z'$ with LFV leptonic decays).
The cross-section for the production of $\mu$ p $\rightarrow$ $\tau$ jet events via the exchange of a R2 or a S3 LQ is shown in Fig.~\ref{fig:xsec_mutau_1}. The generator-level $\eta$ and $p_{\rm T}$ distributions for the final state $\tau$ are shown in Fig.~\ref{fig:xsec_mutau_2} for the case of a 1 TeV S3 LQ. The cross-section for this process is higher than for the previous final states considered, since no specific final state quark flavor is requested. Effective reconstruction of this type of event and background rejection relies on constraining the reconstructed missing transverse energy in the event, on vetoing other final state lepton flavors, and on tagging the final state $\tau$, which is well within the acceptance of the tracking detectors, as Fig.~\ref{fig:xsec_mutau_2} illustrates.

\section{Detector Requirements and Machine-Detector Interface}


\subsection{Detector Design Considerations}
\label{sec:detector}

\begin{figure}[htb]
    \centering
    \includegraphics[width=\textwidth]{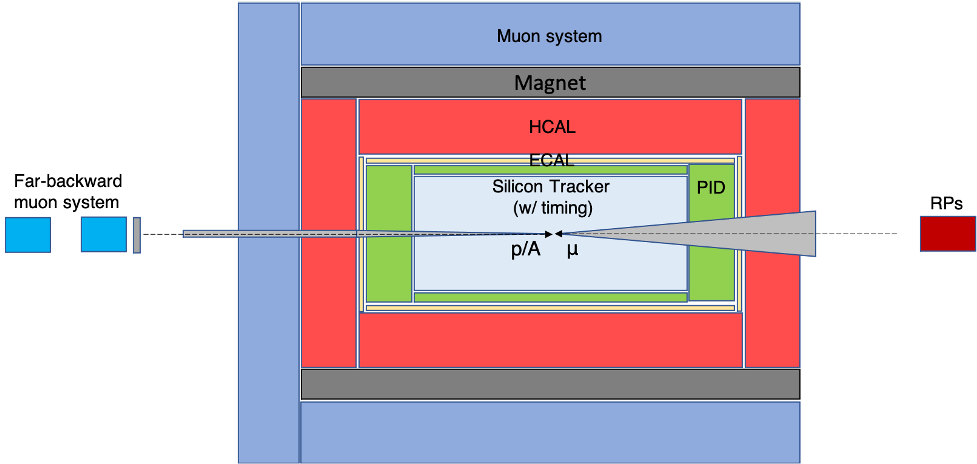}
    \caption{A sketch of a possible detector design at muon-proton
      colliders. The tungsten nozzle is only applied on the incoming
      muon direction. Given the asymmetric nature of the collisions,
      emphasis is on the central and far backward regions. Coverage
    shown is illustrative only.}
    \label{fig:detector}
\end{figure}

As mentioned earlier, the detector design at a muon-ion collider shares
a lot in common with a $\mu^{+}\mu^{-}$ collider but it also has its unique
requirements and challenges. A common challenge for the detector design is
the beam-induced background, for which precise timing measurements
from the detectors will help discriminate but is otherwise not studied
here.
A sketch a conceptual design of a general-purpose 
detector at a muon-ion collider is shown in 
Fig.~\ref{fig:detector}. 
The shielding tungsten nozzle is applied only to the 
muon coming side, resulting in central detector
acceptance of $-5<\eta<2.4$. As shown in Appendix~\ref{sec:dis-kin},
the scattered parton (e.g., jets) and muon are mostly going
toward the backward direction. Therefore, the available
central detector acceptance with the single-sided nozzle configuration
is sufficient to meet most of physics requirements at a muon-ion collider.
The shielding nozzle will mainly limit the access to particles
produced by remnants of the proton in the forward direction.

The center detector consists of silicon tracker system with precision
timing information (e.g., 4-D tracking using low gain 
avalanche diodes, or LGADs), which is essential 
to effectively suppress the muon beam induced background.
The electromagnetic and hadronic calorimeters are needed for 
detecting and identifying objects like jets, photons and electrons.
Similarly, emerging technology that embeds silicon sensors inside 
calorimeters to provide precision timing and position information 
will be particularly beneficial to the detector system at a muon-based
accelerator for suppressing beam induced backgrounds. The particle identification (PID) system for hadrons is crucial for QCD and nuclear
physics at a lepton-hadron collider. Unlike the EIC where the coverage
to the very high momentum (up to 100~GeV) regime by the PID systems 
only needed in the forward direction, high momentum hadrons are 
produced in both forward and backward directions at the MuIC or LHmuC.
Therefore, ring-imaging Cherenkov (RICH) detectors with gas media 
are necessary in both endcaps of the experimental system, which generally
take up to about 1~m in $z$ direction. The gas RICH detectors 
have a lower momentum threshold of about 3 GeV. The PID below 3~GeV
can be well covered by the LGADs-based time-of-flight system or 4-D tracker,
as was designed for the EIC. 

The far-forward direction ($5<\eta<8$)
should be instrumented with Roman Pots for detecting the scattered 
proton in elastic processes. In the far backward region ($-8<\eta<-5$),
a muon spectrometer system is required to detect the scattered muon
and precisely determine its momentum so that the DIS kinematics can be
reconstructed. Considerations and requirements of such a muon spectrometer is discussed
in Section~\ref{sec:muon-spec} below. 

The performance of reconstructing DIS kinematic variables, $Q^{2}$ and $x$,
at the MuIC with parameterized detector resolutions is presented in
Ref.~\cite{Acosta:2021qpx}. Detailed simulations of detector system, beam-induced
background, and MDI will further inform the design and R\&D activites.

Finally, luminosity measurements at a MuIC could be a challenge. The
analogue of the process used at electron-ion colliders, i.e. $\mu p \to \mu
p \gamma$,  would be challenging in the presence of high-energy electrons from the 
beam induced background.  Other normalization processes or methods need to be investigated.

\subsection{Far-backward Muon Spectrometer}
\label{sec:muon-spec}

A key experimental aspect of a TeV-scale muon-ion collider experiment, as explained in Appendix~\ref{sec:dis-kin}, is the need for a muon spectrometer able to measure muons with reasonable resolution down to very small scattering angles ($\eta \gtrsim -7$) and at momenta up to the beam energy (${\approx }1$~TeV). We note that this need  is in common with that for a TeV $\mu^+\mu^-$ collider experiment as well, if the tagging of forward muons from vector-boson fusion processes is considered a physics priority. If one places a measurement plane downstream of the collision point in the muon beam direction at $z=-5$~m (i.e. just beyond the central region of the experiment) with detectors placed outside of the beampipe radius of ${\approx}5$~cm, then this measurement station could cover the pseudorapidity range $-5< \eta < -1$. The radial coverage of the detectors would need to extend to $r\approx 3.5$~m to reach $\eta=-1$. To cover smaller angle scattered muons with $\eta = -7$ and $r>5$~cm requires a station to be placed 30~m or more downstream of the collision point. For example, a measurement station placed at $z=-40$~m, with a radius extending to $r=1$~m, could cover the pseudorapidity range $-7< \eta < -4.5$.

The measurement of the momentum of these small angle muons will require a dedicated magnet spectrometer with the field provided from one or more dipole  (or perhaps toroid) magnets. The $\int \vec{B}\cdot d\vec{\ell}$ will need to be substantial for TeV muons, likely of order 10~T$\cdot$m. The measurement of the saggita along with some vertexing capability from the tracking system will require more than the two stations noted above, as more than one measurement of the same muon would be needed. This would argue for a telescope arrangement of measurement stations stretching down the beamline, with a transverse radius of each station that does not need to be so large (of order a meter). However, the detailed design and performance of a plausible muon spectrometer is beyond the scope of this paper.

\appendix

\section{Deep Inelastic Scattering Kinematics and Resolution}
\label{sec:dis-kin}

\begin{figure}[!htb]
    \centering
    \includegraphics[width=0.43\textwidth]{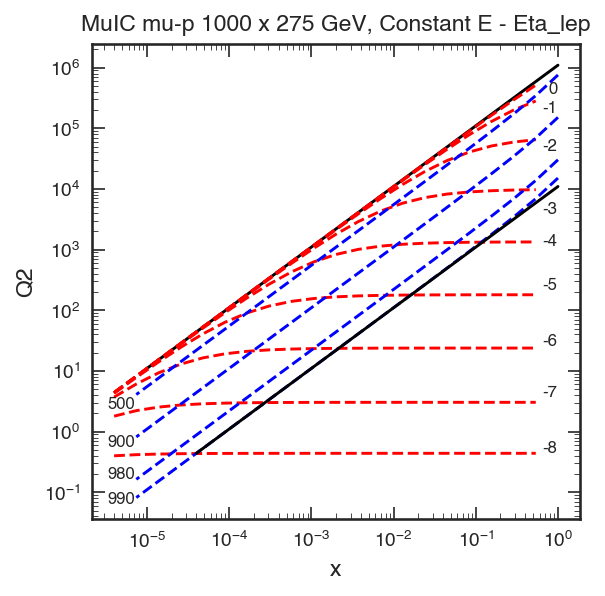}
    \includegraphics[width=0.45\textwidth]{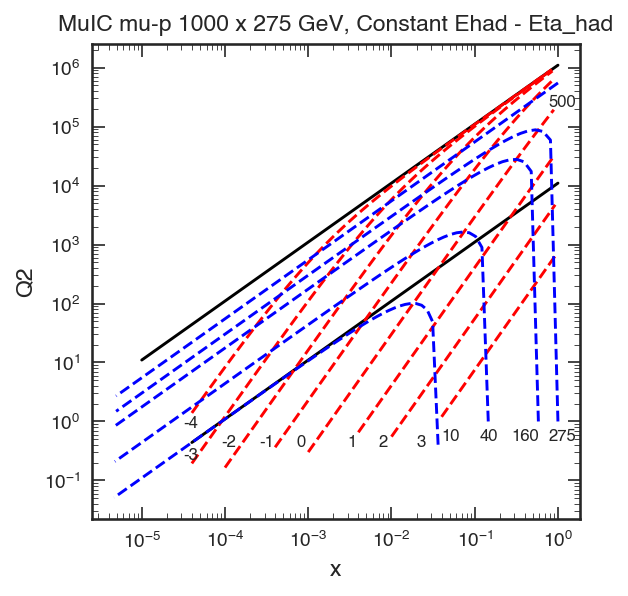}
    \caption{Kinematics of the scattered muon (left) and final-state hadrons (right) in the $Q^2$-$x$ plane for muon-proton deep inelastic scattering with a 1000~GeV muon beam colliding with a 275~GeV proton beam. The dashed blue lines correspond to constant  energy in GeV and the dashed red lines to constant pseudorapidity, respectively. The lower diagonal solid line corresponds to the inelasticity $y=0.01$.}
    \label{fig:MuICkinematics}
\end{figure}

The final state kinematics of the scattered lepton and produced hadrons in deep inelastic scattering are important considerations for the design of an experiment. Figure~\ref{fig:MuICkinematics} shows the kinematics of the scattered muon and final-state hadrons in the $Q^2$-$x$ plane for muon-proton deep inelastic scattering with a 1~TeV muon beam colliding with the 275~GeV proton beam of the BNL EIC. We define the initial proton direction as the forward ($+z$)  direction. 
To cover the full $Q^2$-$x$ plane in the perturbative QCD regime ($Q^2 \gtrsim 1$~GeV$^2$),  it is important to measure the scattered muons with pseudorapidities as small as $\eta=-7$ and with momenta approaching 1~TeV. For inelasticities $y>0.01$, the energy of the scattered hadron system is at least 10~GeV, and it can reach 500~GeV or more at high $y$. The average pseudorapidity of the hadronic system is central or forward at high $Q^2$, but trends toward the muon beam direction at low $Q^2$ with $\eta > -4$ or even smaller. 

\begin{figure}[!htb]
    \centering
    \includegraphics[width=0.43\textwidth]{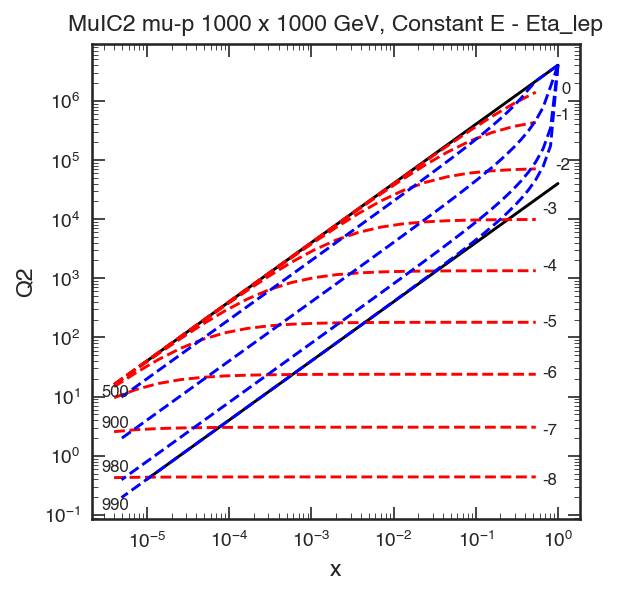}
    \includegraphics[width=0.45\textwidth]{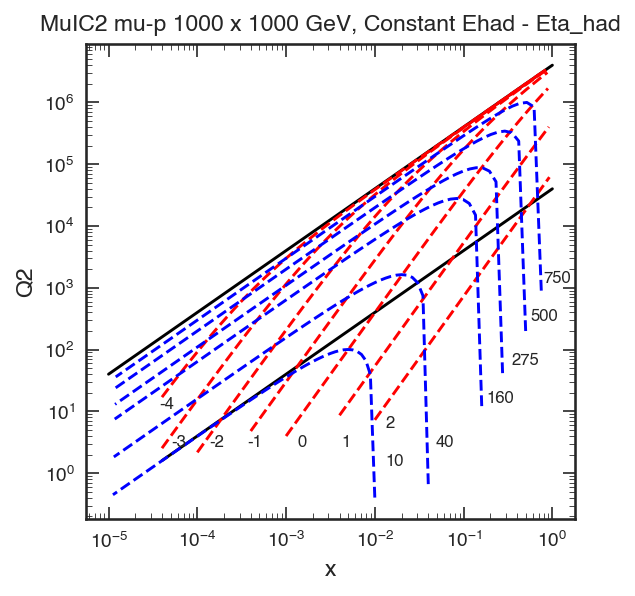}
    \caption{Kinematics of the scattered muon (left) and final-state hadrons (right) in the $Q^2$-$x$ plane for muon-proton deep inelastic scattering with a 1000~GeV muon beam colliding with a 1000~GeV proton beam. The dashed blue lines correspond to constant  energy in GeV and the dashed red lines to constant pseudorapidity, respectively. The lower diagonal solid line corresponds to the inelasticity $y=0.01$.}
    \label{fig:MuIC2kinematics}
\end{figure}

Figure~\ref{fig:MuIC2kinematics} shows the corresponding scattering kinematics for a more symmetric muon-proton collider where both beams have an energy of 1~TeV, such as might be achieved by upgrading the bending magnets of the hadron ring at the BNL EIC or by resurrecting a Tevatron proton beam to collide with a TeV muon beam. The doubling of the center-of-mass energy to 2~TeV increases the reach in $Q^2$ and $x$, but the scattering kinematics are quite similar as for the 1~TeV machine. The energy of the scattered hadron system does see an increase  at the highest $Q^2$, however.

\begin{figure}[!htb]
    \centering
    \includegraphics[width=0.43\textwidth]{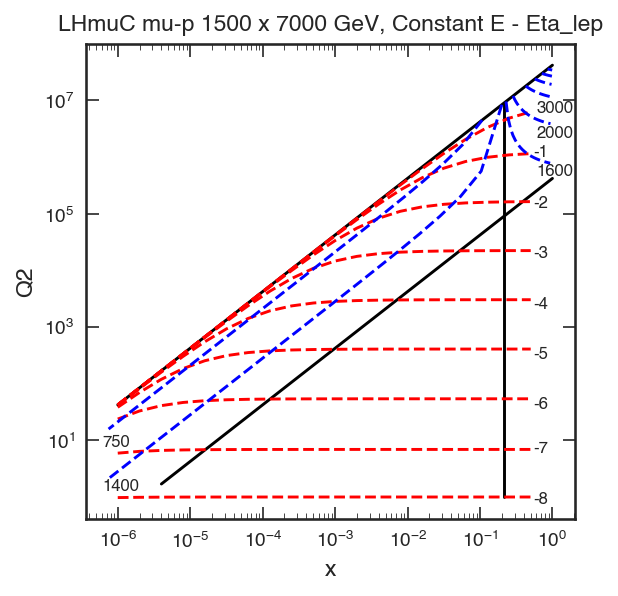}
    \includegraphics[width=0.45\textwidth]{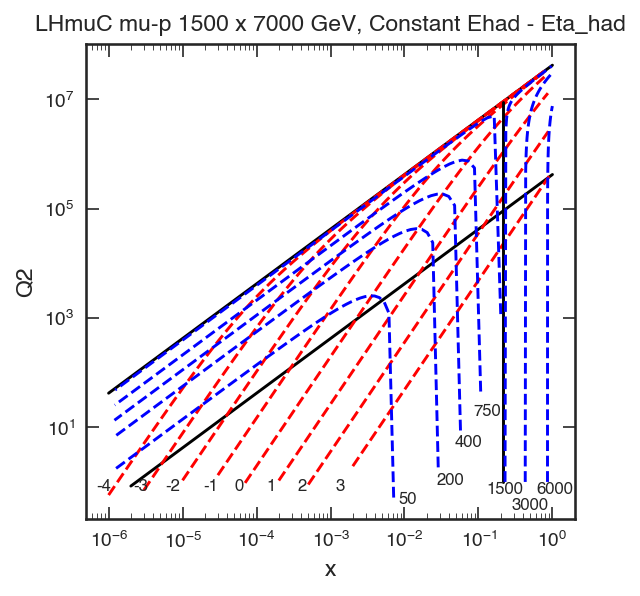}
    \caption{Kinematics of the scattered muon (left) and final-state hadrons (right) in the $Q^2$-$x$ plane for muon-proton deep inelastic scattering with a 1500~GeV muon beam colliding with a 7000~GeV proton beam. The dashed blue lines correspond to constant  energy in GeV and the dashed red lines to constant pseudorapidity, respectively. The lower diagonal solid line corresponds to the inelasticity $y=0.01$.}
    \label{fig:LHmuCkinematics}
\end{figure}

Finally, Fig.~\ref{fig:LHmuCkinematics} shows the corresponding scattering kinematics for a more ambitious collider comprised of a 1.5~TeV muon beam colliding with a 7~TeV proton beam, such as might be achieved by taking one of the muon beams of a 3~TeV $\mu^+\mu^-$ collider and colliding it with one of the CERN LHC proton beams. The center-of-mass energy of such a collider is 6.5~TeV, which vastly increases the reach in $Q^2$ and $x$. Like the previously mentioned options, the scattered muon at low $Q^2$ is in the far backward direction ($\eta \gtrsim -7$) with momenta up to 1.5~TeV. At high $Q^2$, the muon scatters in the central region with multi-TeV momentum. In this collider configuration, the scattered hadron system is more central than the previous two options, and is significantly more energetic.  

\begin{figure}[!htb]
    \centering
    \includegraphics[width=0.43\textwidth]{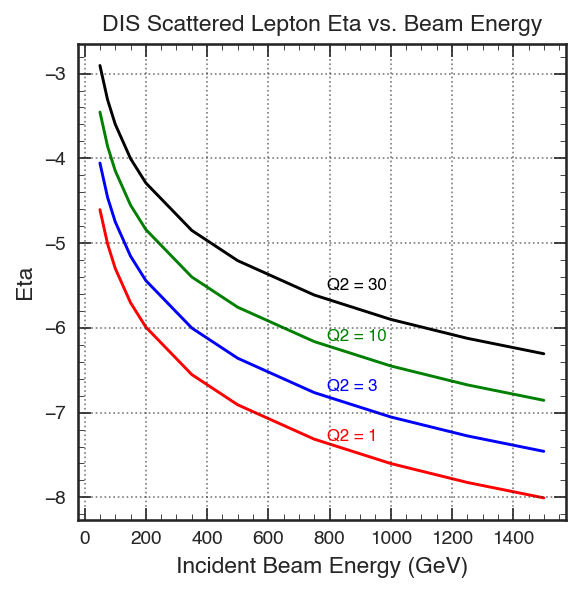}
    \caption{The pseudorapidity of the scattered lepton in deep inelastic scattering as a function of the incident lepton beam energy for $Q^2=1$, 3, 10, and 30~GeV$^2$. }
    \label{fig:DISleptonEta}
\end{figure}

We can summarize the kinematics of the scattered lepton and hadronic system in the following way. For low $x \ll E_\mu/E_{\rm p}$, where $E_\mu$ and $E_{\rm p}$ are the incident lepton and proton beam energies, respectively, we have for the scattered energies of the muon ($E^\prime_\mu$) and hadronic system ($E^\prime_{\rm h}$):
\begin{equation}
    E^\prime_\mu  \approx E_\mu (1-y) 
\end{equation}
\begin{equation}
    E^\prime_{\rm h}  \approx y E_\mu
\end{equation}
where $y$ is the DIS inelasticity variable. Likewise for the pseudorapidities of the scattered muon ($\eta^\prime_\mu$) and hadronic system ($\eta^\prime_{\rm h}$) at low $Q^2$ and $y$, we have:
\begin{equation}
\eta^\prime_\mu \approx -\ln \bigg( {2 E_\mu \over \sqrt{Q^2}} \bigg)
\end{equation}
\begin{equation}
\eta^\prime_{\rm h} \approx -\ln \bigg( {\sqrt{Q^2} \over 2 E_{\rm p}\, x } \bigg)
\end{equation}
Thus, we see that it is important to measure the scattered muon momentum up to the initial muon beam energy (even exceeding it when the proton beam energy exceeds the muon beam energy), which is in the TeV regime for the options considered here, and at pseudorapidities very close to the beamline. One way to illustrate the needed angular acceptance for the scattered muons for a range of options is expressed in Fig.~\ref{fig:DISleptonEta}, which shows  the dependence of the scattered muon pseudorapidity as a function of the incident muon beam energy for several $Q^2$ choices. 

\begin{figure}[!htb]
    \centering
    \includegraphics[width=0.43\textwidth]{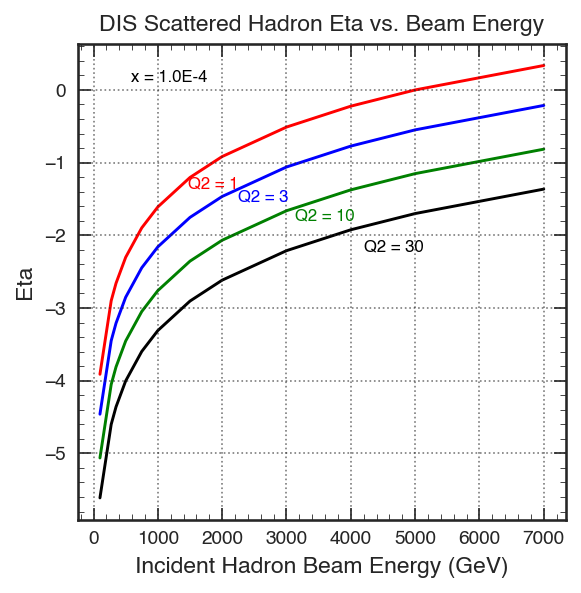}
    \includegraphics[width=0.43\textwidth]{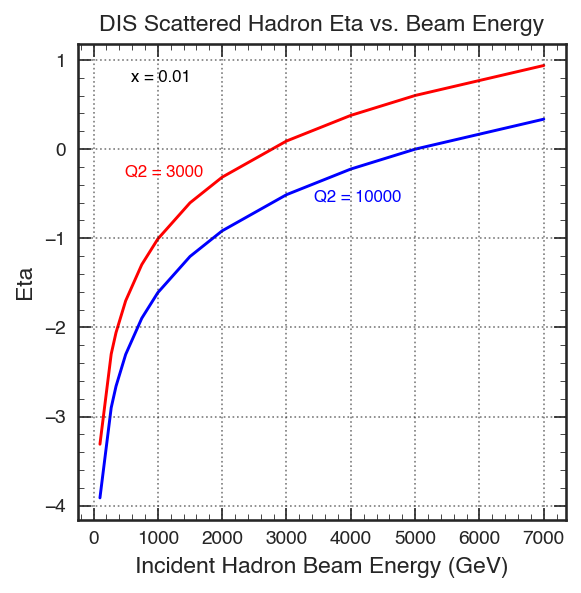}
    \caption{The pseudorapidity of the scattered hadronic system  in deep inelastic scattering as a function of the incident proton beam energy for $x=1.0 \times 10^{-4}$ (left) and $x=0.01$ (right) for several choices of $Q^2$. }
    \label{fig:DISHadronEta}
\end{figure}

Similarly, the scattered hadronic system also can have an energy up to initial muon beam energy, or above when the proton beam energy exceeds the muon beam energy. The pseudorapidity of the scattered hadronic system depends on $x$ and $Q^2$. Figure~\ref{fig:DISHadronEta} shows the hadron pseudorapidity dependence on the incident proton beam energy for $x=1.0\times 10^{-4}$ (left) and for $x=0.01$ (right) for various choices of $Q^2$. 
The detection of hadrons is not required to be as far backward as with the muons, although $\eta \gtrsim -4$ would be good, and does not require their detection very far forward (e.g. $\eta\lesssim 2$ would be acceptable).
The hadron pseudorapidity trends toward to the muon beam direction as $Q^2$ increases. 

\begin{figure*}[t!]
\centering
\includegraphics[width=0.3\textwidth]{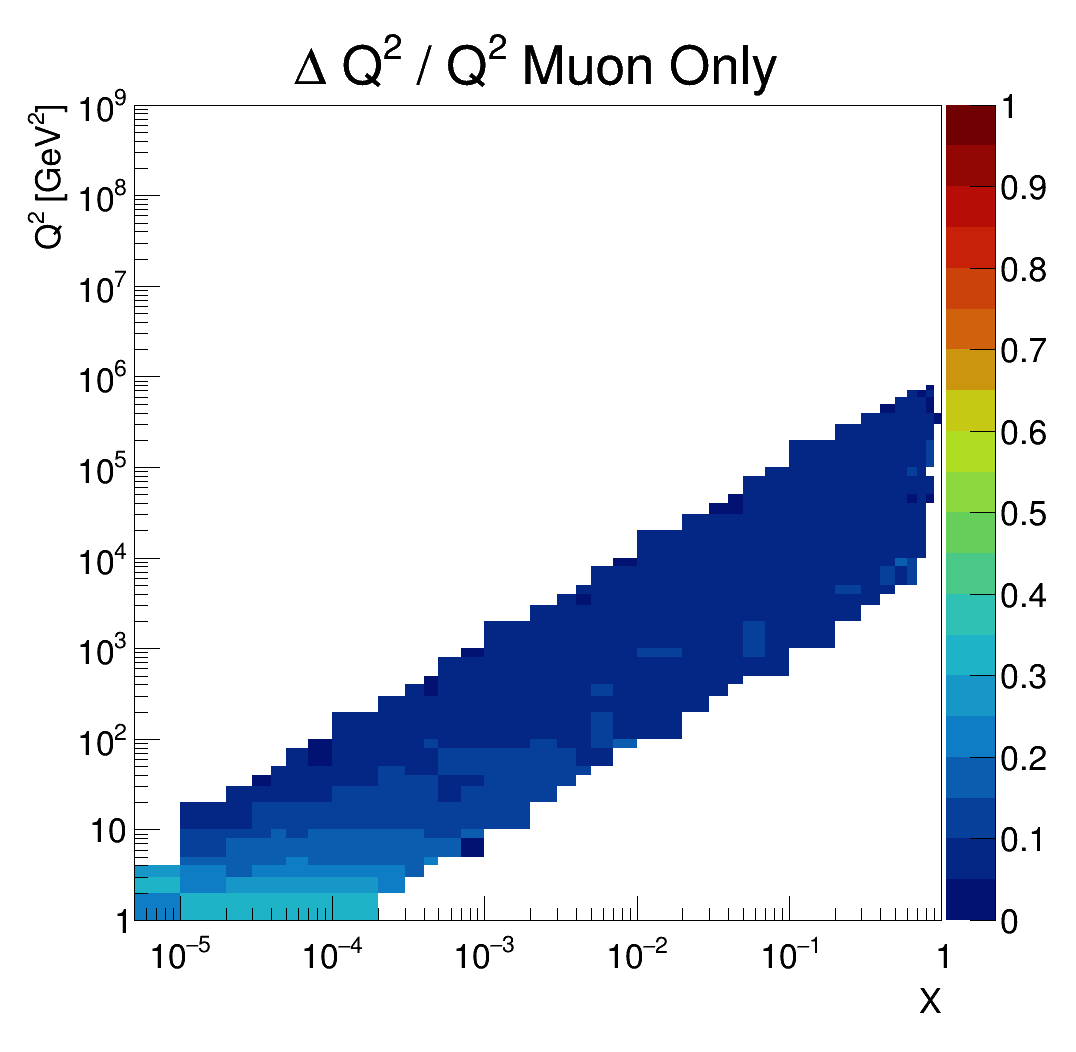}
\includegraphics[width=0.3\textwidth]{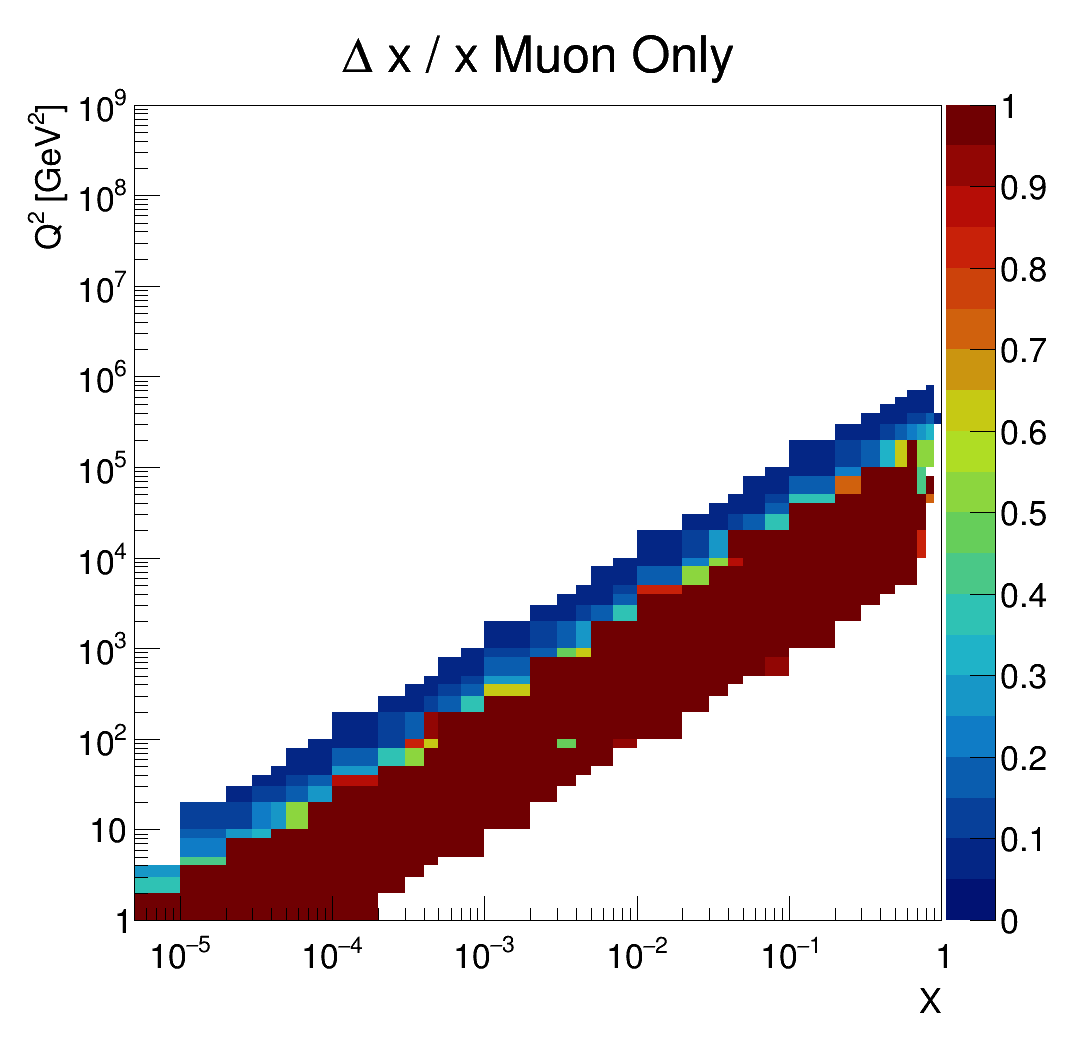}
\includegraphics[width=0.3\textwidth]{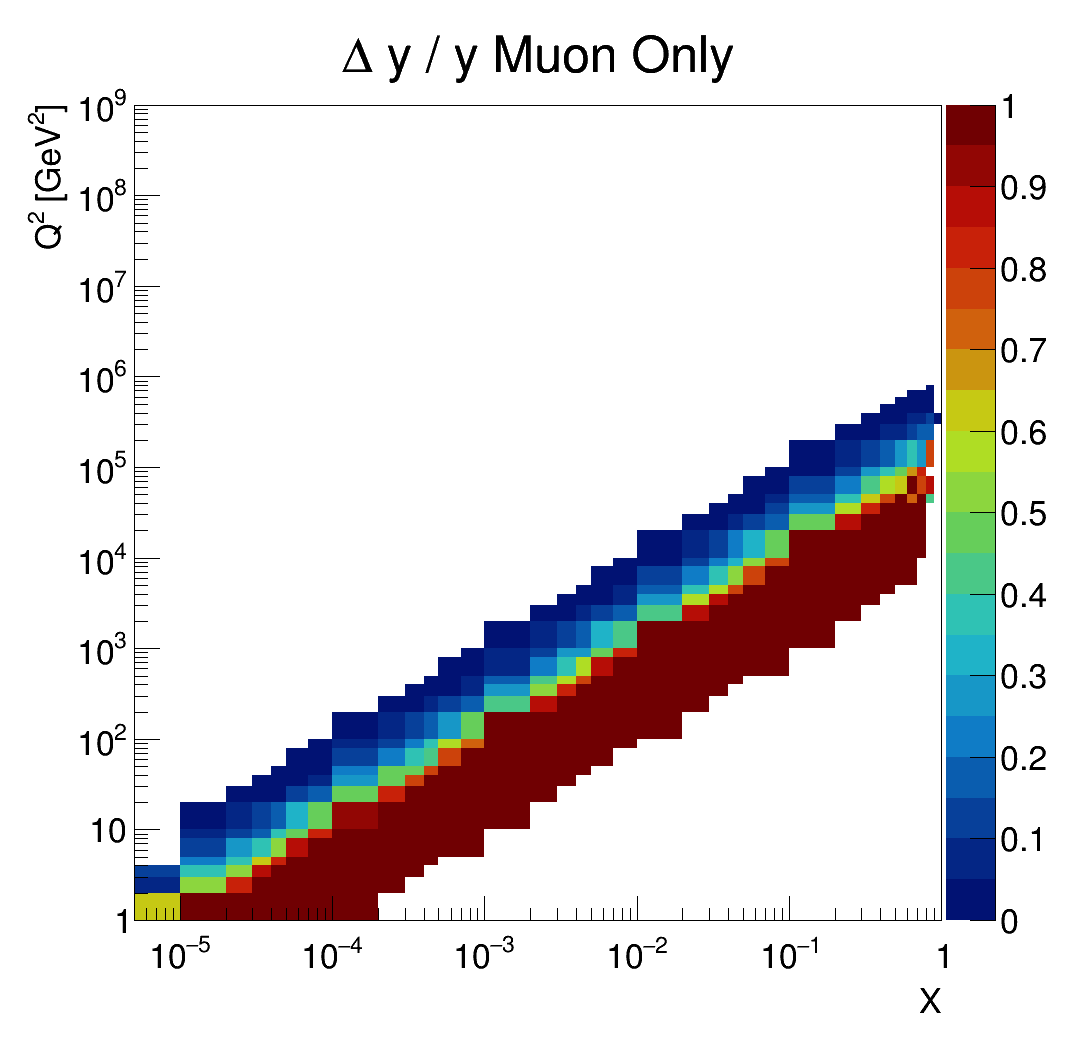}
\includegraphics[width=0.3\textwidth]{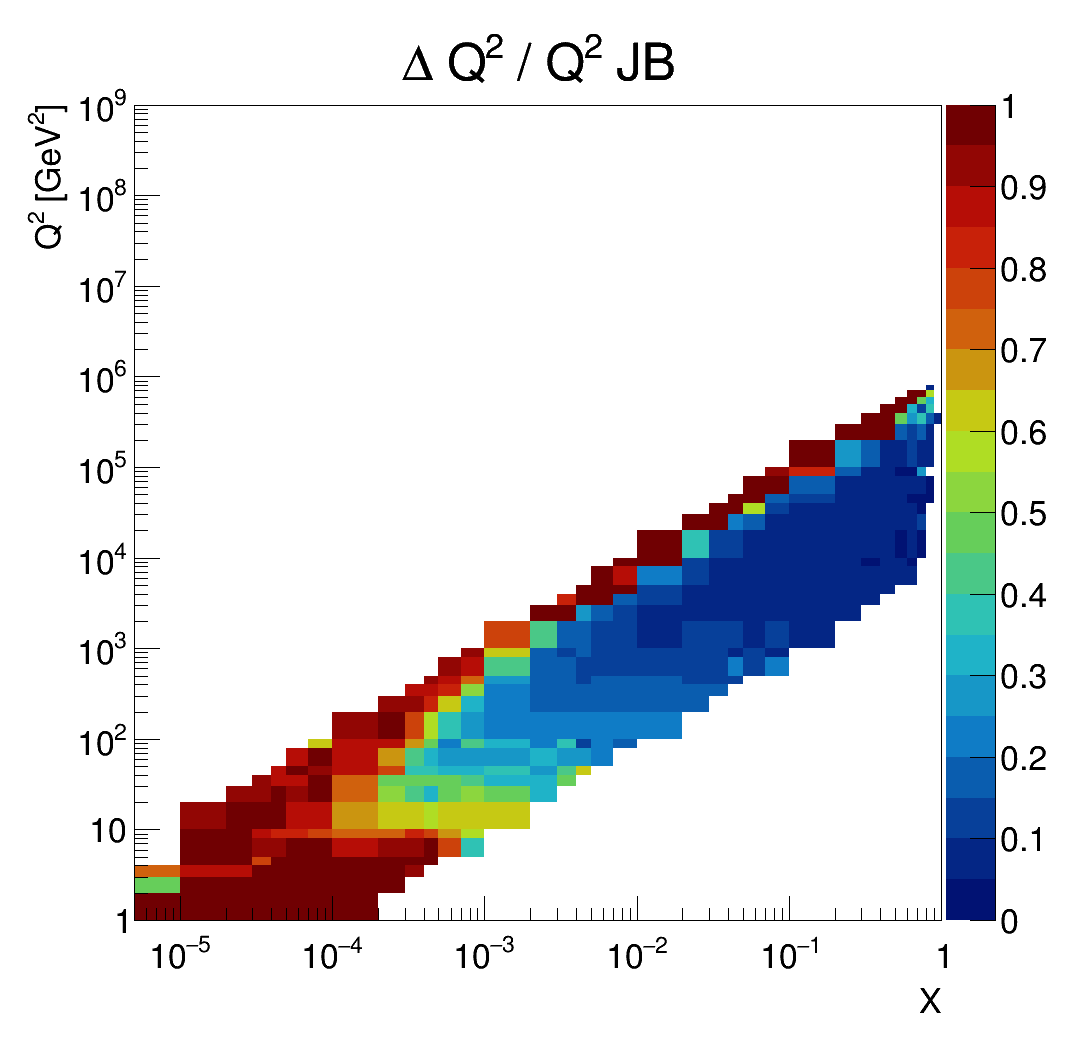}
\includegraphics[width=0.3\textwidth]{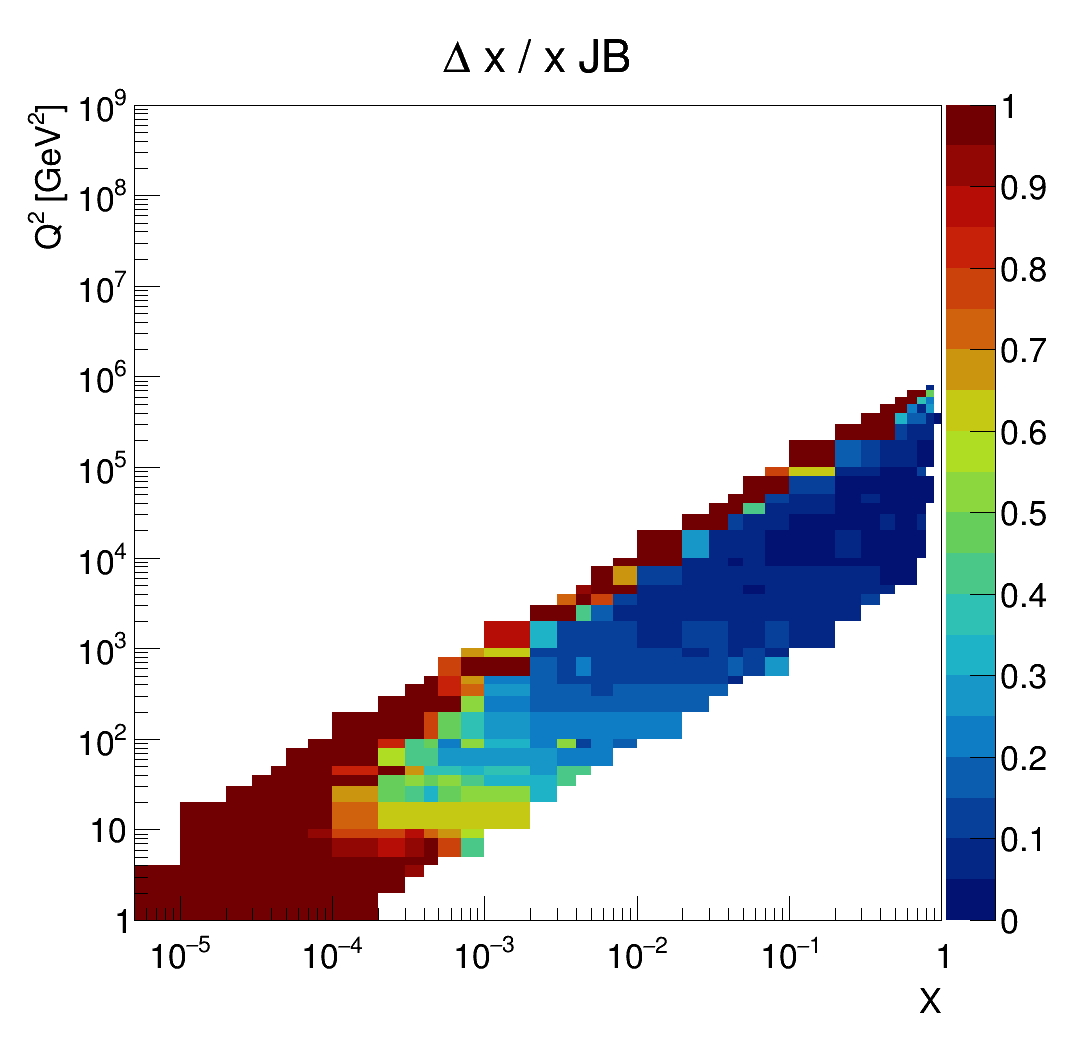}
\includegraphics[width=0.3\textwidth]{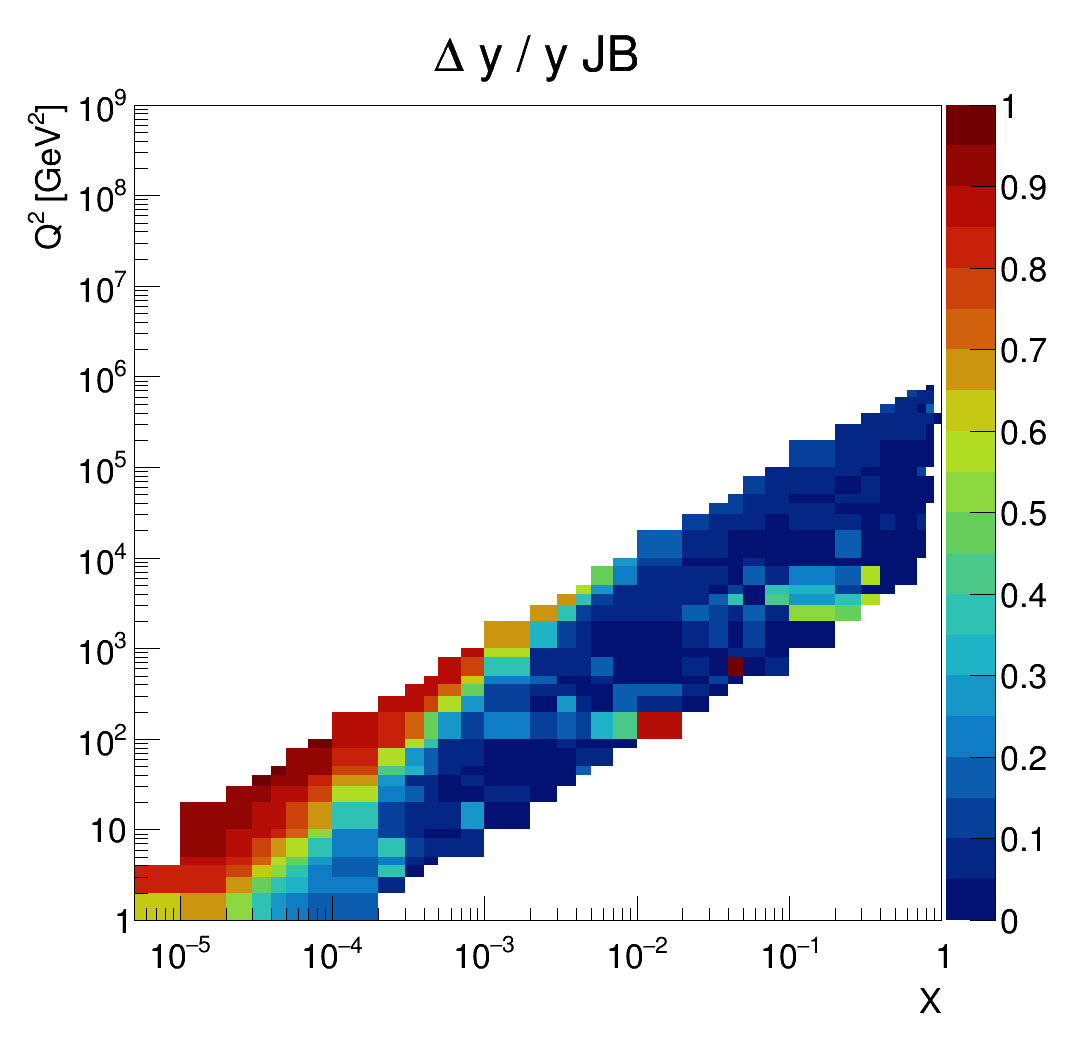}
\includegraphics[width=0.3\textwidth]{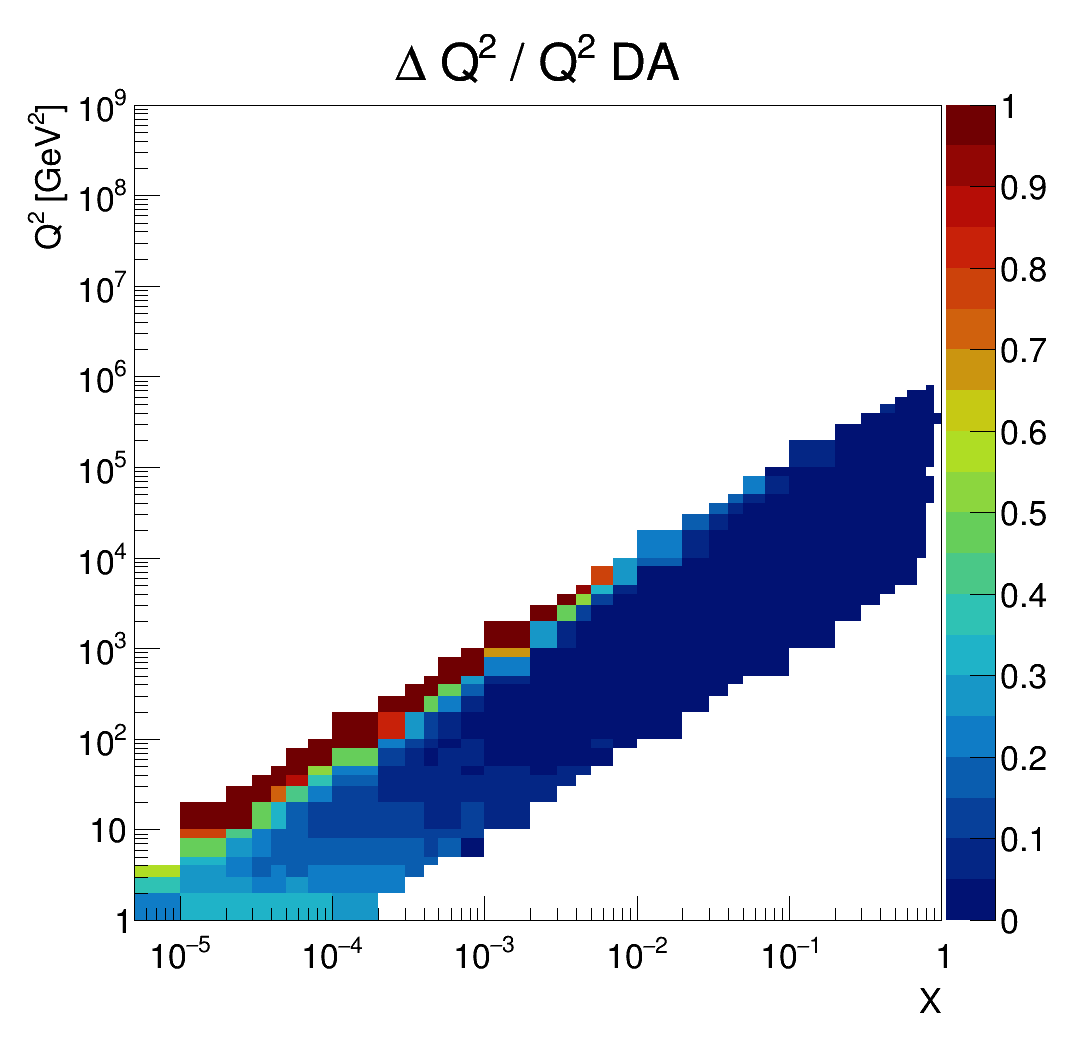}
\includegraphics[width=0.3\textwidth]{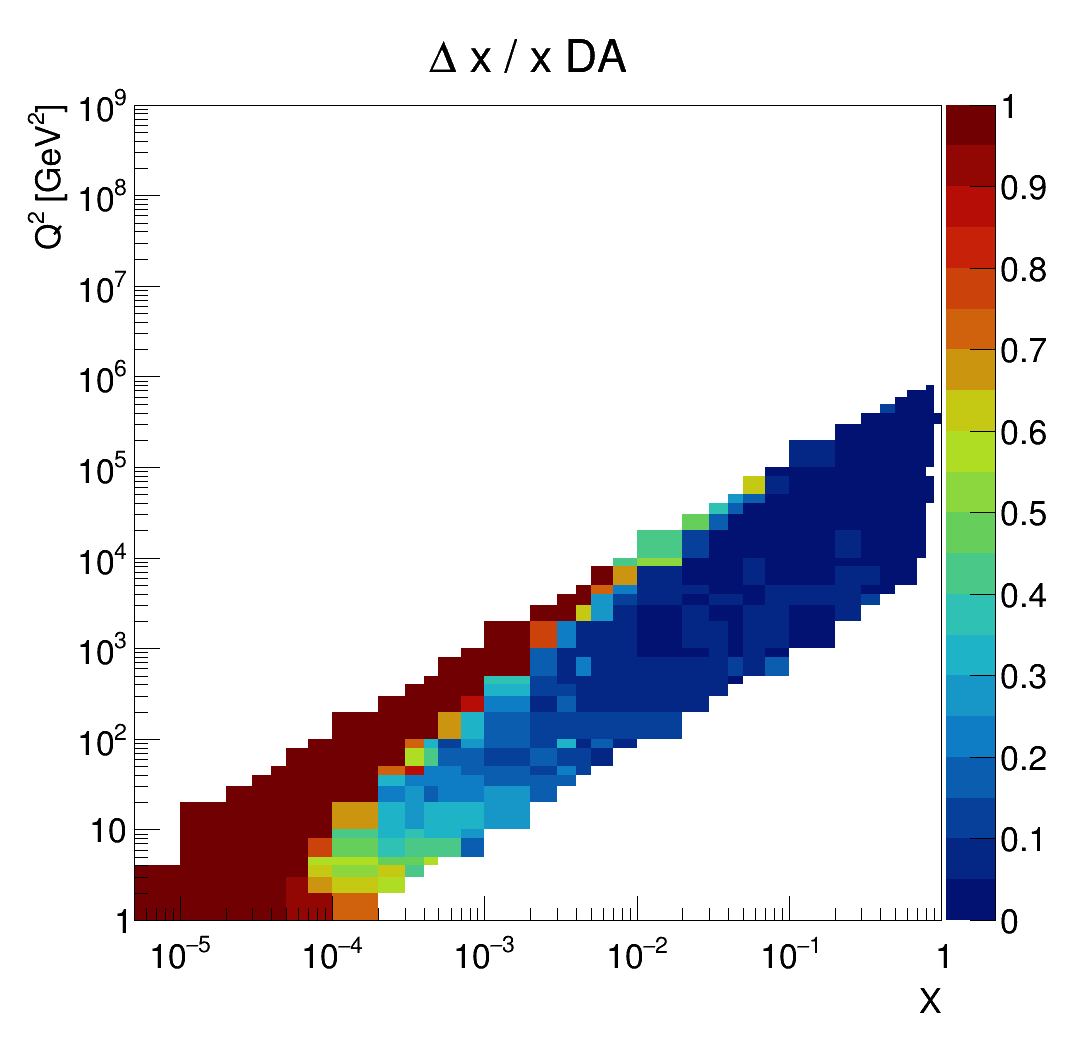}
\includegraphics[width=0.3\textwidth]{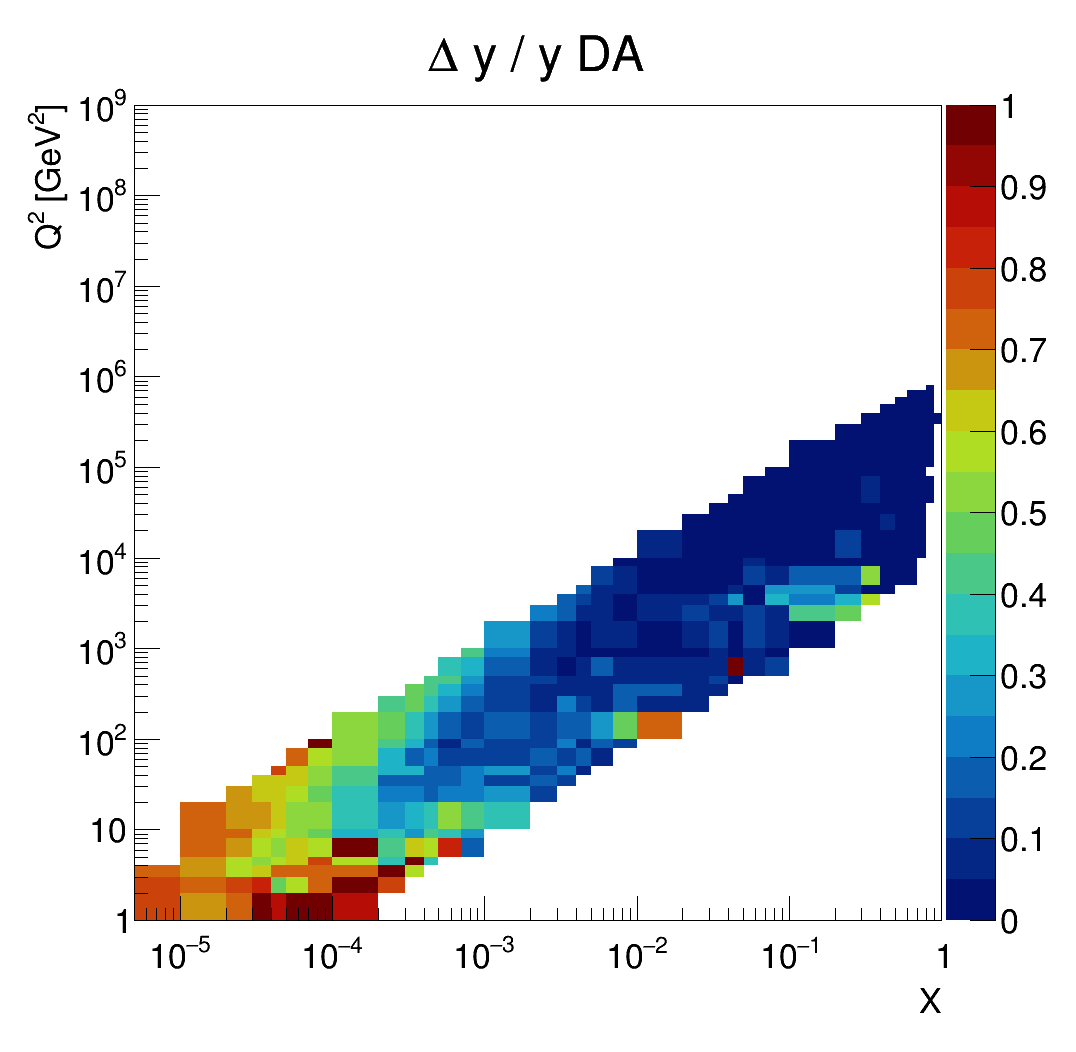}

\caption{Resolutions of DIS variables $Q^{2}$, $x$ and $y$ at the MuIC as functions of 
$Q^{2}$ and $x$ in two dimensions
reconstructed with three methods: lepton-only (top), 
Jacquet-Blondel (JB) using hadronic activities within 
$-4 < \eta_{\rm h}<2.4$ (middle), and Double Angle (DA) within 
$-4 < \eta_{\rm h}<2.4$ (bottom). 
The selected range in $y$ is $0.1 < y < 0.9$.
The assumed detector resolution parameters were summarized 
in Ref.~\cite{Acosta:2021qpx}.
\label{fig:resol-eta4}
}
\end{figure*}

The resolution of the DIS kinematic variables $Q^2$, $x$, and $y$ at the MuIC, reconstructed according to three algorithms, were studied in Ref.~\cite{Acosta:2021qpx} for some target detector performance parameters that were used to smear the Monte Carlo generated particles from Pythia~8 in the final state (Table~\ref{tab:table_resolution} summarizes these resolutions). The muon coverage is expected to extend to $\eta = -7$, and the hadron acceptance to $|\eta|<5$. We checked the impact on limiting the experimental hadron coverage further to $-4 < \eta_{\rm h}<2.4$, and display the corresponding resolution results in Fig.~\ref{fig:resol-eta4}. The upper bound on $\eta_{\rm h}$ is to accommodate a shielding cone that is necessary to absorb beam-induced backgrounds from the muon beam, and the lower bound is to restrict the range to the EIC experiment requirement  \cite{eic_cdr}. The reconstruction algorithms are the lepton-only method (top row, which is unaffected by the hadron restriction), the Jacquet-Blondel method based solely on the measured hadron energy and effective angle measured from energy sums (middle row), 
and the Double-Angle method based on the lepton and the effective hadron angles (bottom row).  
The upper bound on $\eta_{\rm h}$ has little effect, as can be inferred also from Fig.~\ref{fig:MuICkinematics}. The lower bound does lead to some further degradation at low $x$ and $Q^2$. For the Jacquet-Blondel reconstruction of $Q^2$ and $x$, a resolution ${\lesssim }50$\% increases the minimum measured $Q^2$ roughly from 5 to 10~GeV$^2$, and the minimum $x$ from $10^{-4}$ to about $5\times 10^{-4}$. Simulated events only make it into the resolution plots if there is a hadron in the acceptance region. 

\begin{table}[b!]
\centering
\caption{List of assumed detector resolutions
for measuring various species of particles at MuIC for the DIS kinematic resolution study, from Ref.~\cite{Acosta:2021qpx}.}
\vspace{0.5cm}
\includegraphics[width=0.7\linewidth]{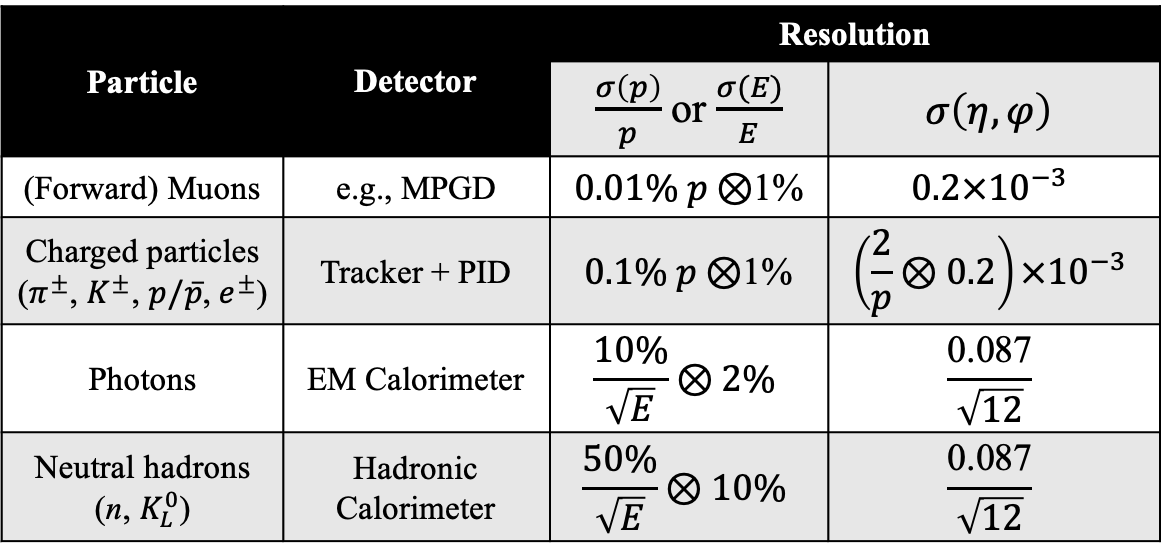}
\label{tab:table_resolution}
\end{table}

\section{Parameterized Detector Simulation}
\label{sec:delphes}

The detector simulation for the studies reported in Section~\ref{sec:higgs_phys} is performed with the Delphes framework~\cite{delphes}, where the particle propagation and the detector responses are calculated based on the true kinematic information of the final state particles and smeared parametrically.

Following the conceptual design shown in Figure~\ref{fig:detector}, the Delphes simulation of MuIC includes a silicon tracker and a RICH detector both covering $-4 < \eta < 2.4$, an ECAL and a HCAL covering $-5 < \eta < 2.4$, a central muon detector covering $-4 < \eta < 0$, and a far-backward muon detector covering $-7 < \eta < -4$. The far-forward RP detector is not included in the simulation.

A momentum smearing is applied to all charged tracks reconstructed by the detector, which differs by particle type and $\eta$ region, as listed in Table~\ref{tab:delphes_reso}. Smearing parameters in the central detector are taken from the CMS Delphes card. These numbers do not align with the resolutions in Table~\ref{tab:table_resolution} as they also aim to capture mis-seeding and mis-matching effects in track reconstruction and are not just resolution from track fits. 

In addition to the momentum smearing, an $\eta$ smearing of $\Delta\eta = 10^{-4} / \text{arctan(exp}(\eta))$, corresponding to a $\Delta\theta = 0.2$~mrad, is applied to the muons within $-7 < \eta < -4$. Angular smearing inside the tracker volume is neglected.

\begin{table}[!htb]
    \centering
    \caption{Relative momentum smearing $\Delta p_{\rm T}/p_{\rm T}$ for different types of tracks in the Delphes simulation. Values of $p_{\rm T}$ and $E$ are in GeV for these formulae.}
\scriptsize
    \begin{tabular}{c|ccc}
         \hline
         $\eta$ regions & Charged hadron & Electron & Muon \\
         \hline
         $|\eta| < 0.5$ & $\sqrt{6\%^2 + (0.13\%*p_{\rm T})^2}$ & $\sqrt{3\%^2 + (0.13\%*p_{\rm T})^2}$ & - \\
         $0.5 < |\eta| < 1.5$ & $\sqrt{10\%^2 + (0.17\%*p_{\rm T})^2}$ & $\sqrt{5\%^2 + (0.17\%*p_{\rm T})^2}$ & - \\
         $-4 < \eta < -1.5$ \& $1.5 < \eta < 2.4$ & $\sqrt{25\%^2 + (0.31\%*p_{\rm T})^2}$ & $\sqrt{15\%^2 + (0.31\%*p_{\rm T})^2}$ & - \\
         \hline
         $-0.5<\eta<0$    & - & - & $\sqrt{1\%^2 + (0.01\%*p_{\rm T})^2}$ \\
         $-1.5<\eta<-0.5$ & - & - & $\sqrt{1.5\%^2 + (0.015\%*p_{\rm T})^2}$ \\
         $-4.0<\eta<-1.5$ & - & - & $\sqrt{2.5\%^2 + (0.035\%*p_{\rm T})^2}$ \\
         $-7.0<\eta<-4.0$ & - & - & $\sqrt{5\%^2 + (0.01\%*E)^2}$ \\
         \hline
    \end{tabular}
    \label{tab:delphes_reso}
\end{table}

The PID assignment from the RICH detector is set to be 95\% efficient for pions, kaons, and protons, and 97\% efficient for muons and electrons. The misidentification rates for pions and kaons to be assigned as muons are set at 1--2\%, while the misidentification rate for protons to be assigned as muons is set at 0.2\%.

The calorimeters are parameterized in grids for different $\eta$ regions. The energy of each particle is assigned to one ECAL/HCAL unit along its path and not shared across neighboring units. All electrons, photons, and pions are assumed to have 100\% of their energy deposited in the ECAL, kaons are assumed to have 30\% of their energy in ECAL and 70\% in HCAL, while other hadrons are assumed to have 100\% energy deposit in HCAL. These ``true'' energy deposits are then smeared into the measured energy deposits in calorimeter units. Parameters of the calorimeter grid and resolution are detailed in Table~\ref{tab:delphes_calo}. These numbers are taken from the CMS Delphes card, which follows the CMS calorimeter geometry and resolution~\cite{Sirunyan_2017, Sirunyan_2021}.

\begin{table}[!htb]
    \centering
    \caption{Grid and resolution parameters used for the ECAL and HCAL in the Delphes simulation.}
\scriptsize
    \begin{tabular}{c|c|cc}
        \hline
        $\eta$ regions & Parameters & ECAL & HCAL \\
        \hline
        Barrel & $\Delta\phi \times \Delta\eta$ grid & $1^\circ \times 0.0174$ & $5^\circ \times 0.087$ \\
        $|\eta| < 1.5$  & Resolution $\Delta E$ & $(1+0.64\eta^2) \times \sqrt{0.008^2 E^2 + 0.11^2 E + 0.4^2}$ & $\sqrt{0.05^2 E^2 + 1.5^2 E}$ \\
        \hline
        Endcap & $\Delta\phi \times \Delta\eta$ grid & $1^\circ \times 0.0174$ & $10^\circ \times 0.175$ \\
        $-3 <\eta < -1.5 \& 1.5 < \eta < 2.4$ & Resolution $\Delta E$ & $(2.16+5.6(|\eta|-2)^2) \times \sqrt{0.008^2 E^2 + 0.11^2 E + 0.4^2}$ & $\sqrt{0.05^2 E^2 + 1.5^2 E}$ \\
        \hline
        Forward & $\Delta\phi \times \Delta\eta$ grid & $10^\circ \times 0.175$ & $20^\circ \times 0.3$ \\
        $-5 < \eta < -3$ & Resolution $\Delta E$ &  $\sqrt{0.107^2 E^2 + 2.08^2 E}$ & $\sqrt{0.13^2 E^2 + 2.7^2 E}$ \\
        \hline
    \end{tabular}
    \label{tab:delphes_calo}
\end{table}

After these steps, each ECAL deposit overlapping with a track is identified as an electron, each HCAL deposit overlapping with a track is identified as an charged hadron, and unassigned ECAL and HCAL deposits are identified as photons and neutral hadrons, respectively. 
Jets are then clustered with the anti-$k_{T}$ algorithm~\cite{Cacciari_2008,fastjet2012} within a $\Delta R$ of 0.5. The b-tagging is parameterized with the CMS b-tagging performance~\cite{Sirunyan_2018}, corresponding to a nominal b-jet efficiency of about 68\% and a fake rate from light flavor jets of about 1\%.

\acknowledgments

This work is in part supported by the Department of Energy 
grant numbers DE-SC0005131 (W.L.) and DE-SC0010266 (D.A.). N.H. acknowledges support from the Office of Undergraduate Research and Fellowships at Northeastern University.


\begin{thebibliography}{99}


\bibitem{NAP25171}
{\em An Assessment of U.S.-Based Electron-Ion Collider Science}.
\newblock The National Academies Press, Washington, DC, 2018.

\bibitem{Accardi:2012qut}
A.~Accardi et~al.
\newblock {\em Electron Ion Collider: The Next QCD Frontier}: {Understanding the
  glue that binds us all}.
\newblock {\em Eur. Phys. J. A}, 52:268, 2016.

\bibitem{Agostini:2020fmq}
P.~Agostini et~al.
\newblock {\em The Large Hadron-Electron Collider at the HL-LHC}.
\newblock {arXiv:2007.14491 [hep-ex]}, 2020.

\bibitem{Abada:2019lih}
A.~Abada et~al.
\newblock {\em FCC Physics Opportunities}: {Future Circular Collider Conceptual
  Design Report Volume 1}.
\newblock {\em Eur. Phys. J. C}, 79:474, 2019.

\bibitem{Shiltsev:1997pv}
V.~D. Shiltsev.
\newblock {\em An Asymmetric muon - proton collider: Luminosity consideration}.
\newblock {\em Conf. Proc. C}, 970512:420, 1997.

\bibitem{Ginzburg:1998yw}
I.~F. Ginzburg.
\newblock {\em Physics at future e p, gamma p (linac-ring) and mu p colliders}.
\newblock {\em Turk. J. Phys.}, 22:607--610, 1998.

\bibitem{doi:10.1063/1.56424}
Kingman Cheung.
\newblock Muon-proton colliders: Leptoquarks and contact interactions.
\newblock {\em AIP Conference Proceedings}, 441(1):338--344, 1998.

\bibitem{Sultansoy:1999na}
S.~Sultansoy.
\newblock {\em The PostHERA era: Brief review of future lepton hadron and photon
  hadron colliders}.
\newblock {arXiv:hep-ph/9911417}, 1999.

\bibitem{Cheung:1999wy}
Kingman Cheung.
\newblock {\em Muon proton colliders: Leptoquarks, contact interactions and extra
  dimensions}.
\newblock {\em AIP Conf. Proc.}, 542(1):160--170, 2000.

\bibitem{Acar:2016rde}
Y.~C. Acar, A.~N. Akay, S.~Beser, A.~C. Canbay, H.~Karadeniz, U.~Kaya, B.~B.
  Oner, and S.~Sultansoy.
\newblock {\em Future circular collider based lepton\textendash{}hadron and
  photon\textendash{}hadron colliders: Luminosity and physics}.
\newblock {\em Nucl. Instrum. Meth. A}, 871:47, 2017.

\bibitem{Canbay:2017rbg}
Ali~C. Canbay, Umit Kaya, Bora Ketenoglu, Bilgehan~Baris Oner, and Saleh
  Sultansoy.
\newblock {\em SppC based energy frontier lepton-proton colliders: luminosity and
  physics}.
\newblock {\em Adv. High Energy Phys.}, 2017:4021493, 2017.

\bibitem{Acar:2017eli}
Yigit~Can Acar, Umit Kaya, and Bilgehan~Baris Oner.
\newblock {\em Resonant production of color octet muons at Future Circular
  Collider-based muon-proton colliders}.
\newblock {\em Chin. Phys. C}, 42(8):083108, 2018.

\bibitem{Caliskan:2018vep}
Abdullatif Caliskan.
\newblock {\em Search for excited muons at the future SPPC-based muon-proton
  colliders}.
\newblock {arXiv:1802.09874 [hep-ph]}, 2018.

\bibitem{Ketenoglu:2018fai}
Bora Ketenoglu.
\newblock {\em Main parameters of SppC-based ''linac-ring eA'' and ''ring-ring
  muA'' colliders}.
\newblock {arXiv:1811.05129 [physics.acc-ph]}, 2018.

\bibitem{Kaya:2019ecf}
U.~Kaya, B.~Ketenoglu, S.~Sultansoy, and F.~Zimmermann.
\newblock {\em Main parameters of HL-LHC and HE-LHC based mu-p colliders}.
\newblock {arXiv:1905.05564 [physics.acc-ph]}, 2019.

\bibitem{Ozansoy:2019rmu}
Aysuhan Ozansoy.
\newblock {\em Investigating doubly charged leptons at future energy frontier
  muon-proton colliders}.
\newblock {\em Communications Faculty of Sciences University of Ankara Series
  A2-A3: Physical Sciences and Engineering}, 61(1):111--128, 2019.

\bibitem{Aydin:2021iky}
Gural Aydin, Yusuf~Oguzhan G\"unaydin, Mehmet Sahin, Saleh Sultansoy, and
  Mehmet~T\"urker Tarakcioglu.
\newblock {\em Contact Interactions at Future Circular Collider based Muon-Proton
  Colliders}.
\newblock {arXiv:2105.09686 [hep-ph]}, 2021.

\bibitem{Cheung:2021iev}
Kingman Cheung and Zeren~Simon Wang.
\newblock {\em Physics potential of a muon-proton collider}.
\newblock {\em Phys. Rev. D}, 103:116009, 2021.

\bibitem{Acosta:2021qpx}
Darin Acosta and Wei Li.
\newblock {\em A muon\textendash{}ion collider at BNL: The future QCD frontier and
  path to a new energy frontier of
  \ensuremath{\mu}+\ensuremath{\mu}\ensuremath{-} colliders}.
\newblock {\em Nucl. Instrum. Meth. A}, 1027:166334, 2022.

\bibitem{Delahaye:2019omf}
Jean~Pierre Delahaye, Marcella Diemoz, Ken Long, Bruno Mansouli\'e, Nadia
  Pastrone, Lenny Rivkin, Daniel Schulte, Alexander Skrinsky, and Andrea
  Wulzer.
\newblock {\em Muon Colliders}.
\newblock {arXiv:1901.06150 [physics.acc-ph]}, 2019.

\bibitem{MCwpPhysics}
D~Stratakis et~al.
\newblock {\em A Muon Collider Facility for Physics Discovery}.
\newblock {arXiv:2203.08033 [physics.acc-ph]}, 2022.

\bibitem{MICE:2019jkl}
M.~Bogomilov et~al.
\newblock {\em Demonstration of cooling by the Muon Ionization Cooling Experiment}.
\newblock {\em Nature}, 578(7793):53, 2020.

\bibitem{Behnke:2013xla}
{The International Linear Collider Technical Design Report - Volume 1:
  Executive Summary}.
\newblock {arXiv:1306.6327 [physics.acc-ph]}, 2013.

\bibitem{Charles:2018vfv}
T.~K. Charles et~al.
\newblock {\em The Compact Linear Collider (CLIC) - 2018 Summary Report}.
\newblock {arXiv:1812.06018 [physics.acc-ph]}, 2018.

\bibitem{CEPCStudyGroup:2018ghi}
Mingyi Dong et~al.
\newblock {\em CEPC Conceptual Design Report: Volume 2 - Physics \& Detector}.
\newblock {arXiv:1811.10545 [hep-ex]}, 2018. 

\bibitem{MCwp3TeV}
J.~de~Blas et~al.
\newblock {\em The physics case of a 3 TeV muon collider stage}.
\newblock {arXiv:2203.07261 [hep-ph]}, 2022.

\bibitem{Palmer:2014nza}
R.~B. Palmer.
\newblock {\em Muon Colliders}.
\newblock {\em Rev. Accel. Sci. Tech.}, 7:137, 2014.

\bibitem{Aschenauer:2014cki}
E.~C. Aschenauer et~al.
\newblock {\em eRHIC Design Study: An Electron-Ion Collider at BNL}.
\newblock {arXiv:1409.1633 [physics.acc-ph]}, 2014.

\bibitem{eic_cdr}
Electron-ion collider at brookhaven national laboratory - conceptual design
  report 2021. {doi:10.2172/1765663}, 2021.

\bibitem{NeutrinoMitigate}
Bruce King.
\newblock {\em Neutrino Radiation Challenges and Proposed Solutions for Many-TeV
  Muon Colliders}.
\newblock {doi:10.1063/1.1361675}, 2000.

\bibitem{Cline:1996yi}
D.~Cline, B.~Norum, and R.~Rossmanith.
\newblock {\em Polarization in a muon collider}.
\newblock {\em Conf. Proc. C}, 960610:867, 1996.

\bibitem{Norum:1996mi}
B.~Norum and R.~Rossmanith.
\newblock {\em Polarized beams in a muon collider}.
\newblock {\em Nucl. Phys. B Proc. Suppl.}, 51:191, 1996.

\bibitem{Neuffer:1999aw}
David Neuffer.
\newblock {\em $\mu^+ \mu^-$ Colliders}.
\newblock {doi:10.5170/CERN-1999-012}, 1999.

\bibitem{Ankenbrandt:1999cta}
Charles~M. Ankenbrandt et~al.
\newblock {\em Status of muon collider research and development and future plans}.
\newblock {\em Phys. Rev. ST Accel. Beams}, 2:081001, 1999.

\bibitem{Adolphsen:2022ibf}
C.~Adolphsen et~al.
\newblock {\em European Strategy for Particle Physics -- Accelerator R\&D Roadmap}.
\newblock {\em CERN Yellow Rep. Monogr. \textbf{1}}, 1-270 (2022).
\newblock {arXiv:2201.07895 [physics.acc-ph]}.

\bibitem{AbdulKhalek:2021gbh}
R.~Abdul~Khalek et~al.
\newblock {\em Science Requirements and Detector Concepts for the Electron-Ion
  Collider: EIC Yellow Report}.
\newblock {arXiv:2103.05419 [physics.ins-det]}, 2021.

\bibitem{GolecBiernat:1998js}
Krzysztof~J. Golec-Biernat and M.~Wusthoff.
\newblock {\em Saturation effects in deep inelastic scattering at low Q**2 and its
  implications on diffraction}.
\newblock {\em Phys. Rev. D}, 59:014017, 1998.

\bibitem{AlAli:2021let}
Hind Al~Ali et~al.
\newblock {\em The Muon Smasher's Guide}.
\newblock {arXiv:2103.14043 [hep-ph]}, 2021.

\bibitem{Abdallah:2021aut}
Mohamed Abdallah et~al. (STAR Collaboration)
\newblock {\em Longitudinal double-spin asymmetry for inclusive jet and dijet
  production in polarized proton collisions at $\sqrt{s}=200$ GeV}.
\newblock {\em Phys. Rev. D}, 103:L091103, 2021.

\bibitem{Aschenauer:2012ve}
Elke~C. Aschenauer, Rodolfo Sassot, and Marco Stratmann.
\newblock {\em Helicity Parton Distributions at a Future Electron-Ion Collider: A
  Quantitative Appraisal}.
\newblock {\em Phys. Rev. D}, 86:054020, 2012.

\bibitem{McLerran:1993ni}
Larry~D. McLerran and Raju Venugopalan.
\newblock {\em Computing quark and gluon distribution functions for very large
  nuclei}.
\newblock {\em Phys. Rev. D}, 49:2233, 1994.

\bibitem{CMS:2010ifv}
Vardan Khachatryan et~al. (CMS Collaboration)
\newblock {\em Observation of Long-Range Near-Side Angular Correlations in
  Proton-Proton Collisions at the LHC}.
\newblock {\em JHEP}, 09:091, 2010.

\bibitem{CMS:2012qk}
Serguei Chatrchyan et~al. (CMS Collaboration)
\newblock {\em Observation of Long-Range Near-Side Angular Correlations in
  Proton-Lead Collisions at the LHC}.
\newblock {\em Phys. Lett. B}, 718:795, 2013.

\bibitem{Dusling:2015gta}
Kevin Dusling, Wei Li, and Bj\"orn Schenke.
\newblock {\em Novel collective phenomena in high-energy proton\textendash{}proton
  and proton\textendash{}nucleus collisions}.
\newblock {\em Int. J. Mod. Phys. E}, 25:1630002, 2016.

\bibitem{ZEUS:2019jya}
I.~Abt et~al. (ZEUS Collaboration)
\newblock {\em Two-particle azimuthal correlations as a probe of collective
  behaviour in deep inelastic $ep$ scattering at HERA}.
\newblock {\em JHEP}, 04:070, 2020.

\bibitem{ATLAS:2021jhn}
Georges Aad et~al. (ATLAS Collaboration)
\newblock {\em Two-particle azimuthal correlations in photonuclear ultraperipheral
  Pb+Pb collisions at 5.02 TeV with ATLAS}.
\newblock {\em Phys. Rev. C}, 104(1):014903, 2021.

\bibitem{CMS-PAS-HIN-18-008}
CMS Collaboration
\newblock {\em Search for elliptic azimuthal anisotropies in $\gamma$p interactions within
  ultra-peripheral $\mathrm{p}\mathrm{Pb}$ collisions at
  $\sqrt{s_{NN}}=8.16~\mathrm{TeV}$}.
\newblock {http://cds.cern.ch/record/2725477}, Technical report, CERN, Geneva, 2020.

\bibitem{Badea:2019vey}
Anthony Badea, Austin Baty, Paoti Chang, Gian~Michele Innocenti, Marcello
  Maggi, Christopher Mcginn, Michael Peters, Tzu-An Sheng, Jesse Thaler, and
  Yen-Jie Lee.
\newblock {\em Measurements of two-particle correlations in $e^+e^-$ collisions at
  91 GeV with ALEPH archived data}.
\newblock {\em Phys. Rev. Lett.}, 123:212002, 2019.

\bibitem{Baty:2021ugw}
Austin Baty, Parker Gardner, and Wei Li.
\newblock {\em Collective evolution of a parton in the vacuum: the ultimate
  partonic ''droplet'', non-perturbative QCD and quantum entanglement}.
\newblock   {arXiv:2104.11735 [hep-ph]}, 2021.

\bibitem{Sjostrand:2014zea}
Torbj\"orn Sj\"ostrand, Stefan Ask, Jesper~R. Christiansen, Richard Corke,
  Nishita Desai, Philip Ilten, Stephen Mrenna, Stefan Prestel, Christine~O.
  Rasmussen, and Peter~Z. Skands.
\newblock {\em An introduction to PYTHIA 8.2}.
\newblock {\em Comput. Phys. Commun.}, 191:159--177, 2015.

\bibitem{NNPDF:2017mvq}
Richard~D. Ball et~al. (NNPDF)
\newblock {\em Parton distributions from high-precision collider data}.
\newblock {\em Eur. Phys. J. C}, 77(10):663, 2017.

\bibitem{H1:2012qti}
F.~D. Aaron et~al. (H1)
\newblock {\em Inclusive Deep Inelastic Scattering at High $Q^2$ with
  Longitudinally Polarised Lepton Beams at HERA}.
\newblock {\em JHEP}, 09:061, 2012.

\bibitem{ZEUS:2012zcp}
H.~Abramowicz et~al. (ZEUS Collaboration)
\newblock {\em Measurement of high-Q2 neutral current deep inelastic $e^+p$
  scattering cross sections with a longitudinally polarized positron beam at
  HERA}.
\newblock {\em Phys. Rev. D}, 87(5):052014, 2013.

\bibitem{ZEUS:2009W}
H.~Abramowicz et~al. (ZEUS Collaboration)
\newblock {\em Search for events with an isolated lepton and missing transverse
  momentum and a measurement of {W} production at HERA.}
\newblock {\em Physics Letters B}, 672(2):106–115, Feb 2009.

\bibitem{H1:2009W}
F.~D. Aaron et~al. (H1 Collaboration)
\newblock {\em Events with isolated leptons and missing transverse momentum
  and measurement of {W} production at HERA.}
\newblock {\em The European Physical Journal C}, 64(2):251–271, Oct 2009.

\bibitem{H1:ZEUS:2010W}
F.~D. Aaron et~al. (H1 and ZEUS Collaborations)
\newblock {\em Events with an isolated lepton and missing transverse momentum and
  measurement of {W} production at HERA.}
\newblock {\em Journal of High Energy Physics}, 2010(3), Mar 2010.

\bibitem{GFitter:2018}
J.~Haller, A.~Hoecker, R.~Kogler, K.~Mönig, T.~Peiffer, and J.~Stelzer.
\newblock {\em Update of the global electroweak fit and constraints on
  two-higgs-doublet models.}
\newblock {\em The European Physical Journal C}, 78(8), Aug 2018.

\bibitem{Madgraph}
J.~Alwall, R.~Frederix, S.~Frixione, V.~Hirschi, F.~Maltoni, O.~Mattelaer,
  H.-S. Shao, T.~Stelzer, P.~Torrielli, and M.~Zaro.
\newblock {\em The automated computation of tree-level and next-to-leading order
  differential cross sections, and their matching to parton shower simulations.}
\newblock {\em JHEP}, 2014(7), Jul 2014.

\bibitem{Butterworth:2059563}
Jon Butterworth, Stefano Carrazza, Amanda Cooper-Sarkar, Albert De~Roeck, Joel
  Feltesse, Stefano Forte, Jun Gao, Sasha Glazov, Joey Huston, Zahari Kassabov,
  Ronan McNulty, Andreas Morsch, Pavel Nadolsky, Voica Radescu, Juan Rojo, and
  Robert Thorne.
\newblock {\em PDF4LHC recommendations for LHC Run II}.
\newblock {\em J. Phys. G}, 43:023001. 65 p, Oct 2015.

\bibitem{Costantini:2020stv}
Antonio Costantini, Federico De~Lillo, Fabio Maltoni, Luca Mantani, Olivier
  Mattelaer, Richard Ruiz, and Xiaoran Zhao.
\newblock {\em Vector boson fusion at multi-TeV muon colliders}.
\newblock {\em JHEP}, 09:080, 2020.

\bibitem{atlas2018hbb}
M.~Aaboud et~al. (ATLAS Collaboration)
\newblock {\em Observation of $h\to b\overline{b}$ decays and $vh$ production with the {ATLAS}
  detector.}
\newblock {\em Physics Letters B}, 786:59--86, Nov 2018.

\bibitem{cms2018hbb}
  A. M.Sirunyan et~al. (CMS Collaboration)
  \newblock {\em Observation of higgs boson decay to bottom quarks.}
\newblock {\em Physical Review Letters}, 121(12), Sep 2018.

\bibitem{delphes}
J.~de~Favereau, C.~Delaere, P.~Demin, A.~Giammanco, V.~Lemaître, A.~Mertens,
  and M.~Selvaggi.
\newblock {\em Delphes 3: a modular framework for fast simulation of a generic
  collider experiment.}
\newblock {\em JHEP}, 2014(2), Feb 2014.

\bibitem{Cepeda:2650162}
M.~Cepeda et~al.
\newblock {\em Report from Working Group 2: Higgs Physics at the HL-LHC and
  HE-LHC}.
\newblock {\em CERN Yellow Rep. Monogr.}, 7:221--584. 364 p, Dec 2018.

\bibitem{lhcblu} R.~Aaij et al. (LHCb)
\newblock {\em Test of lepton universality in beauty-quark decays.}
\newblock {arXiv:2103.11769 [hep-ex]}, 2021.

\bibitem{g-2}
{Muon $g-2$~Collaboration}.
\newblock {\em Final report of the e821 muon anomalous magnetic moment measurement
  at BNL}
\newblock {\em Phys. Rev. D}, 73:072003, Apr 2006.

\bibitem{LQBox}
Ilja Dorsner and Admir Greljo.
\newblock {\em Leptoquark toolbox for precision collider studies}.
\newblock {arXiv:1801.07641 [hep-ph]}, 2018.

\bibitem{Sirunyan_2017}
A.M. Sirunyan et~al. (CMS Collaboration)
\newblock {\em Particle-flow reconstruction and global event description with the
  {CMS} detector.}
\newblock {\em Journal of Instrumentation}, 12(10):P10003--P10003, oct 2017.

\bibitem{Sirunyan_2021}
A.M. Sirunyan et~al. (CMS Collaboration)
\newblock {\em Electron and photon reconstruction and identification with the {CMS}}
  experiment at the {CERN} {LHC}.
\newblock {\em Journal of Instrumentation}, 16(05):P05014, may 2021.

\bibitem{Cacciari_2008}
Matteo Cacciari, Gavin~P Salam, and Gregory Soyez.
\newblock {\em The anti-ktjet clustering algorithm.}
\newblock {\em Journal of High Energy Physics}, 2008(04):063--063, apr 2008.

\bibitem{fastjet2012}
Matteo Cacciari, Gavin~P. Salam, and Gregory Soyez.
\newblock {\em Fastjet user manual.}
\newblock {\em The European Physical Journal C}, 72(3), Mar 2012.

\bibitem{Sirunyan_2018}
A.M. Sirunyan et~al. (CMS Collaboration)
\newblock {\em Identification of heavy-flavour jets with the {CMS} detector in pp
  collisions at 13 {TeV}.}
\newblock {\em Journal of Instrumentation}, 13(05):P05011--P05011, may 2018.


  

\end{thebibliography}


\end{document}